\documentclass[%
 reprint,
 amsmath,amssymb,
 aps,
]{revtex4-2}

\usepackage{graphicx}
\usepackage{dcolumn}
\usepackage{bm}

\begin{document}

\preprint{APS/123-QED}

\title{Spin Elasticity}

\author{Zhong-Chen Gao }
 \email{Corresponding author: gaozc0129@shnu.edu.cn }
\affiliation{Mathematics and Science College, Shanghai Normal University, Shanghai 200234, China }%

\author{Tianyi Zhang}%
\affiliation{Beijing National Laboratory for Condensed Matter Physics, University of Chinese Academy of Sciences, Chinese Academy of Sciences, Beijing 100190, China }%
\author{Feifei Wang }
\affiliation{Key Laboratory of Optoelectronic Material and Device, Department of Physics, Shanghai Normal University, Shanghai 200234, China }%
\author{Jingguo Hu }
\affiliation{School of Physical Science and Technology (School of Integrated Circuits), Yangzhou University, Yangzhou 225002, China }%
\author{Peng Yan }
\affiliation{School of Physics and State Key Laboratory of Electronic Thin Films and Integrated Devices, University of Electronic Science and Technology of China, Chengdu 611731, China }%
\author{Xiufeng Han }
\affiliation{Beijing National Laboratory for Condensed Matter Physics, University of Chinese Academy of Sciences, Chinese Academy of Sciences, Beijing 100190, China }%
\affiliation{Center of Materials Science and Optoelectronics Engineering, University of Chinese Academy of Sciences, Beijing 100049, China }%
\affiliation{Songshan Lake Materials Laboratory, Dongguan, Guangdong 523808, China }%

\date{\today}

\begin{abstract}
Elasticity has long been regarded as a property exclusive to material media. Here we uncover its hidden existence in the spin degree of freedom. We introduce \textit{spin elasticity}—an intrinsic mechanism that governs recoverable deformation of spin morphology. This discovery reveals a previously unrecognized universality: elasticity operates in both matter and spin spaces, underpinning structural integrity across physical realms. By establishing the missing spin counterpart, this work completes the elastic picture and points toward a broader paradigm where elasticity transcends its conventional boundaries.  
\begin{description}

\item[Subject area]
Spintronics, Condensed Matter Physics 
\end{description}
\end{abstract}

\maketitle


\section{\label{sec:level1}INTRODUCTION}

Without elasticity, the physical world as we know it would cease to exist. It is elasticity that underpins the stability and structure of all matter around us. More than that, the trajectory of human civilization itself is deeply intertwined with the mastery of this property. From the taut bowstrings of the Pleistocene [1] and the spring-dampened chariots of Tutankhamun [2] to the bronze alloy springs powering water clocks and catapults in Ctesibius’s Alexandria [3], empirically designed elastic components had already become integral to daily life millennia ago. A pivotal turning point came with the formulation of Hooke’s law—Ut tensio, sic vis—in the late 17th century [4], followed by the formal establishment of elasticity theory by Cauchy, Saint-Venant, and others in the 19th century [5,6]. These advances propelled the Industrial Revolution, providing critical theoretical foundations for precision engineering innovations such as the high-pressure steam engine [7] and the Brooklyn bridge [8]. Today, elastic analysis and design permeate every scale of human endeavor—from the soles of our shoes to the aircraft overhead and the Voyager probe now traversing interstellar space. 

At its core, the canonical theory of elasticity describes bodies—whether metals with atomic lattices or rubbers with polymer chains—structured by charge and mass. In this framework, elasticity originates from the spatial arrangement of massive particles governed by charge-mediated intermolecular electromagnetic forces, which share two distinct characteristics: (1) they are attractive at large distances, decaying to zero at infinity; and (2) they are strongly repulsive at short range. Linear elasticity, in turn, emerges as a valid approximation only for small displacements. Spin, the third fundamental attribute of particles alongside charge and mass, has remained conspicuously absent from this narrative. 

Could elasticity, then, manifest in the spin degree of freedom?

\section{\label{sec:level2}SPIN ELASTOMER}

Here, we propose the concept of “spin elastomer” as a general designation for any spin texture capable of transfiguring under an applied load and recovering its initial configuration upon removal. In this picture, the fundamental entities are spin states instead of particles. Accordingly, spin torque—rather than force—serves as the fundamental load, driving collective rotations of local spins that manifest as textural deformations. 

\section{\label{sec:level3}ASSEMBLY PRINCIPLE}

Creating a spin elastomer requires two key features from its constituent unit textures: (i) the unit texture must itself be locally elastic, and (ii) it must be capable of shifting to accommodate macroscopic shape changes. Far from rare, locally elastic textures are commonplace in magnetic systems. They typically emerge near inhomogeneities—such as defects, impurities, and interfaces [9]—where local symmetry breaking reconstructs the magnetic energy landscape [10]. While present, they invariably remain trapped, precluding their use as mobile building blocks. In homogeneous media, as the ground state opposes any deviation from uniform magnetization (considering, for simplicity, a ferromagnetic system), one can only rely on metastable solitons—such as domain walls, vortices, skyrmions [11] —which are mobile and topologically protected, endowing them with an inherent resilience to perturbations and distortions. The units then need be closely packed without annihilation—a condition governed by magnetization homotopy. In the simplest case of a Heisenberg spin chain, due to the dominant exchange interaction, system forbids parallel alignment except within domains. To preserve any noncollinearity, spins must therefore rotate continuously along the chain while avoiding identical orientation. Mathematically, this corresponds to a winding trajectory in the order parameter space \ $ \mathbb{V} = \mathbb{S}^2 \ $ that neither self-intersects nor forms a closed loop. The situation becomes more permissive when additional interactions—such as dipole-dipole, Zeeman, or magnetocrystalline couplings—are taken into account, as their competition can topologically pin the order parameters of a loop trajectory. \textbf{Fig.1} illustrates a simple realization: a 360° domain wall stabilizing within a narrow nanostrip in which the strong geometrical anisotropy confines magnetization to the plane and a continuous in-plane spin rotation (winding number  $Q=1$) leads to topological pinning of the order parameters along a great circle in the \textit{xy}-plane (highlighted in red). In contrast, two adjacent 180° walls with the same polarization yield zero net winding, leaving the order parameters topologically unpinned (gray loop, deformable to a point) and ultimately vulnerable to structural annihilation. 

In more complex cases where the solitonic magnetization varies in three dimensions—such as the adjacent vortex walls shown in \textbf{Fig. A.3}—the condition for structural stability becomes more stringent. At least one path in real space must exist along which the corresponding order parameters in $\mathbb{S}^2$ undergo continuous winding, forming a topologically nontrivial trajectory. The remaining magnetization can then be stabilized as an appendage to this topological backbone. 

\begin{figure}
    \centering
    \includegraphics[width=0.5\linewidth]{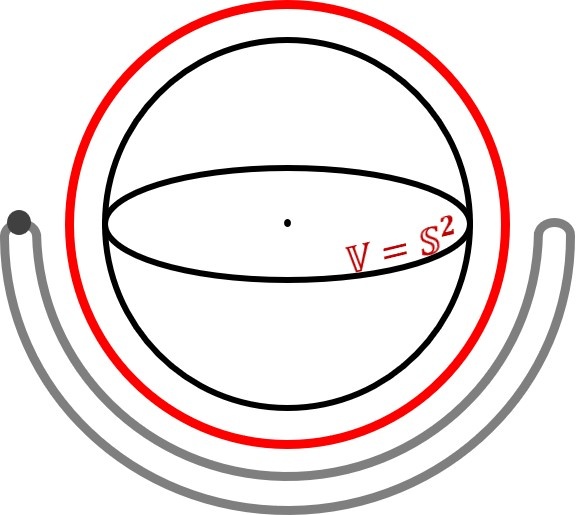}
    \caption{Trajectories in order parameter space $ \mathbb{V} = \mathbb{S}^2 \ $. A 360° wall (red) traces a topologically nontrivial path (winding number $Q=1$) , leading to topological pinning of the order parameters and hence metastability. By contrast, two adjacent 180° walls of the same polarization (gray) follow a trivial trajectory ($Q=0$) ; lacking topological pinning, they are unstable. }
    \label{fig:placeholder}
\end{figure}

\section{\label{sec:level4}\textbf{INTERSOLITONIC INTERACTION DIAGRAM}
}
Elasticity in a spin elastomer hinges on interactions between its constituent units that engender properties (1) and (2). In \textbf{Fig. 2}, we present the first comprehensive interaction diagram of magnetic solitons—exemplified by two 180° walls—in measure of Oersted field. Without topological protection, the interaction is purely magnetostatic: an attraction that decays with separation. With topological protection, by contrast, a strong repulsion emerges, competing with the magnetostatic attraction. Their competition yields an interaction curve akin to that of atomic potentials, thereby enabling spin elasticity. The pronounced scalability of the inter-solitonic separation over a moderate field range further points to robust elastic performance and facile tunability in future engineering.

\begin{figure}
    \centering
    \includegraphics[width=1\linewidth]{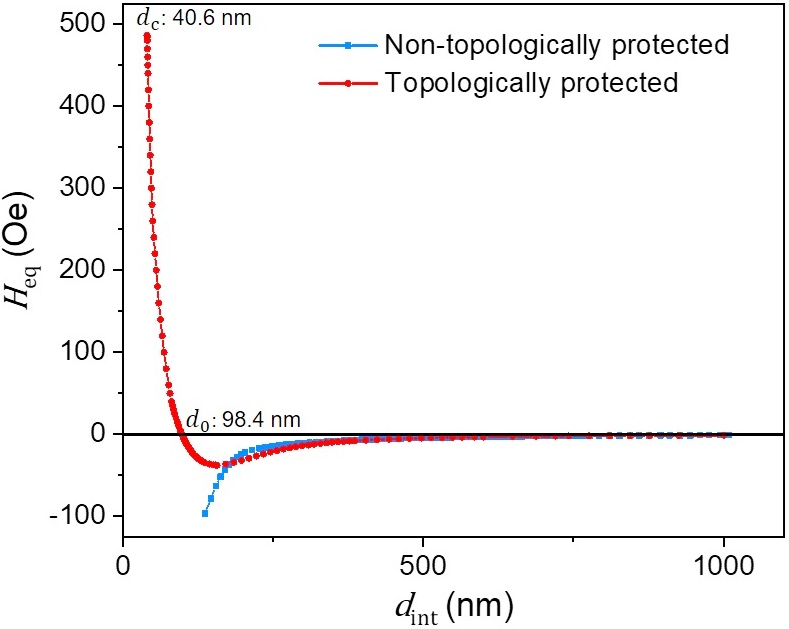}
    \caption{Intersolitonic interaction curves for two 180° walls (magnetic setups detailed in \textbf{Fig. A.4}). The red (blue) curve corresponds to walls with opposite (same) polarization, i.e., the topologically protected (non-protected) case. The equivalent field $H_{\text{eq}}$ is defined as the negative of the externally applied field at which the system equilibrates at a given separation $d_{\text{int}}$. The critical separation $d_{\text{c}}$ marks the onset of annihilation; the equilibrium separation $d_{\text{0}}$ denotes the energy minimum. }

    \label{fig:placeholder}
\end{figure}

Among various types of spin elastomer, those whose geometry varies in only one dimension merit particular attention. We term this 1D realization a \textit{spin spring} (SS). The study of spin springs is motivated by two considerations: (i) their potential as basic elements in future spin-based technologies, and (ii) their role as a conceptual gateway to the broader field of spin elasticity—linking spin torque to the geometry of spin elastomers. \textbf{Fig. 3} presents a concrete example: a spin spring assembled from a pair of domain walls, denoted $T_{t\downarrow}T_{h\uparrow}$ (hereafter referred to as a DWSS; see \textbf{Appendix A} for assembly principles, fabrication, characterization, and advantages). In this configuration, spins are nearly uniformly oriented along the transverse direction, closely approximating a one-dimensional spin chain. The magnetic interactions involved are limited to the essential exchange and dipole-dipole coupling. Given this simplicity, the DWSS will serve as the primary model system throughout the remainder of this paper to elucidate the signature properties of spin elastomers. 

\begin{figure}
    \centering
    \includegraphics[width=0.9\linewidth]{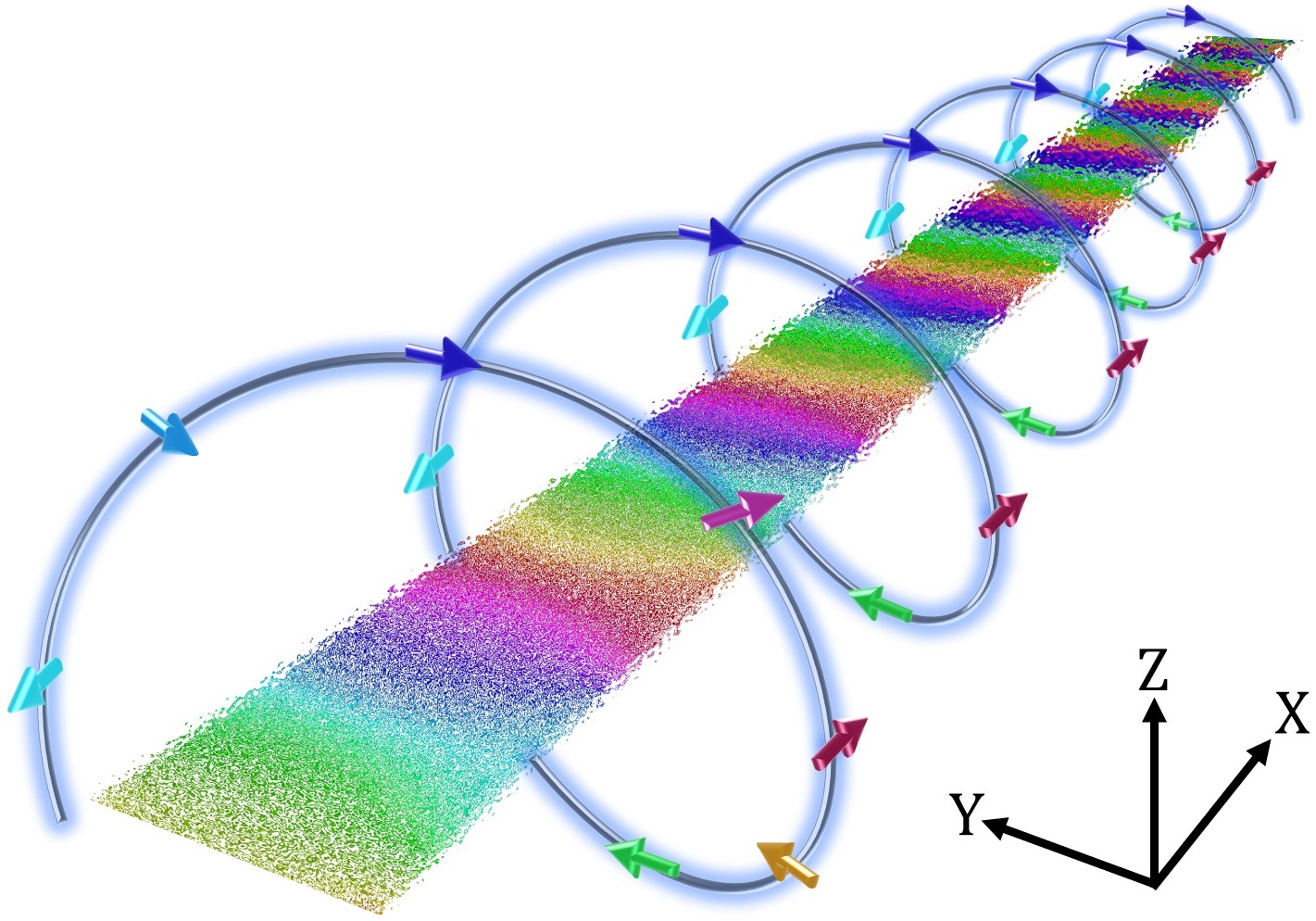}
    \caption{Schematic of a domain wall spin spring (DWSS) comprising domain wall pair $T_{t\downarrow}T_{h\uparrow}$. The local spin orientations along the longitudinal direction are mapped onto a figurative helix for visual clarity. }
    \label{fig:placeholder}
\end{figure}

\section{\label{sec:level5}\textbf{UT TENSIO, SIC $\textbf{T}_{\text{st}}$}
}
A defining characteristic of any elastic body is its response to tensile loading—a ubiquitous condition in engineering practice. We begin by examining a DWSS with \(N_{\text{DW}} =100\) (hereafter $\text{DWSS}_{\text{100}}$) subjected to a steady spin torque $\textbf{T}_{\text{st}}$ applied at its two ends. Geometric confinement dictates that its axial response depends on the sense of spin rotation. For the counter-clockwise in-plane rotation inherent to our $T_{t\downarrow}T_{h\uparrow}$ DWSS , an out-of-plane spin torque stabilizes contraction when directed outward and elongation when directed inward (\textbf{Fig. 4(a)}). The opposite holds for clockwise rotation. \textbf{Fig. 4(b)} plots the restoring spin torque $T_{\text{rest}}$ against the spring length $L$. A robust linear relationship emerges over a wide range of elongation and contraction, revealing a Hooke’s law in the spin degree of freedom: 
\begin{equation}
T_{\text{rest}} =-k\Delta L,  
\end{equation}
where $\Delta L$ denotes the change in spring length. The effective spring constant $k$ is negative for counter-clockwise spin rotation and positive for clockwise rotation. This relation lays a solid foundation for spin-based elasticity theory—charting a pathway to spin elastic devices and spin engineering. Notably, \textbf{Eq. (1)} operates without requiring collinearity between $T_{\text{rest}}$ and $\Delta L$; the restoring torque does not directly oppose the length change. Instead, the underlying mechanism involves minute out-of-plane tilting of spins owing to torque imbalance, which in turn drives in-plane texture reconstruction (\textbf{Fig. B.1}). 

\begin{figure*}
    \centering
    \includegraphics[width=0.7\linewidth]{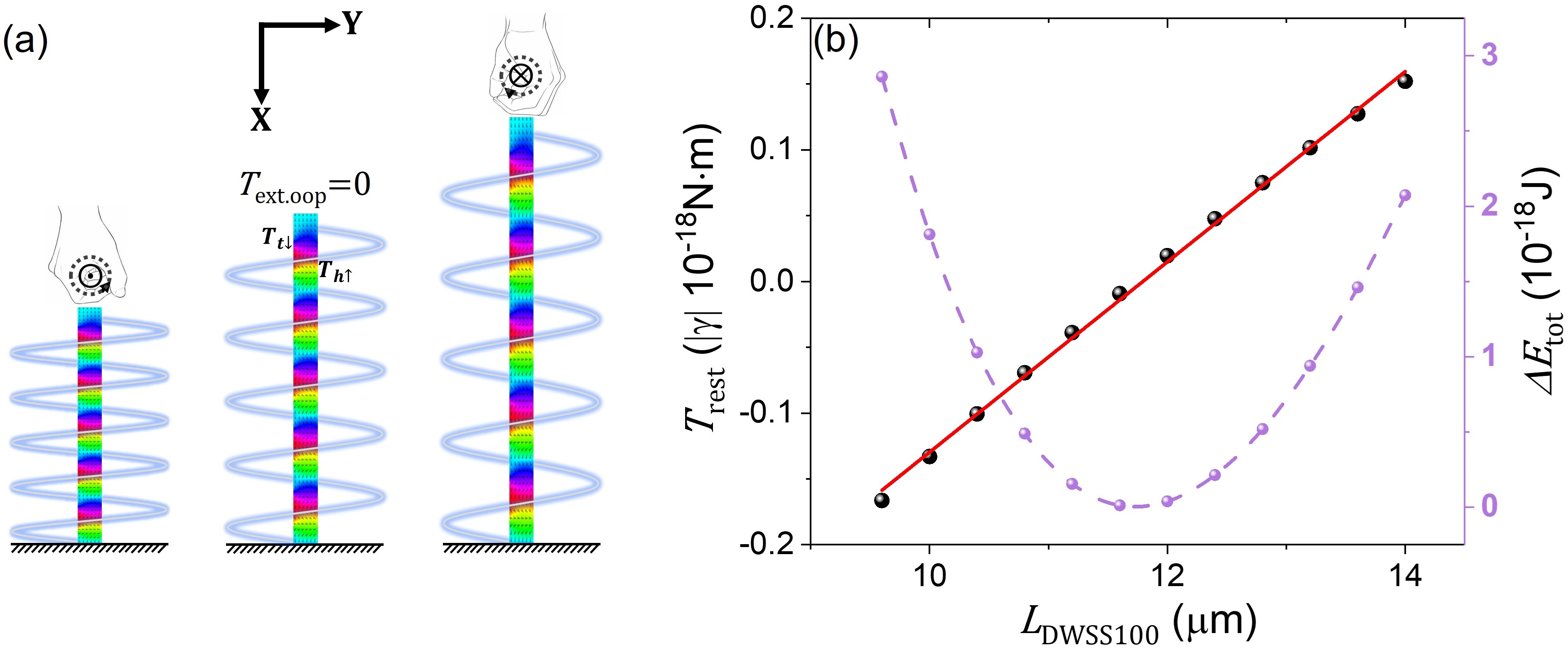}
    \caption{(a) Elongation and contraction of a DWSS under an external out-of-plane spin torque $T_{\text{ext.oop}}$. (b) Restoring torque $T_{\text{rest}}$ (red curve) and total energy gain $\Delta E_{\text{tot}}$ (blue curve) as functions of  $\text{DWSS}_{\text{100}}$ length $L_{\text{DWSS100}}$. The equilibrium length of the unperturbed $\text{DWSS}_{\text{100}}$ is \(L_0 =11.73\,\mathrm{\mu m} \).}
    \label{fig:placeholder}
\end{figure*}

\section{\label{sec:level6}\textbf{SPIN ELASTIC POTENTIAL ENERGY} }

The deformation of a DWSS—and, by extension, spin elastomers in general—also enables energy storage, which we term \textit{spin elastic potential energy}. As shown in \textbf{Fig. 4(b)}, both compression and elongation store energy in the system. The underlying mechanism can be understood as follows. In a compressed (or stretched) DWSS without rigid boundary constraints, the imbalance of spin torque at the boundaries induces minute out-of-plane tilting of the spins. In terms of precessional torque $- \lvert \gamma \rvert \textbf{m} \times \textbf{H}$, the strong in-plane effective field thereby drives counter-clockwise (or clockwise) spin rotation, leading to expansion (or contraction) of the DWSS. During this relaxation, the external torque $\textbf{T}_{\text{ext.oop}}$ does negative work (\(\delta \mathcal{W} = -\frac{\textbf{T}_{\text{ext.oop}}}{\lvert \gamma \rvert } \cdot (\textbf{m} \times d\textbf{m})\)), releasing energy. Conversely, deforming the DWSS requires an opposing in-plane torque, with $T_{\text{ext.oop}}$ doing positive work and storing energy. Quantitatively, the work done by the boundary torque is 
\begin{equation}
\mathcal{W} =\int -\frac{1}{\lvert \gamma \rvert }  \textbf{T}_{\text{ext.oop}} \cdot d \bm{\theta} \approx -\frac{k}{2\lvert \gamma \rvert}(\partial_x \theta|_{\text{bound}}) (\Delta L)^2,  
\end{equation}
where \(\partial_x \theta|_{\text{bound}} \approx \partial_x \theta|_{\text{eq.bound}}=0.0377\,\mathrm{rad/nm}\) is the spatial derivative of in-plane spin orientation $\theta$ at the DW boundary (\textbf{Fig. A.5(a)}), consistent with the nearly parabolic energy curve observed in \textbf{Fig. 4(b)}. The slight deviation from a perfect parabola can be attributed to the expected fact that in compressed state \(\partial_x \theta|_{\text{bound}} \ge 0.0377\,\mathrm{rad/nm}\) and in stretched state \(\partial_x \theta|_{\text{bound}} \le 0.0377\,\mathrm{rad/nm}\).

The stored energy has different physical origins in the two regimes. In compression, energy resides in the exchange interaction: the exchange energy decreases with DWSS length (\textbf{Fig. 5}), as expected from the volumetric exchange energy density \(E_{\text{exch.den}} =\frac{A}{2}(\partial_x \theta)^2\) where \(A \sim J/a\) is the order-parameter stiffness [12]. In elongation, energy is stored in the dipole-dipole interaction: stretching separates the opposite magnetic charges of the two domain walls, raising the demagnetization energy. 

In the absence of external torque, the stored energy can eventually be released to produce useful work in a spintronic circuit by spin-motive forces reciprocal to the spin torque [12, 13]. Because of the nonvolatility and the endurance of the magnetic system, the spin elastic potential energy essentially last indefinitely, which goes beyond traditional chemical battery technology. 

\begin{figure}
    \centering
    \includegraphics[width=0.9\linewidth]{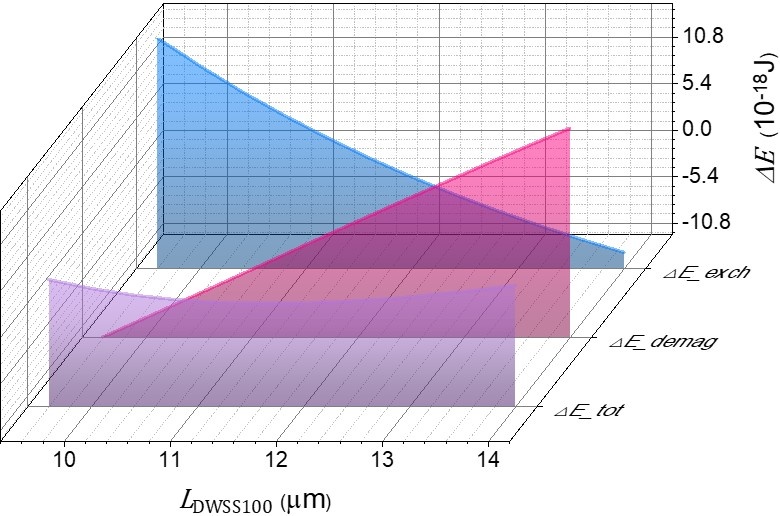}
    \caption{Exchange energy change $\Delta E_{\text{exch}}$}, demagnetization energy change $\Delta E_{\text{demag}}$, and total energy change $\Delta E_{\text{tot}}$ as functions of $\text{DWSS}_{\text{100}}$ length $L_{\text{DWSS100}}$.
    \label{fig:placeholder}
\end{figure}

\section{\label{sec:level7}\textbf{TOPOLOGICAL ELASTICITY} }

Owing to the topology and mobility of magnetic solitons, spin elastomers are endowed with an inherent capacity to adapt their morphology as they traverse different magnetic elements. \textbf{Fig. 6(a)} illustrates this adaptability, showing a DWSS stabilized in nanostrips of varying width and curvature. Despite these geometric variations, correlation between the restoring spin torque and the DWSS length persists (\textbf{Figs. 6(b)} and \textbf{6(c)}). In particular, the effective stiffness $k$ of the $\text{DWSS}_{\text{100}}$ remains nearly unchanged with nanostrip width. Nevertheless, the spin torque per unit area required to achieve a given elongation ratio decreases with increasing width (\textbf{Fig. 6(d)}), indicating a width-dependent elastic modulus $E_{\text{Y}}$. For DWSS in curved nanostrips, a simple Hooke's law no longer suffices. Compression follows approximately the same behavior as in straight nanostrips; however, under expansion, the stiffness $k$ (and hence $E_{\text{Y}}$) exhibits a sharp increase (\textbf{Fig. 6(c)}), which we attribute to the formation of a potential well that traps the domain walls in regions of finite curvature [14]. These observations point to a generalized, topology-aware Hooke's law: 
\begin{equation}
\Delta \Gamma  =T_{\text{ext}}\int_{0}^{\Gamma}\frac{\xi_{r}}{S_{r}E_{Y}}\, dr,  
\end{equation}
where $\Delta \Gamma$ is the axial length change along a curvilinear coordinate $\Gamma$, $T_{\text{ext}}$ is the external spin torque applied at the boundary, and $S_{r}$, $\xi_{r}$ and $E_{\text{Y}}$ denote, respectively, the cross-sectional area, spin-torque transfer coefficient, and elastic modulus at position $r$. Two remarks are in order: 

$\bullet$ The axis $\Gamma$ can assume an arbitrary curvilinear form. 

$\bullet$ The parameters  $S_{r}$, $\xi_{r}$ and $E_{\text{Y}}$ are not constant but vary with the local topological deformation. 

\begin{figure*}
    \centering
    \includegraphics[width=0.8\linewidth]{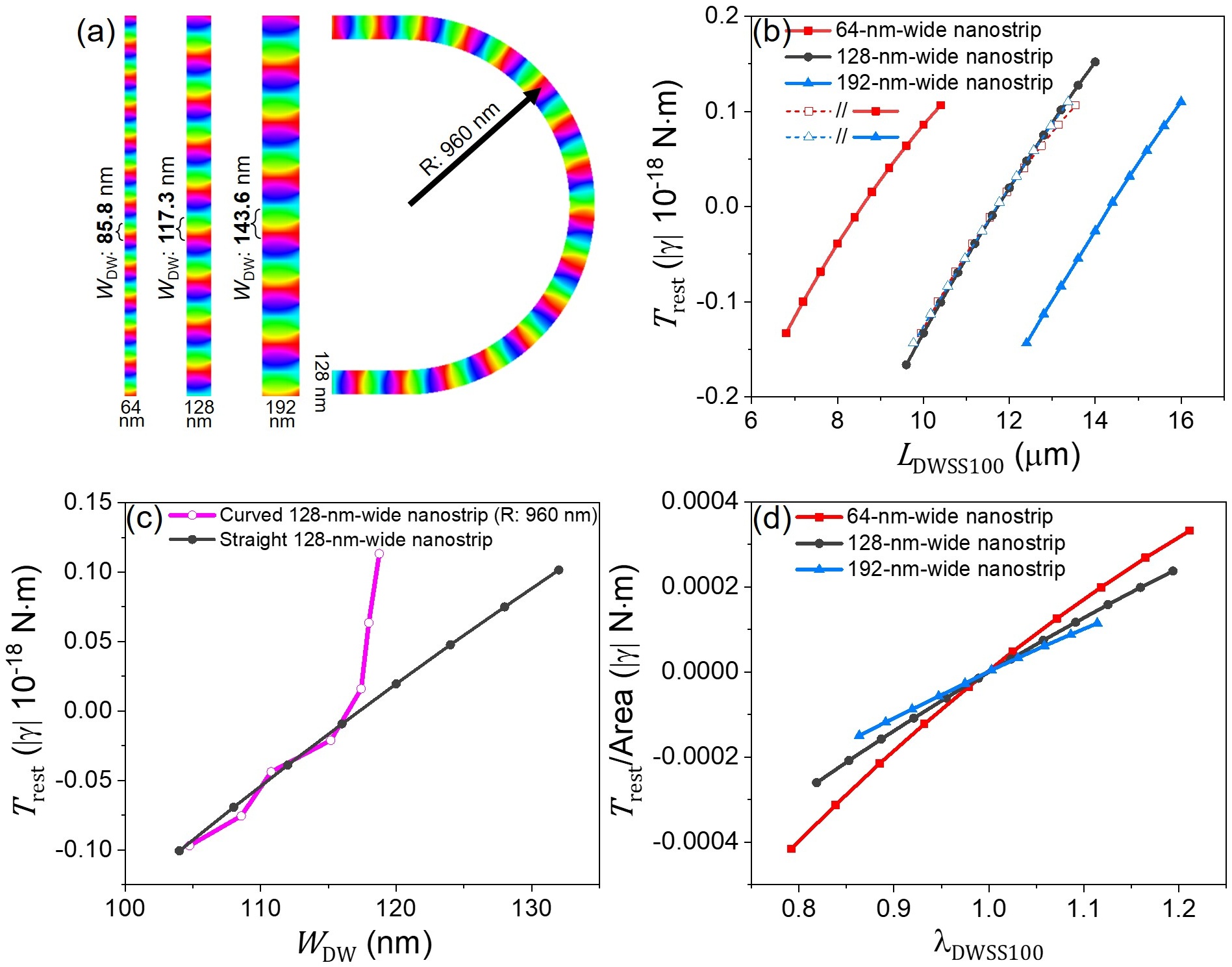}
    \caption{(a) Schematics of DWSS in nanostrips of different geometries. In straight nanostrips of width 64 nm, 128 nm and 192 nm, the DW width $W_{\text{DW}}$ is 85.8 nm, 117.3 nm and 143.6 nm, respectively. In a curved nanostrip (radius 960 nm), the DWs exhibit radially nonuniform deformation. (b) Restoring torque $T_{\text{rest}}$ versus $\text{DWSS}_{\text{100}}$ length $L_{\text{DWSS100}}$ for three straight nanostrips. (c) Restoring torque $T_{\text{rest}}$ versus DW width $W_{\text{DW}}$ for straight and curved nanostrips of 128 nm. (d) Restoring torque per unit area versus elongation ratio $\lambda _{\text{DWSS100}}$ for three straight nanostrips.}
    \label{fig:placeholder}
\end{figure*}

\section{\label{sec:level8}\textbf{POISSON EFFECT} }

For more general 2D and 3D spin elastomers, beyond axial deformation, a further concern in linear statics lies in the lateral response under load. Here we report that DWSS exhibits a phenomenon analogous to the conventional Poisson effect—strictly speaking, a DWSS in a nanostrip is not truly one-dimensional; a noncollinear texture spontaneously develops in the transverse direction, rendering the domain walls slightly spindle-shaped to minimize demagnetization energy [15] (see \textbf{Fig. 6(a)}). Under longitudinal elongation (compression), we observe lateral contraction (expansion) of the domain-wall cores within the DWSS (\textbf{Fig. 7(a)}). In contrast to conventional materials, the effective Poisson’s ratio of the domain-wall cores, defined as \(\nu _{\text{DW core}} =-\frac{\delta_y}{\delta_x}\) (where $\delta_y$ and $\delta_x$ are the percentage elongations in the transverse and longitudinal directions, respectively), is not constant. As \textbf{Fig. 7(b)} shows, $\delta_y$ as a function of $\delta_x$ fluctuates, indicating that the Poisson's ratio $\nu_{\bm{\mu}}$ at a given state $\bm{\mu}$ varies with deformation. These observations can be understood via spin stress-strain analysis as introduced in the following section. Notably, the Poisson's ratio in spin elastomers may exceed the conventional upper bound of 0.5. Two factors can account for this departure: (i) spin elastomers are often two-dimensional, and (ii) as topological assemblies, they are not constrained by the incompressibility hypothesis. 

\begin{figure*}
    \centering
    \includegraphics[width=0.8\linewidth]{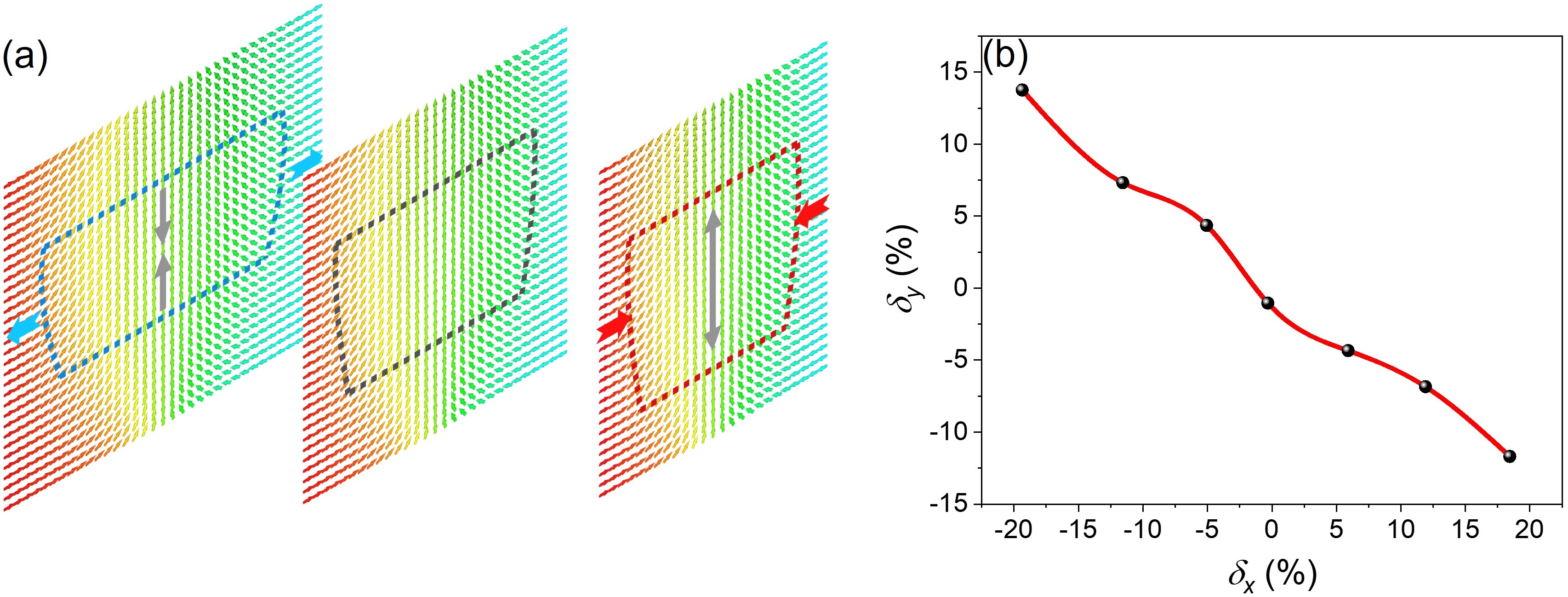}
    \caption{(a) Schematic of the Poisson effect in a DWSS. Upon longitudinal elongation (compression), the domain-wall core region—\(\{(\mu_x,y)|\mu_x\in [\frac{\sqrt{3}}{2}, -\frac{\sqrt{3}}{2}], y\in[34 \,\mathrm{nm}, 94 \,\mathrm{nm}]\} =100[\frac{\sqrt{3}}{2}, -\frac{\sqrt{3}}{2}]\times[34 \,\mathrm{nm}, 94 \,\mathrm{nm}]\)enclosed by dashed lines—exhibits clear lateral contraction (expansion). In this illustration, the upper edge of the core is held fixed, and the longitudinal scaling is assumed uniform across the core. (b) \textit{y}-direction percentage elongation $\delta_y$ versus \textit{x}-direction percentage elongation $\delta_x$ for the domain-wall core. Here \(\delta_y =(\frac{\Delta y_{\text{lower edge}}}{94-34})\times100\%, \delta_x=(L_{\text{upper edge,stressed}}/L_{\text{upper edge,relaxed}}-1)\times100\%\).}
    \label{fig:placeholder}
\end{figure*}

\section{\label{sec:level9}\textbf{SPIN ELASTIC THEORY} }

To fully characterize the elasticity of a spin elastomer, one may adopt the classical continuum approach by introducing the notions of \textit{strain} and \textit{stress}. At the mesoscale, magnetization can be treated as a differentiable continuous vector field, providing a basis for defining strain. By assigning spatial coordinates to the spin states within the elastomer and tracking their positions before and after loading, one can obtain a spin-deformation gradient tensor $\textbf{F}$ or $\textbf{F}_{\text{spin}}$,  whose components are given by \(F_{ij} =\frac{\partial x_i}{\partial X_j}\) where $X$ denotes the coordinate of a spin state in the initial (Lagrangian) configuration and $x$ its coordinate after deformation (Eulerian configuration). Crucially, only spin states residing in noncollinear textures need to be considered in this mapping. Within a connected domain of uniform magnetization, spin states cannot be assigned definite coordinates. Since each spin state $\bm{\mu}$ is parametrized by two angular coordinates on the order-parameter space $\mathbb{S}^2$,  full spatial coordinatization is achievable only for solitons of dimension up to two. This includes objects such as skyrmions, hedgehogs, and merons [11]. For an assembly of solitons—e.g., a skyrmion crystal [16]—full coordinatization remains possible despite the presence of identical spin states, as they are segregated into distinct topological entities. In contrast, three-dimensional textures such as Hopfions [17] and skyrmion bundles [18] admit only partial coordinatization, as they inevitably contain connected domains of uniform magnetization. To describe the spatial variation of the spin field and map infinitesimal relative changes in spin orientation, we adopt the displacement gradient tensor \(\nabla \textbf{u}_{\bm{\mu}} =\textbf{F}_{\text{spin}}-\textbf{I}\), \(d  \textbf{u}_{\bm{\mu}} =\nabla  \textbf{u}_{\bm{\mu}} d\textbf{X}\) (where \(\textbf{u}_{\bm{\mu}} =x-X\) is the displacement of a specific spin state and $\textbf{I}$ the identity matrix) from classical continuum mechanics as our strain measure for spin elastomers—hereafter termed \textit{spin strain} and denoted by $\textbf{D}_{\text{spin}}$ or $\textbf{D}$. Adoption of the displacement gradient over the conventional Green–Lagrange strain tensor \(\bm{\varepsilon} =\frac{1}{2}(\textbf{F}^{\text{T}}\textbf{F}-\textbf{I})\) for defining spin strain rests on the following grounds: 

$\bullet$ The Green–Lagrange strain is a second-order tensor designed to measure changes in length and angle of material line elements—a natural choice for describing lattice deformation. In contrast, spin texture distortion involves no alteration in lattice spacing; it consists solely of relative variations in spin orientation between neighboring points, a quantity directly encoded by the spin-gradient tensor $\nabla \textbf{m}$.

$\bullet$ Spatial variation of spin orientation includes both symmetric (strain-like deformation) and antisymmetric (spin rotation) components, which does not need to satisfy the frame indifference condition required for classical mechanical strain. 

$\bullet$ The displacement gradient \(\nabla \textbf{u} =\textbf{F}-\textbf{I}\) (via \(d  \textbf{u} =\nabla  \textbf{u} d\textbf{X}\)) is perfectly isomorphic to the spin-gradient tensor $\nabla \textbf{m}$ (via \(d  \textbf{m} =\nabla  \textbf{m} d\textbf{r}\)). By contrast, the Green–Lagrange strain involves quadratic operations that are superfluous for describing the linear, first-order variation of spin orientation. 

$\bullet$ The spin-gradient tensor—and by extension, the displacement gradient—naturally captures the inhomogeneity of the spin field, which directly governs the magnetic energy contributions (e.g., exchange, anisotropy, Dzyaloshinskii–Moriya interaction) relevant to spin textures.

For a two-dimensional spin texture that admits complete spatial coordinatization, the spin strain tensor reduces to: 
\begin{equation}
\bm{\textbf{D}}  =\begin{bmatrix}
    \frac{\partial u_1}{\partial X_1} & \frac{\partial u_2}{\partial X_1} \\
    \frac{\partial u_1}{\partial X_2} & \frac{\partial u_2}{\partial X_2}
\end{bmatrix},D_{ij} =\frac{\partial u_i}{\partial X_j},  
\end{equation}
where the indices \(i,j =1,2\) correspond to the two spatial dimensions. 

\begin{figure*}
    \centering
    \includegraphics[width=0.7\linewidth]{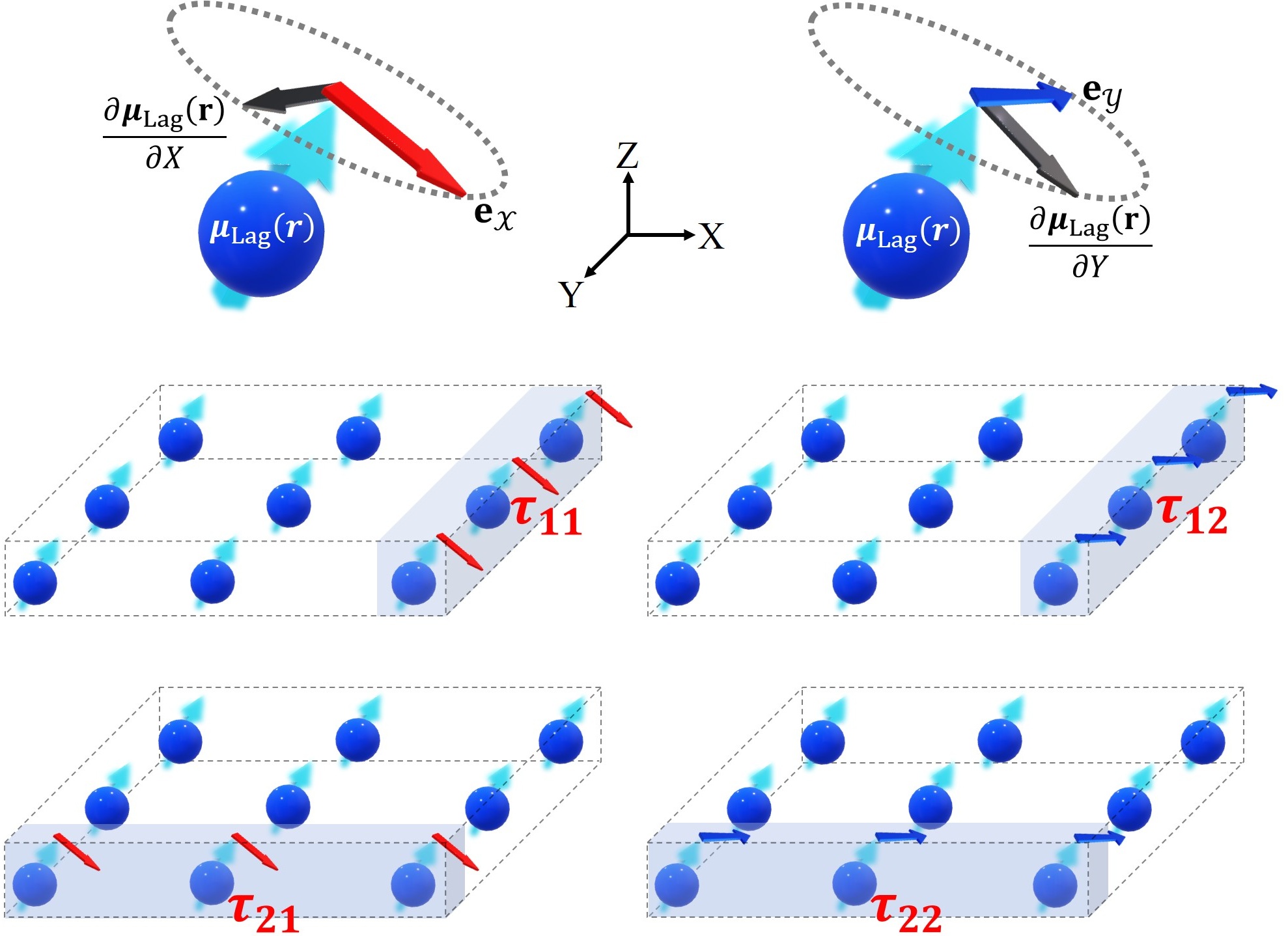}
    \caption{Schematic of spin stress tensor $\tau_{ij}$ where \textit{i} denotes the normal direction of the surface on which the stress acts, and \textit{j} the direction of the stress component. The two basis vectors for decomposing the spin stress are chosen as \(\bm{e}_{\mathcal{X}} =\bm{\mu}_{\text{Lag}}(\bm{r})\times \frac{\partial \bm{\mu}_{\text{Lag}}(\bm{r})}{\partial X}/\left|\frac{\partial \bm{\mu}_{\text{Lag}}(\bm{r})}{\partial X} \right|, \bm{e}_{\mathcal{Y}} =\bm{\mu}_{\text{Lag}}(\bm{r})\times \frac{\partial \bm{\mu}_{\text{Lag}}(\bm{r})}{\partial Y}/\left|\frac{\partial \bm{\mu}_{\text{Lag}}(\bm{r})}{\partial Y} \right|\) where $\bm{\mu}_{\text{Lag}}(\bm{r})$ denotes the spin orientation at position $\bm{r}$ in the Lagrangian (reference) configuration. }
    \label{fig:placeholder}
\end{figure*}

Deformation in a spin elastomer, in turn, signifies a modified system-derived effective field at the boundary, necessitating an external spin torque for stabilization. For the spin torque to be balanced at every point—known as the Brown equation—additional countervailing torques must arise in the interior (hence a “stressed” spin state) relative to the unloaded state, which underpins elastic recovery. To quantify this stressed state, we introduce the \textit{spin stress} $\tau_i$ defined on a surface $S_i$ (with normal along the \textit{i}-axis) as the countervailing spin torque density. This stress can be decomposed along two basis vectors: \(\bm{e}_{\mathcal{X}} =\bm{\mu}_{\text{Lag}}(\bm{r})\times \frac{\partial \bm{\mu}_{\text{Lag}}(\bm{r})}{\partial X}/\left|\frac{\partial \bm{\mu}_{\text{Lag}}(\bm{r})}{\partial X} \right|, \bm{e}_{\mathcal{Y}} =\bm{\mu}_{\text{Lag}}(\bm{r})\times \frac{\partial \bm{\mu}_{\text{Lag}}(\bm{r})}{\partial Y}/\left|\frac{\partial \bm{\mu}_{\text{Lag}}(\bm{r})}{\partial Y} \right|\) (where $\bm{\mu}_{\text{Lag}}(\bm{r})$ denotes the spin orientation in the Lagrangian (reference) configuration), for which we find that on the positive \textit{i}-face of a volume element, a locally enhanced (or reduced) gradient $\frac{\partial \bm{\mu}_{\text{Lag}}(\bm{r})}{\partial j},(j:X,Y)$ gives rise to an increased restoring spin torque along the $\pm \bm{e}_j$ direction. Equivalently, an external spin torque applied along the opposite direction ($\mp \bm{e}_j$) stabilizes compression (or expansion) along the \textit{j}-axis, refer back to \textbf{Fig. 4(a)}. The complete stress state is therefore described by a second-order tensor:
\begin{equation}
\tau_{ij}  =\begin{bmatrix}
    \tau_{11} & \tau_{12} \\
    \tau_{21} & \tau_{22}
\end{bmatrix},  
\end{equation}
where, in analogy with solid mechanics, $\tau_{11}$ and $\tau_{22}$ can be termed \textit{normal spin stresses} and $\tau_{12}$ and $\tau_{21}$ \textit{shear spin stresses}. This tensor pairs naturally with the spin strain tensor $D_{ij}$ defined in \textbf{Eq. (4)}. Each component has its independent magnitude and unique effect on the volume element (\textbf{Fig. 8}). 

With the spin strain $D_{ij}$ and spin stress $\tau_{ij}$ defined, their explicit forms in a general two-dimensional setting can be derived from the deformation of the spin texture. The spin strain tensor follows from the displacement gradient: 
 \begin{subequations}
     \begin{equation}
         \begin{bmatrix}
    D_{11} & D_{12} \\
    D_{21} & D_{22}
\end{bmatrix}=\begin{bmatrix}
    \frac{1}{a_1}-1 & -a_2 \\
    -b_1 & \frac{1}{b_2}-1
\end{bmatrix},
     \end{equation}
where the coefficients $a_i$, $b_i$ are defined by the transformation of the Eulerian gradients relative to the Lagrangian configuration: 
     \begin{equation}
         \frac{\partial \bm{\mu}_{\text{Euler}}}{\partial i} =a_i \cdot \frac{\partial \bm{\mu}_{\text{Lag}}}{\partial X}+b_i \cdot \frac{\partial \bm{\mu}_{\text{Lag}}}{\partial Y}, i=1,2. 
     \end{equation}
 \end{subequations}
The spin stress tensor, in turn, is expressed as
\begin{subequations}
\begin{equation}
\tau_{ij}  =\begin{bmatrix}
    A_1 & A_2 \\
    B_1 & B_2
\end{bmatrix},  
\end{equation}
with its components representing the torque density on a surface element. On a surface with normal along the \textit{i}-axis, the spin stress vector $\bm{\tau}_{i}$ (i.e., the torque per unit area acting on that face) can be decomposed along the two basis vectors introduced in \textbf{Fig. 8}: 
\begin{equation}
\bm{\tau}_{i}(\bm{\mu})  =A_i\cdot\bm{e}_{\mathcal{X}}+B_i\cdot\bm{e}_{\mathcal{Y}}.  
\end{equation}
This torque density is directly related to the external spin torque applied on the surface (poss. incl. external exchange-interaction spin-torque variation $\Delta \textbf{T}_{\text{ext.exch}}(\bm{\mu})$, Zeeman-interaction spin-torque variation $\Delta \textbf{T}_{\text{Zee}}(\bm{\mu})$ due to spatial displacement of the $\bm{\mu}$ state, spin-transfer torque $\left|\Delta \textbf{T}_{\text{STT}} \right|$ from spin polarized current, magnon-transfer torque $\left|\Delta \textbf{T}_{\text{MTT}} \right|$ from spin wave, etc.):
\begin{equation}
\bm{\tau}_{i}(\bm{\mu}) =\frac{d\bm{\textbf{T}}_i{}_{\text{,ext}}(\bm{\mu})}{dS_i},\Delta \bm{\textbf{T}}_i{}_{\text{,ext}}(\bm{\mu}) =-\Delta \bm{\textbf{T}}_i{}_{\text{,rest}}(\bm{\mu}).
\end{equation}
The incremental restoring torque $\Delta \bm{\textbf{T}}_i{}_{\text{,rest}}(\bm{\mu})$ arises from the passive deformation and may receive contributions: 
\begin{equation}
\Delta \bm{\textbf{T}}_i{}_{\text{,rest}}(\bm{\mu})  =\Delta\bm{\textbf{T}}_{\text{dipo}}(\bm{\mu})+\Delta\bm{\textbf{T}}_{\text{int.exch}}(\bm{\mu})+\Delta\bm{\textbf{T}}_{\text{ani}}(\bm{\mu}),  
\end{equation}
\end{subequations}
corresponding, respectively, to changes in the dipolar field, the internal exchange field, and the magnetocrystalline anisotropy field. Each increment can be evaluated as:
\begin{subequations}
\begin{equation}
\Delta \textbf{T}_{\text{dipo}}(\bm{\mu}) =-\lvert \gamma \rvert\cdot\Delta \textbf{M}(\bm{\mu})\times\Delta \textbf{H}_{\text{dipo}}(\bm{\mu}),  
\end{equation}
\begin{equation}
\Delta \textbf{T}_{\text{int.exch}}(\bm{\mu}) =-\lvert \gamma \rvert\cdot\Delta \textbf{M}(\bm{\mu})\times\Delta \textbf{H}_{\text{int.exch}}(\bm{\mu}),  
\end{equation}
\begin{equation}
\Delta \textbf{T}_{\text{ani}}(\bm{\mu}) =-\lvert \gamma \rvert\cdot\Delta \textbf{M}(\bm{\mu})\times\Delta \textbf{H}_{\text{ani}}(\bm{\mu}),  
\end{equation}
\end{subequations}
where $\Delta \textbf{M}(\bm{\mu})$ is the outer shell magnetization within volume $\Delta S_i\cdot a$ (with $a$ the spin lattice constant) on which the external torque directly acts, and $\Delta \textbf{H}_{\text{dipo}}(\bm{\mu})$, $\Delta \textbf{H}_{\text{int.exch}}(\bm{\mu})$, $\Delta \textbf{H}_{\text{ani}}(\bm{\mu})$ are the corresponding changes in the local effective fields at the spin state $\bm{\mu}$.

Clearly, in equilibrium, the nonzero $\Delta \bm{\textbf{T}}_i{}_{\text{,rest}}(\bm{\mu})$ and $\Delta \bm{\textbf{T}}_i{}_{\text{,ext}}(\bm{\mu})$ exactly counterbalance one another. This balance defines the stressed state in the vicinity of a spin state $\bm{\mu}(\bm{r})$ and, by extension, the local spin stress $\bm{\tau}$ and spin strain $\textbf{D}$. A mismatch between the torques drives further deformation until a new equilibrium is reached, following the mechanism illustrated in \textbf{Fig. B.1}. In particular, the removal of the external torque leads to elastic recovery. It is important to note that spin stress is a collective property of an assembly of spins; the concept has no meaning for an isolated spin.

By measuring the equilibrium spin strain $\textbf{D}$ under a prescribed spin stress, one may construct a constitutive relation \(\begin{bmatrix}
    \tau_{11} & \tau_{12} \\
    \tau_{21} & \tau_{22}
\end{bmatrix}=\textbf{C}\begin{bmatrix}
    D_{11} & D_{12} \\
    D_{21} & D_{22}
\end{bmatrix}\) where $\textbf{C}$ is a tensor that connects the two quantities. However, it must be emphasized that such a relation is strictly local: it describes the response near a specific spin state $\bm{\mu}$ while the rest of the system remains fixed. Unlike in classical elasticity, an invariant, universally applicable constitutive tensor $\textbf{C}$ does not exist for the system as a whole. This is because spin textures are inherently inhomogeneous, and the long-range dipole-dipole interaction couples every spin to the global configuration. In general, $\textbf{C}$ depends not only on the material type but also on the sample geometry, the texture configuration, the specific spin state $\bm{\mu}$, and the detailed deformation landscape. Spin $\bm{\tau}-\textbf{D}$ analysis is therefore considerably more intricate than classical stress-strain analysis (For infinite periodic structures, the problem simplifies considerably, as it reduces to the analysis of a single unit cell). 

Despite this complexity, the $\bm{\tau}-\textbf{D}$ framework offers a viable theoretical approach. The tensors $\bm{\tau}$ and $\textbf{D}$ encode complete information about the magnetic state. In principle, given the initial distribution of $\textbf{D}$ (and hence the temporal constitutive behavior) and the loading history, the evolution of $\textbf{D}$—and therefore of the magnetization—can be determined seamlessly. More importantly, this framework provides a novel analytical perspective that renders the deformation of spin elastomers—elongation, compression, distortion—amenable to direct visualization and physical insight. This stands in sharp contrast to conventional micromagnetic calculations, which, being centered on energy minimization, often lack transparent concepts and tools for predictive intuition.

Finally, a crucial qualification: spin $\bm{\tau}-\textbf{D}$ analysis is restricted to the purely elastic deformation of existing spin textures. It does not apply to regimes involving nucleation, transformation, or annihilation of topological solitons—i.e., where vector states are created or destroyed. 

\section{\label{sec:level10}\textbf{\textbf{SPIN STRESS-STRAIN ANALYSIS} } }
We now apply the $\bm{\tau}-\textbf{D}$ framework to quantify the internal state of a DWSS under various geometric confinements. \textbf{Fig. 9} presents calculated distributions of the longitudinal spin strain $D_{11}$ and spin stress $\tau_{11}$ within the $\text{DWSS}_{\text{100}}$.

\begin{figure*}
    \centering
    \includegraphics[width=0.75\linewidth]{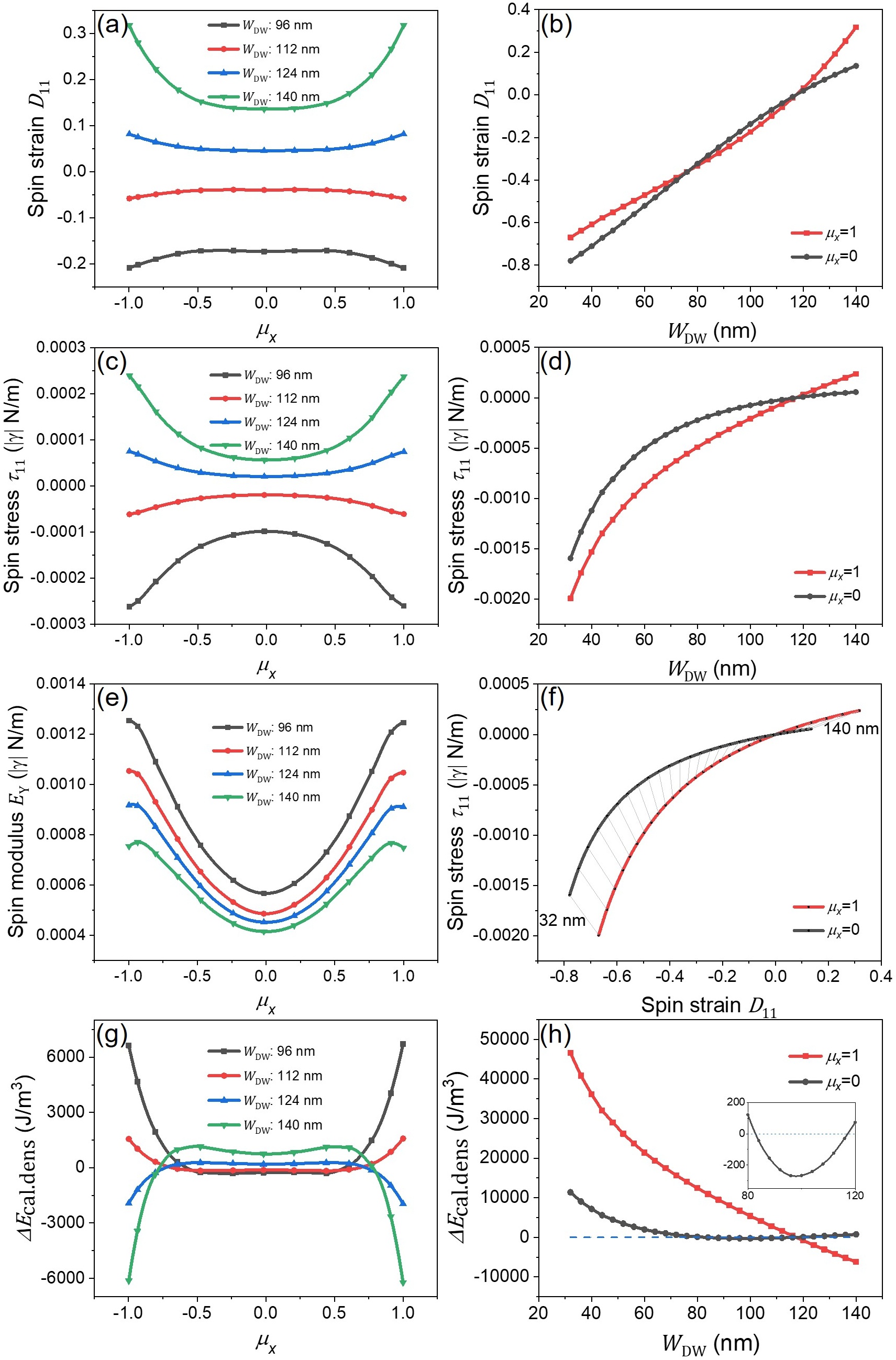}
    \caption{Spatial distribution and parametric dependence of spin strain, spin stress, spin modulus, and energy density in a $\text{DWSS}_{\text{100}}$. (a) Spin strain $D_{11}$, (c) spin stress $\tau_{11}$, (e) spin modulus $E_{\text{Y}}$, and (g) calibrated energy density increment $\Delta E_{\text{cal.dens}}$ as functions of spin-state coordinate $\mu_x$ for four domain-wall widths: \(W_{\text{DW}} =96,112,124,140\,\mathrm{nm}\). (b) Spin strain $D_{11}$ and (d) Spin stress $\tau_{11}$ at \(\mu_x =1\) and \(\mu_x =0\) versus DW width $ W_{\text{DW}}$. (f) Spin stress $\tau_{11}$ versus spin strain $D_{11}$ at \(\mu_x =1\) and \(\mu_x =0\). (h) Calibrated energy density increment $\Delta E_{\text{cal.dens}}$ at \(\mu_x =1\) and \(\mu_x =0\) versus DW width $ W_{\text{DW}}$. The spin modulus $E_{\text{Y}}$ is defined as \(E_{\text{Y}} =\frac{\tau_{11}}{D_{11}}\). The calibrated energy density increment $\Delta E_{\text{cal.dens}}$ is computed as \(\Delta E_{\text{cal.dens}}=E_{\text{dens.stressed}}\cdot C_{\text{vem}}-E_{\text{dens.relaxed}}\), where \(C_{\text{vem}} =1+D_{11}(\mu_x)\) is the volume expansion multiple. }
    \label{fig:placeholder}
\end{figure*}

\subsection{\label{sec:level2}Spin strain distribution and the origin of Poisson effect}
As in conventional elastic bodies, the longitudinal strain $D_{11}(\bm{\mu})$ at a given spin state increases monotonically with elongation (or, equivalently, with domain-wall width $W_{\text{DW}}$), as shown in \textbf{Fig. 9(b)}. However, the strain is not uniform along the longitudinal direction. Under modest deformation, $\lvert D_{11} \rvert$ typically rises toward the domain-wall boundaries (\textbf{Fig. 9(a)}); under severe compression, the $D_{11}$ profile develops a protuberance near the center—a feature corroborated by the intersection near \(W_{\text{DW}} =76\,\mathrm{nm}\) in \textbf{Fig. 9(b)}. Comparing $D_{11}(W_{\text{DW}})$ for different lateral positions (\textbf{Fig. B.2}) readily explains the Poisson effect reported in \textbf{Fig. 7}. Under both tension and compression, $\lvert D_{11} \rvert$ at the bottom flank (core) of the DW is consistently larger (smaller) than at the top flank (core), indicating that the lower part of the DW core is more rigid. Consequently, under longitudinal elongation (compression), spin states in the lower core tend to shift upward (downward), producing lateral contraction (expansion). Thus, Poisson's ratio in spin systems originates from the laterally nonuniform distribution of longitudinal strain. On the other hand, the variation of $\nu_{\bm{\mu}}$ with deformation arises because the functions of $D_{11}(W_{\text{DW}})$ at different lateral positions vary without fixed proportionality (e.g., the $D_{11}$ curves at the bottom core and the top core in \textbf{Fig. B.2}) —a nonuniformity that is likely generic. 

\subsection{\label{sec:level2}Non-conservation of spin stress}
A striking departure from classical mechanics is that spin stress is not conserved across the structure: $\tau_{11}$ varies with position and lacks simple transferability (\textbf{Fig. 9(c)}), in apparent violation of Newton's third law. The essential reason is the nonlocal character of the dipole-dipole interaction. Upon application of a spin-transfer torque, the counteracting spin torque at a given site arises not solely from its immediate neighbors but from a distributed ensemble of spins throughout the system. Hence, the torque flow is not fully transmitted to the next spin. Other interactions—magnetocrystalline anisotropy and Zeeman coupling—behave similarly: part of the counteracting spin torque is supplied locally by the anisotropy field or the external field rather than by neighboring spins. In our DWSS, $\tau_{11}$ increases with distance from the domain-wall center (\textbf{Fig. 9(c)}). The same trend holds for its spatial derivative, implying a stronger dipolar field near the boundaries (a flat $\tau_{11}$ profile indicates constant stress transfer and thus a weak dipolar field; a large derivative reflects highly variable transfer and a strong dipolar field), a prediction verified in \textbf{Fig. B.3}. 

\subsection{\label{sec:level2}\textbf{Strain \& site-dependent spin modulus} }
From $\tau_{11}$ and $D_{11}$, we define the \textit{spin modulus} \(E_{\text{Y}} =\frac{\tau_{11}}{D_{11}}\) (\textbf{Fig. 9(e)}). Even for the same material, $E_{\text{Y}}$ varies with both the spin state $\mu_x$ and the DW width $ W_{\text{DW}}$ (equivalently, $D_{11}$). In general, the smaller $ W_{\text{DW}}$, the larger $E_{\text{Y}}$. The functional form of $E_{\text{Y}}(\mu_x)$, however, remains largely preserved: larger values arise farther from the DW center. 

\subsection{\label{sec:level2}\textbf{Microscopic energy storage} }
\textbf{Fig. 9(g)} shows the calibrated energy density increment $\Delta E_{\text{cal.dens}}$ as a function of $\mu_x$. Energy is not stored uniformly: under tension, the DW core stores energy while the periphery releases it; under compression, the reverse occurs. This implies that in the relaxed state, the core is already somewhat pre-tensioned and the periphery pre-compressed. Only under sufficiently large compression does the entire DWSS store energy; the critical width is \(W_{\text{DW}} =84\,\mathrm{nm}\) (\textbf{Fig. 9(h)}). 

\subsection{\label{sec:level2}\textbf{\textbf{Theoretical validity} } }
The preceding analyses demonstrate that the $\bm{\tau}-\textbf{D}$ framework provides a quantitative description of subtle elastic behaviors in DWs and, more generally, in spin elastomers under load. As a stringent test for the theory, we examine domain-wall morphology under external stimuli. While it is well known that DW deformation occurs during driven motion, conventional theoretical treatments have typically assumed rigid profiles or relied on a limited set of distortion parameters [19-21]—assumptions that are often inconsistent with the observed behavior and hinder a rigorous description of DW dynamics. In \textbf{Fig. B.4}, we show that under spin-polarized currents, depending on the spin stress transfer, DWs exhibit a rich variety of morphologies. Most notably, abnormally asymmetric strain profiles—a scenario beyond the reach of earlier treatments—are captured by the $\bm{\tau}-\textbf{D}$ analysis based on the spin stress transfer curve. Excellent agreement is shown between the $\bm{\tau}-\textbf{D}$ calculations and direct micromagnetic simulations for both pinned and unpinned DWs. 

\subsection{\label{sec:level2}\textbf{\textbf{$\tau-D$ diagram} } }
In structural mechanics, the stress-strain curve is the fundamental characterization of a material’s mechanical response. \textbf{Fig. 9(f) }shows the spin $\tau-D$ curves for two representative spin states, \(\mu_x =1\) and \(\mu_x =0\), within the DWSS. The curves diverge, reflecting the site-dependent nature of the elastic response—though the strains at different sites are not independent but interrelated, as indicated by the gray connecting lines. To capture the full $\tau-D$ correspondence of the system, a comprehensive sweep over spin states is required. Within the limits of our computational resources, we have computed curves for a set of representative positions (\textbf{Fig. 10}): the top edge (TE), middle edge (ME), and bottom edge (BE) at \(\mu_x =1\); and the top center (TC), middle center (MC), and bottom center (BC) at \(\mu_x =0\). Several key observations emerge: 

\begin{figure*}
    \centering
    \includegraphics[width=0.6\linewidth]{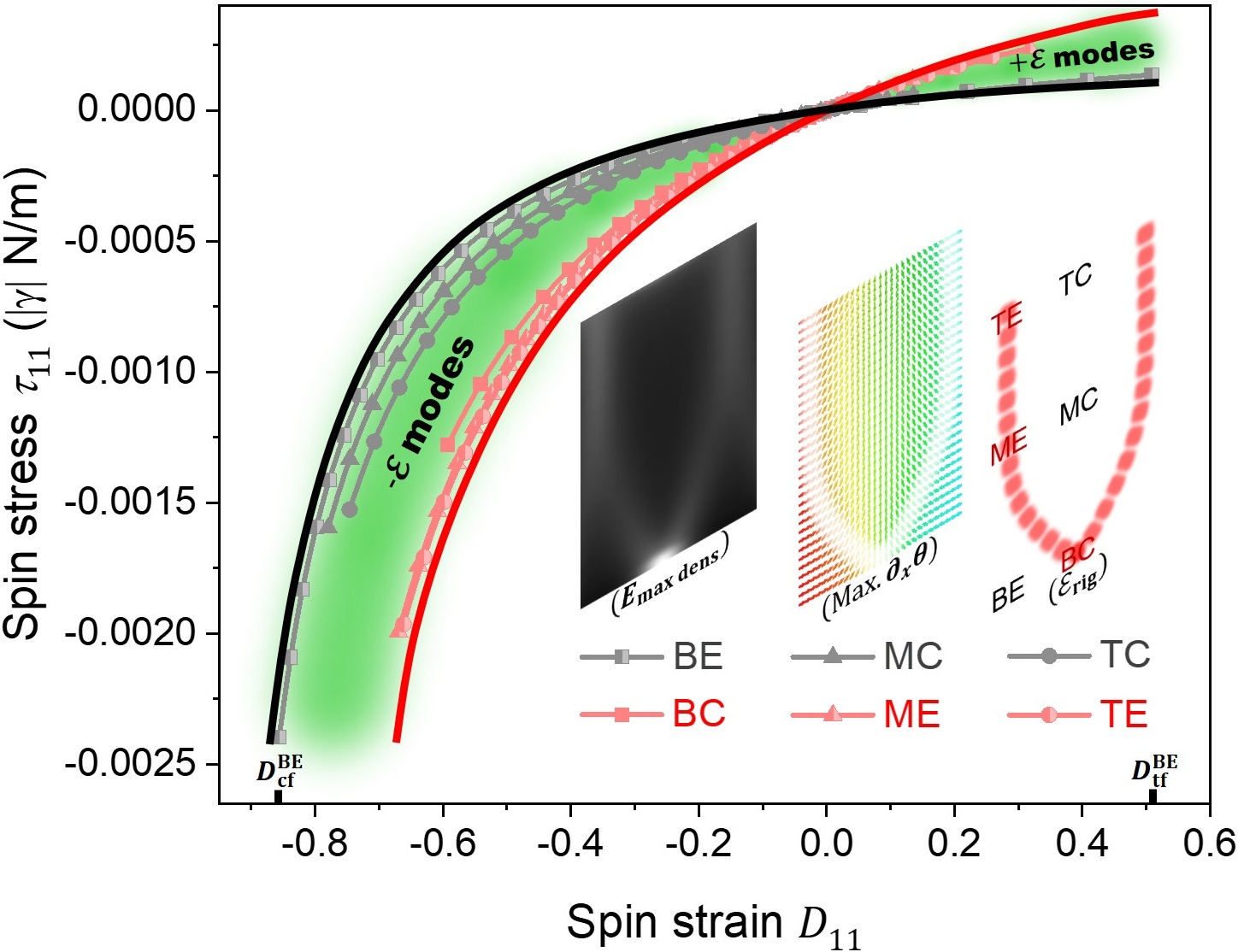}
    \caption{Spin stress-strain ($\tau-D$) curves for six representative spin states within the domain wall: \(\mu_x =1\) at the top edge (TE), middle edge (ME), and bottom edge (BE); and \(\mu_x =0\) at the top center (TC), middle center (MC), and bottom center (BC). The black and red solid curves outline the spectrum of possible elastic modes within the DW. The critical compressive and tensile failure strains, extracted from the BE curve, are denoted $D_{\text{cf}}^{\text{BE}}$ and $D_{\text{tf}}^{\text{BE}}$, respectively. Inset: Correlation between regions of elevated energy density ($E_{\text{max dens}}$), abrupt magnetization switching events (Max. $\partial_x \theta$), and higher-rigidity $\tau-D$ modes ($\mathcal{E} _{\text{rig}}$). The trajectory of abrupt magnetization switching is constructed from the maximum gradient $\frac{\partial \bm{\mu}_y}{\partial x}$ in each spin row (\textbf{Fig. B.8}).}
    \label{fig:placeholder}
\end{figure*}

$\bullet$ For all spin states, $\tau_{11}$ increases monotonously with $D_{11}$.

$\bullet$ Linear elasticity holds within the strain interval \(D_{11} \in (-0.1,0.1)\).

$\bullet$ Beyond this range, the DWSS exhibits strain hardening under compression and strain softening under tension.

$\bullet$ The most pronounced hardening (under compression) and softening (under tension) occur at the bottom edge (BE), indicating that this site is most resistant to perturbation under compression (as confirmed by \textbf{Fig. B.5}, which records strain oscillations under a propagating spin stress wave, discussed in \textbf{Sec. XIV}) and most compliant under tension.

$\bullet$ The consistently larger strain magnitude at BE across the entire deformation range implies more severe contraction under compression and more pronounced expansion under tension (\textbf{Fig. B.6}).

$\bullet$ Abnormally larger strain at ME and TE under tension, compared to MC and TC, arises (via inspection of site interrelations)—consistent with the aforementioned fact that these edge sites are already pre-compressed in the relaxed state, and further strain facilitates energy release.

$\bullet$ The critical compressive strain for topological failure at BE is \(D_{\text{cf}}^{\text{BE}} \approx -0.85\), beyond which the spin texture annihilates. Under tension, analogous to fracture in conventional materials, the DWSS ultimately snaps at \(D_{\text{tf}}^{\text{BE}} \approx -0.51\), most likely due to energy minimization of the system (\textbf{Fig. B.7}). Notably, owing to the stability of topological solitons and the long-range magnetostatic interaction, elastic deformation in spin elastomers remains reversible not only within the linear regime but over the entire range $(D_{\text{cf}},+\infty)$.

Importantly, from diagrams of this kind, one can access similar information in general cases.

\subsection{\label{sec:level2}\textbf{\textbf{\textbf{Spectrum of elastic modes} } } }
A striking feature of \textbf{Fig. 10} is the splitting of the $\tau-D$ curves into two distinct branches, outlining a spectrum of allowed elastic modes—hereafter termed $\tau-D$ modes (or $\mathcal{E}$ modes). The red-colored branch, which comprises mostly edge sites, corresponds to higher rigidity (The mode exchange between BE and BC can be attributed to limited spin stress transferability across BC due to the enhanced dipolar field there; see the green curve in \textbf{Fig. B.3}). Identifying the more rigid modes is of particular importance: in magnetic systems, regions of high rigidity are typically associated with high energy density and abrupt magnetization switching. The inset of \textbf{Fig. 10} confirms a strong correlation between elevated energy density, sharp magnetization transitions, and the higher-rigidity $\mathcal{E}$ modes.

In summary, spin $\bm{\tau}-\textbf{D}$ analysis yields a comprehensive picture of the mechanical response of spin elastomers, quantifying key characteristics such as the proportionality limit, site-dependent modulus, strain distribution, spectrum of elastic modes, regions of elevated energy density (or enhanced nonlinearity), annihilation limit, and ultimate tensile stress. These insights provide a valuable foundation for the rational design and engineering of spin microstructures.

\section{\label{sec:level11}\textbf{\textbf{\textbf{ELECTRICAL MANIPULATION} } } }

\begin{figure*}
    \centering
    \includegraphics[width=0.8\linewidth]{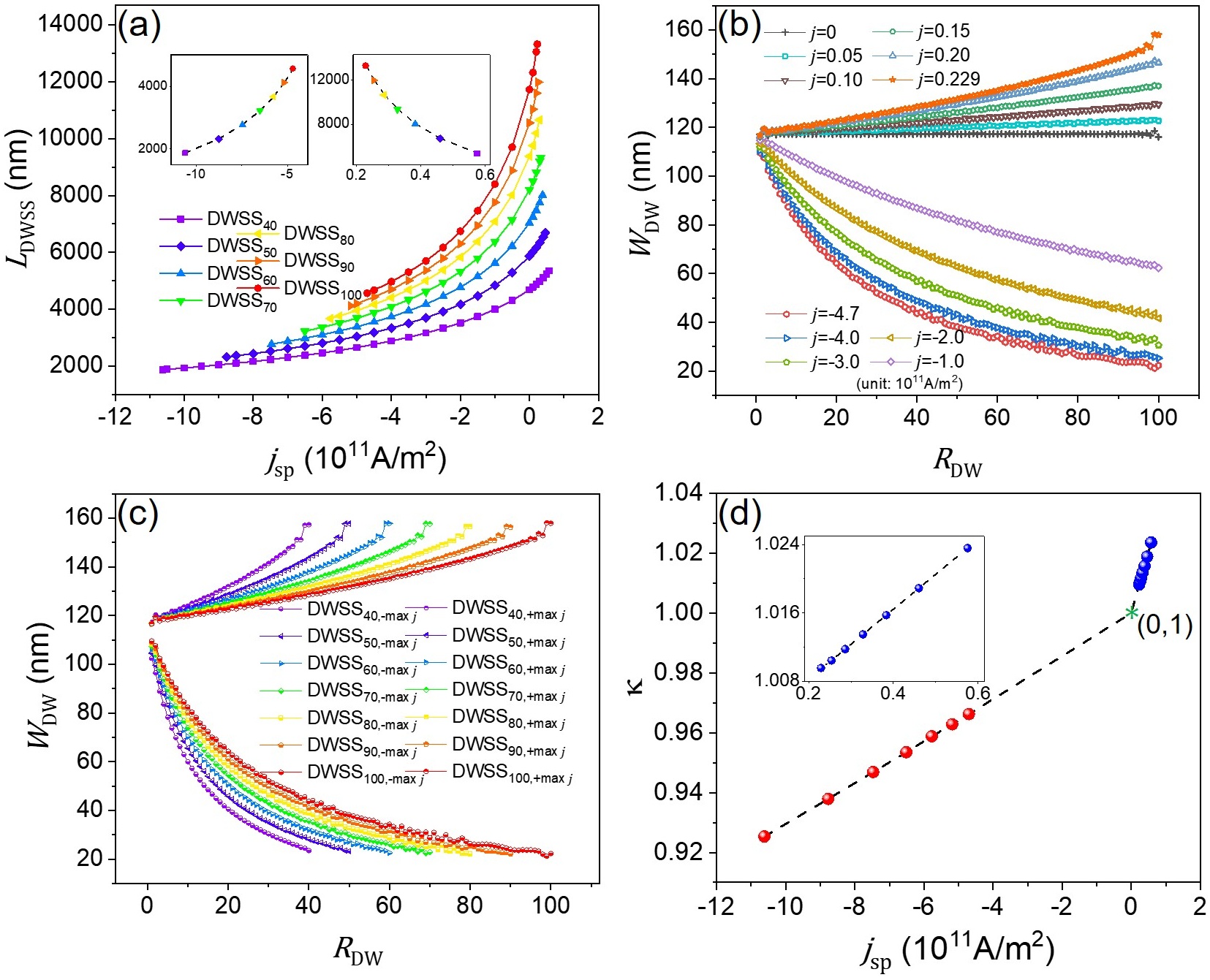}
    \caption{Electrical manipulation of a DWSS via spin-polarized current. (a) Total length $L_{\text{DWSS}}$ versus current density $j_{\text{sp}}$ for different numbers of domain walls $N_{\text{DW}}$. Insets: Critical length versus critical current density $j_{\text{crit}}$ for each $N_{\text{DW}}$. (b) Domain-wall width $W_{\text{DW}}$ versus DW sequence order $R_{\text{DW}}$ in a $\text{DWSS}_{\text{100}}$ for various $j_{\text{sp}}$. (c) $W_{\text{DW}}$ versus $R_{\text{DW}}$ for different $N_{\text{DW}}$ at critical current density $j_{\text{crit}}$. (d) Scale factor $\kappa$ for the deformable portion of the DW versus $j_{\text{sp}}$, extracted from exponential fits to the data using \(W_{\text{DW}} =(117.3-W_{\text{0}})\kappa^{R_{\text{DW}}}+W_{\text{0}}\).}
    \label{fig:placeholder}
\end{figure*}

Spin elastomers can be controlled by any stimulus carrying angular momentum—magnetic fields (field-like torque), spin waves (magnonic transfer torque, MTT) [22], or circularly polarized light (optical spin-transfer torque, OSTT) [23]. Here we focus on electrical manipulation via spin-polarized currents, which exert conventional spin-transfer torque (STT) [24,25] (notably, spin-orbit torque (SOT) via the spin Hall effect [26] or the Rashba effect [27] provides an equally promising alternative) and offer distinct advantages in spatial precision, energy efficiency, speed, and compatibility with next-generation electronics.

\subsection{\label{sec:level2}\textbf{Current-controlled length modulation} }

As shown in \textbf{Fig. B.9 }and \textbf{Fig. 11 (a)}, the overall length $L_{\text{DWSS}}$ varies strongly with the current density $j_{\text{sp}}$. Depending on the current direction, the DWSS can be either extended or compressed. The response $L_{\text{DWSS}}(j_{\text{sp}})$ is monotonic and nonlinear, and can be tuned by the number of constituent domain walls $N_{\text{DW}}$ (see \textbf{Fig. B.10} for variations with material parameters, geometry, and STT characteristics). For fixed $j_{\text{sp}}$, the length change $\Delta L$ increases with $N_{\text{DW}}$, analogous to the elongation of a conventional spring under load. However, the critical current density $j_{\text{crit}}$ for structural failure—both compressive and tensile—decreases monotonically with increasing $N_{\text{DW}}$ (inset of \textbf{Fig. 11 (a)}).

\begin{figure*}
    \centering
    \includegraphics[width=0.8\linewidth]{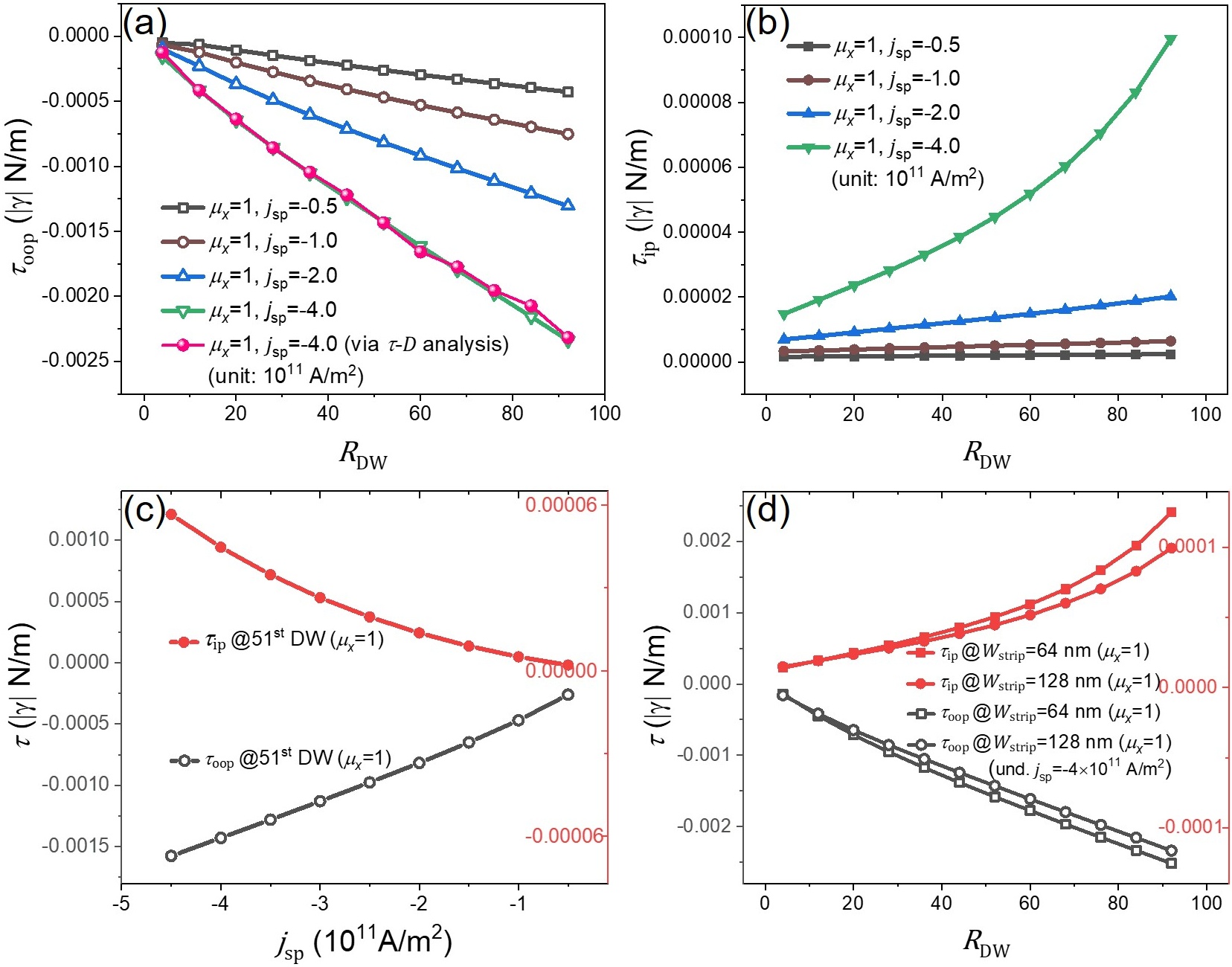}
    \caption{Spin stress accumulation under current. (a) Out-of-the-plane spin stress $\tau_{\text{oop}}$ of \(\mu_x =1\) versus DW sequence order $R_{\text{DW}}$ within a $\text{DWSS}_{\text{100}}$ for various negative current densities $j_{\text{sp}}$. For \(j_{\text{sp}} =-4.0\times 10^{11} \,\mathrm{A/m^2}\), results from the full $\tau-D$ analysis—based on the DW width distribution, the $\tau-D$ relation, and the $D(W_{\text{DW}})$ profile at \(\mu_x =1\)—are overlaid for comparison. (b) In-plane spin stress $\tau_{\text{ip}}$ at \(\mu_x =1\) versus $R_{\text{DW}}$ for the same currents (positive $\tau_{\text{ip}}$ denotes compression). (c) $\tau_{\text{oop}}$ and $\tau_{\text{ip}}$ at \(\mu_x =1\) within the $51^{\text{st}}$ DW versus $j_{\text{sp}}$. (d) $\tau_{\text{oop}}$ and $\tau_{\text{ip}}$ at \(\mu_x =1\) versus $R_{\text{DW}}$ for a $\text{DWSS}_{\text{100}}$ in nanostrips of width 128 nm and 64 nm, both at \(j_{\text{sp}} =-4.0\times 10^{11} \,\mathrm{A/m^2}\).}
    \label{fig:placeholder}
\end{figure*}

\subsection{\label{sec:level2}\textbf{Nonuniform domain-wall width distribution} }
To understand this, we examine the distribution of individual DW widths $W_{\text{DW}}$ within the assembly across a range of $j_{\text{sp}}$ and $N_{\text{DW}}$ (\textbf{Fig. 11 (b)} and \textbf{(c)}). The widths are not uniform: for negative (positive) $j_{\text{sp}}$, $W_{\text{DW}}$ decreases (increases) exponentially with the DW sequence order, implying a propensity for early failure at the right end of the DWSS. Correspondingly, for $W_{\text{DW}}$ with well-defined upper and lower bounds, the critical current density $j_{\text{crit}}$ decreases with increasing $N_{\text{DW}}$, consistent with the inset of \textbf{Fig. 11(a)}. 

The origin of this distribution can be understood qualitatively as follows. In dynamic equilibrium, a negative (positive) current injected into the DWSS induces an STT that must be balanced by a field-like torque arising from into-the-plane (out-of-plane) tilting of the receptor spin. Such tilting can only be sustained through local contraction (expansion) of the DWSS (\textbf{Fig. 4(a)} and \textbf{Fig. B.11}). The corresponding dynamical processes are depicted in \textbf{Fig. B.12}. As the current permeates the assembly, progressively tighter geometric constraints develop from left to right, causing the DW width to decay (grow) exponentially with sequence order. A larger current magnitude accelerates this exponential trend (\textbf{Fig. 11(b)}).

\subsection{\label{sec:level2}\textbf{Phenomenological model of elastic domain wall} }
A simple exponential function captures all curves in \textbf{Fig. 11(c)}: 
\begin{equation}
W_{\text{DW}} =(W_{\text{eq}}-W_{\text{0}})\kappa^{R_{\text{DW}}}+W_{\text{0}},  
\end{equation}
with $R_{\text{DW}}$ indexing the DW sequence order, $\kappa$ a scale factor, and \(W_{\text{eq}} =117.3\,\mathrm{nm}\) the equilibrium DW width. The fits reveal two remarkable features. First, $W_{\text{0}}$ manifests as a constant—approximately 22 nm under compression and 95 nm under tension—revealing the existence of a rigid core that resists further deformation. Second, and most remarkably, the scale factor $\kappa$ for the deformable component is directly proportional to the applied current density $j_{\text{sp}}$ (\textbf{Fig. 11(d)}).

\begin{figure*}
    \centering
    \includegraphics[width=0.8\linewidth]{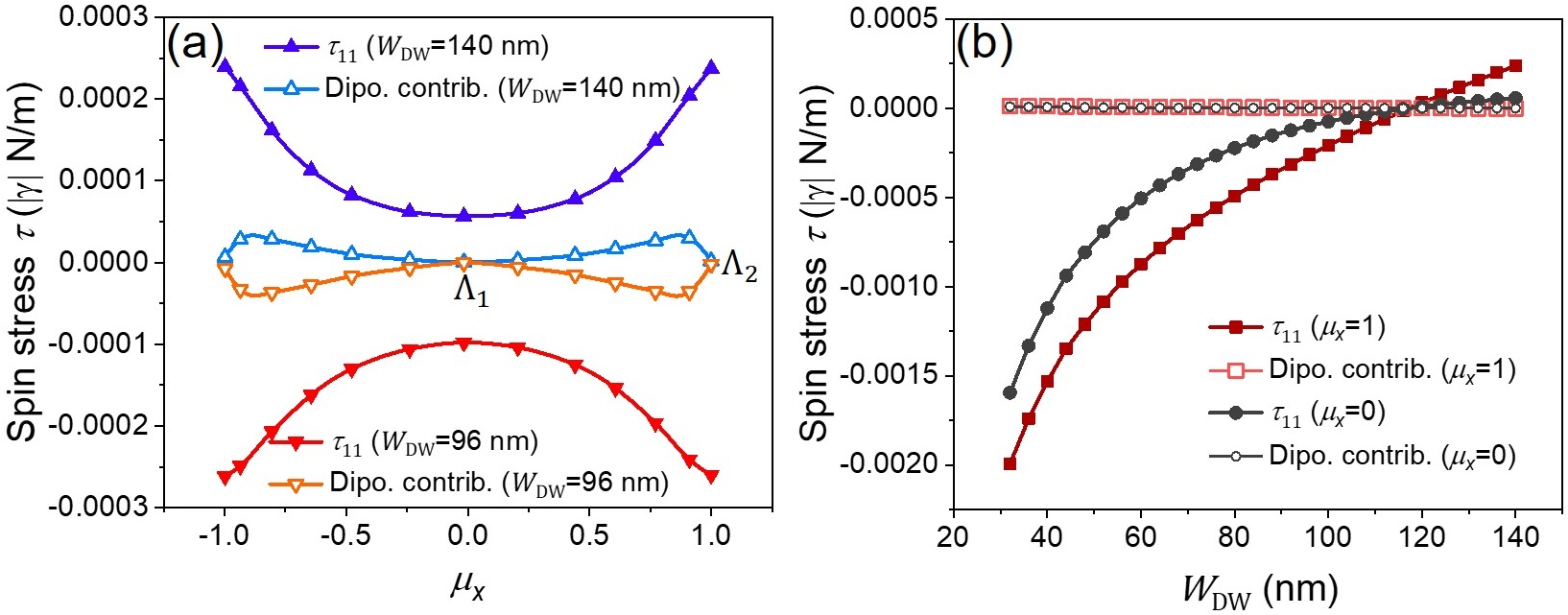}
    \caption{Dipolar contribution to spin stress and its cancellation at high-symmetry points. (a) Spin stress $\tau_{11}$ and its dipolar component as functions of the spin-state coordinate $\mu_x$ for two domain-wall \(W_{\text{DW}} =96\,\mathrm{nm}\) and \(W_{\text{DW}} =140\,\mathrm{nm}\). The high-symmetry points $\Lambda_{\text{1}}$ and $\Lambda_{\text{2}}$ within the DW are indicated. (b) Dipolar contribution to $\tau_{11}$ versus $W_{\text{DW}}$ evaluated at $\Lambda_{\text{1}}$ and $\Lambda_{\text{2}}$.}
    \label{fig:placeholder}
\end{figure*}

\subsection{\label{sec:level2}\textbf{Spin stress accumulation under current} }

Beyond macroscopic geometric changes, the internal stress state of a spin elastomer is equally amenable to electrical control. \textbf{Figs. 12(a)} and \textbf{12(b)} show the spin stress at \(\mu_x =1\) along the DWSS axis under various negative $j_{\text{sp}}$. In addition to the out-of-plane component $\tau_{\text{oop}}$ discussed earlier, a small but distinct in-plane component $\tau_{\text{ip}}$ emerges, acting to counter the spin-transfer torque (STT) induced by the current. Although $\tau_{\text{ip}}$ is modest in magnitude, the out-of-plane tilting it entails—a necessary prerequisite for its generation—plays a critical role: it governs the nucleation of vortices and ultimately the collapse of the DWSS under high currents (\textbf{Figs. B.13} and \textbf{B.14}). Because the angle between adjacent spins (and hence the STT) decreases with increasing DW width, a larger gradient of $\tau_{\text{ip}}$ is expected for higher $R_{\text{DW}}$ (here we take \(\tau_{\text{ip}} >0\) to denote contraction), consistent with the numerical results in \textbf{Fig. 12(b)}. The out-of-plane spin stress $\tau_{\text{oop}}$, by contrast, exhibits an approximately linear dependence on $R_{\text{DW}}$ and originates purely from geometric constraints. As shown in \textbf{Fig. 12(a)}, direct micromagnetic result of $\tau_{\text{oop}}$ for \(j_{\text{sp}} =-4.0\times 10^{11} \,\mathrm{A/m^2}\) agrees remarkably well with calculations based on the DW width distribution, the $\tau-D$ relation, and the $D(W_{\text{DW}})$ profile at \(\mu_x =1\). To summarize, both $\tau_{\text{oop}}$ and $\tau_{\text{ip}}$ accumulate with increasing DW number $N_{\text{DW}}$ (or winding number)—the former due to progressively tighter geometric constraints, the latter from the amplified STT effect.

\subsection{\label{sec:level2}\textbf{Current and geometry dependence of spin stress} }
\textbf{Fig. 12(c)} plots $\tau_{\text{oop}}(j_{\text{sp}})$ and $\tau_{\text{ip}}(j_{\text{sp}})$ for \(\mu_x =1\) within the $51^{\text{st}}$ DW of $\text{DWSS}_{\text{100}}$. Larger current densities produce larger stresses, demonstrating efficient electrical modulation. How these stresses evolve with topological deformation under fixed winding number and current, however, remains an open question. \textbf{Fig. 12(d)} compares the spin stresses in a $\text{DWSS}_{\text{100}}$ housed in a 128-nm-wide strip with those in a topologically identical but geometrically narrower (64 nm) strip, both at \(j_{\text{sp}} =-4.0\times 10^{11} \,\mathrm{A/m^2}\). A narrower strip (hence narrower DW width) yields larger stresses. The dependence of $\tau_{\text{oop}}$ and $\tau_{\text{ip}}$ on $W_{\text{DW}}$, $j_{\text{sp}}$ and $R_{\text{DW}}$ can be distilled to a simple principle: enhanced STT between adjacent spins imposes tighter geometric constraints, increasing $\tau_{\text{oop}}$; concurrently, a larger field-like torque is required to balance the STT, elevating $\tau_{\text{ip}}$.

\section{\label{sec:level12}\textbf{EQUATIONS OF SPIN ELASTIC EQUILIBRIUM AND SPIN ELASTODYNAMICS} }

To complete the theoretical framework, we now extend the analysis to the elastic dynamics of spin elastomers. For brevity, we focus on the motion of constituent solitons as quasiparticles—their positions and velocities—while omitting internal deformations. This mesoscopic dynamics can be captured by targeting a representative spin state within each soliton and deriving its equation of motion from the fundamental Landau-Lifshitz-Gilbert (LLG) equation. 

A key simplification emerges from the symmetry of topological solitons. Within a typical spin elastomer, there exist high-symmetry points $\Lambda_n$ at which the dipole-dipole interaction cancels out exactly, making no net contribution to the spin stress (see the DW example in \textbf{Fig. 13}). Exclusion of the dipole-dipole interaction at these points yields a fixed, local constitutive relation, enabling a closed-form description of spin elastodynamics within the $\bm{\tau}-\textbf{D}$ framework.

In the simplest case involving only exchange interaction—where, in the continuum limit, the exchange field takes the form \(\textbf{H}_{\text{exch}}=-\frac{2A}{\mu_{\text{0}}M_{\text{S}}} \nabla ^2 \bm{\mu}\)—the constitutive relation reduces to: 
\begin{equation}
\bm{\tau}=-\frac{2\lvert \gamma \rvert A}{\mu_{\text{0}}}\bm{\mu}\times \bigg(\textbf{D}\begin{bmatrix}
    \frac{\partial \bm{\mu}_{\text{Lag}}}{\partial X_1} & 0 \\
    0 & \frac{\partial \bm{\mu}_{\text{Lag}}}{\partial X_2}
\end{bmatrix}\bigg).  
\end{equation}
Combined with the geometric \textbf{Eq. (4)}: \(\bm{\textbf{D}}  =\begin{bmatrix}
    \frac{\partial u_1}{\partial X_1} & \frac{\partial u_2}{\partial X_1} \\
    \frac{\partial u_1}{\partial X_2} & \frac{\partial u_2}{\partial X_2}
\end{bmatrix},D_{ij} =\frac{\partial u_i}{\partial X_j}\) and the equations of motion for the spin states at high-symmetry points:
\begin{subequations}
\begin{equation}
(\textbf{v}\cdot \nabla)\bm{\mu}=-\frac{1}{(1+\alpha^2)M_{\text{S}}}(\alpha \bm{\mu}\times \bm{\mathcal{T}}+\bm{\mathcal{T}}),  
\end{equation}
\begin{equation}
\bm{\mathcal{T}}=\nabla \times \bm{\tau}+\bm{\mathcal{T}}_{\text{ext}},  
\end{equation}
\end{subequations}
where \(\textbf{v}=\frac{\partial \textbf{u}}{\partial t}\) is the instantaneous velocity of the spin state at a high-symmetry point, $\bm{\mathcal{T}}$ is the volumetric spin torque density, and \(\bm{\mathcal{T}}_{\text{ext}}=p_1\frac{\partial \bm{\mu}_{\text{Lag}}}{\partial X_1}+p_2\frac{\partial \bm{\mu}_{\text{Lag}}}{\partial X_2}\) represents external contribution, we thus obtain the spin elastodynamic equation: 
\begin{equation}
\begin{aligned}
    \frac{\partial u_i}{\partial t} & = \frac{2\lvert \gamma \rvert A}{(1+\alpha^2)\mu_{0}M_{\text{S}}}[-(\alpha+\cot\theta)\nabla^2 u_i+\frac{1}{\sin \theta}\nabla^2 u_j] \\
    & -\frac{1}{(1+\alpha^2)M_{\text{S}}}\frac{(\sin \theta-\alpha\cos \theta)p_i+\alpha p_j}{\sin \theta}
\end{aligned}
\end{equation}
where $\theta$ is the included angle between $\frac{\partial \bm{\mu}_{\text{Lag}}}{\partial X_1}$ and $\frac{\partial \bm{\mu}_{\text{Lag}}}{\partial X_2}$. The first term on the right-hand side represents the system-derived elastic restoration, while the second term captures the external driving. \textbf{Eq. (12)} reveals a key insight: magnetic solitons can indeed possess inertia, which originates from internal deformation—more precisely, from the strain gradient. Finally, the spin elastic equilibrium equation takes the simple form 
\begin{equation}
\nabla \times \bm{\tau}=-\bm{\mathcal{T}}_{\text{ext}},  
\end{equation}
which balances the internal stress divergence with the external torque density. 

\begin{figure*}
    \centering
    \includegraphics[width=0.78\linewidth]{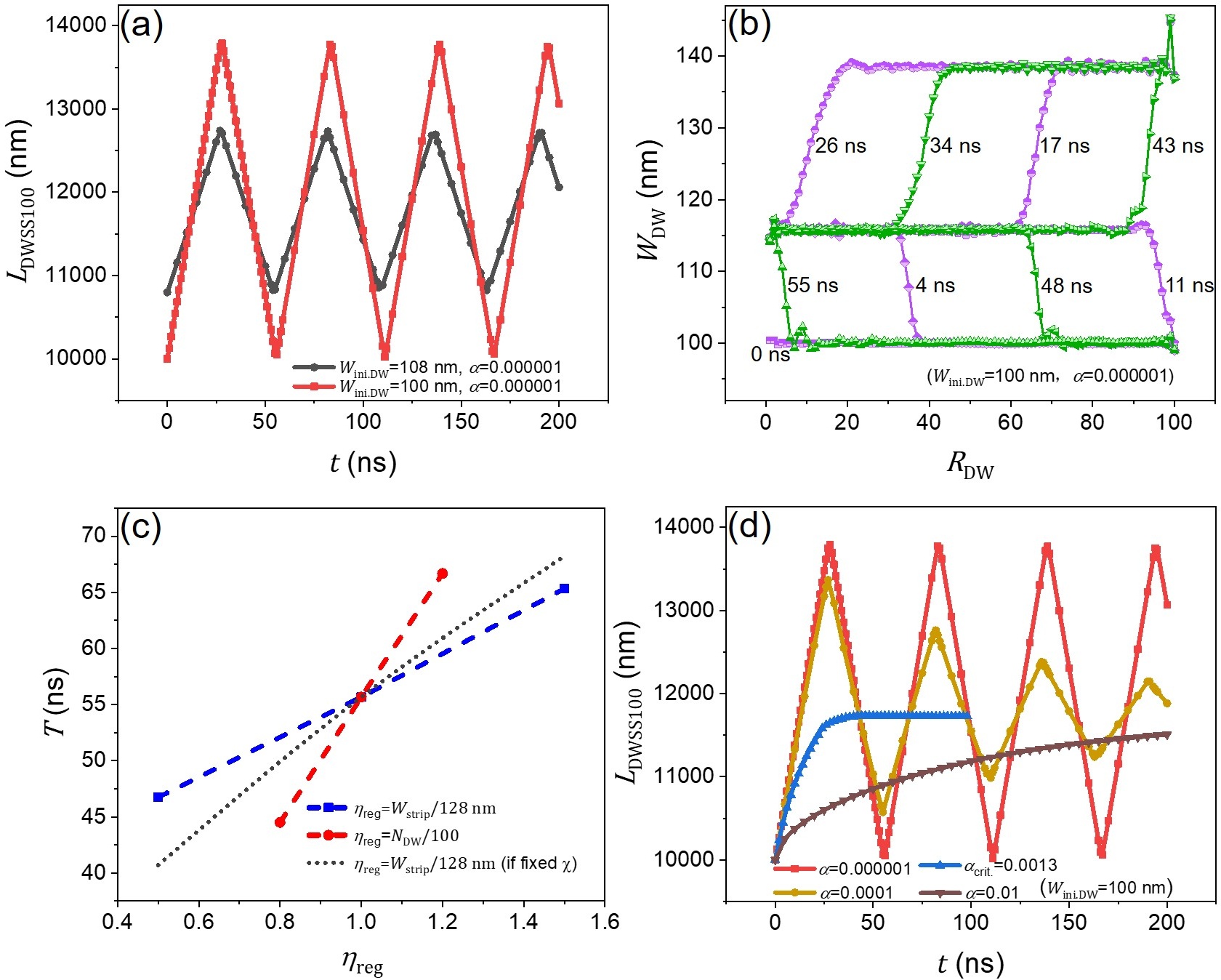}
    \caption{Spontaneous oscillations of a DWSS. (a) Time evolution of a $\text{DWSS}_{\text{100}}$ under initial compression for two different initial DW widths,  \(W_{\text{ini.DW}} =100\,\mathrm{nm}\) and \(W_{\text{ini.DW}} =108\,\mathrm{nm}\). To isolate the intrinsic undamped dynamics, the Gilbert damping is set to an ultralow value \(\alpha=10^{-6}\). (b) Distribution of DW width $W_{\text{DW}}$ within the $\text{DWSS}_{\text{100}}$ at various times over one oscillation period, for initial width \(W_{\text{ini.DW}} =100\,\mathrm{nm}\). The stepwise hopping deformation is clearly visible. (c) Oscillation period as a function of the regulation ratio $\eta_{\text{reg}}$ for normalized strip width $\frac{W_{\text{strip}}}{128\,\mathrm{nm}}$ (with 128 nm the default width) and normalized DW number $\frac{N_{\text{DW}}}{100}$ (with 100 the default number). The gray dashed line shows the period predicted from the modified DW width after strip-width tuning, assuming constant proportionality $\chi$ between DW velocity and initial DW compression/expansion (as defined for the default strip width). (d) Oscillations of $\text{DWSS}_{\text{100}}$ under initial DW compression (\(W_{\text{ini.DW}} =100\,\mathrm{nm}\)) for different damping constants. The critical damping coefficient is \(\alpha=0.0013\).}
    \label{fig:placeholder}
\end{figure*}

\section{\label{sec:level13}\textbf{\textbf{SPIN ELASTIC OSCILLATION AND RESONANCE} } }

A fundamental question in spin elastodynamics is whether elastic oscillations are at all possible—another defining signature of spin elasticity. In ferromagnets, spin dynamics is governed by the Landau-Lifshitz-Gilbert (LLG) equation, a first-order differential equation in time. This stands in stark contrast to the second-order dynamics of antiferromagnets [28,29] or Newtonian mechanics. As a result, magnetic solitons and spin elastomers possess no intrinsic inertia or linear momentum, and should not, in principle, oscillate spontaneously about equilibrium.

\subsection{\label{sec:level2}\textbf{Triangular oscillation} }
Yet, \textbf{Fig. 14(a)} reveals clear oscillations of a DWSS under initial compression—remarkably, in a regular triangular waveform seldom seen in nature. To understand this behavior, we compare the DW profiles at rest and in steady motion, guided by the inertia mechanism predicted in \textbf{Eq. (12)}. As shown in \textbf{Fig. B.15}, a moving DW adopts a distinct configuration: higher velocity enhances out-of-plane tilting (averaged $m_z$ in \textbf{Fig. B.15(a)}; the spatially oscillatory distribution of $m_z$ reflects an adjustment that enables steady propagation of DW)—consistent with the fact that out-of-plane tilting and the concomitant out-of-plane effective field generate the torque required for in-plane spin rotation and thus DW displacement—and narrows the DW width (\textbf{Fig. B.15(b)}). This narrowing not only produces spin strain but also amplifies the strain gradient for a given strain difference, in agreement with \textbf{Eq. (12)} at higher velocities. As the strain gradient is persistently maintained, the DW retains its inertial motion, rendering elastic oscillations possible.

\subsection{\label{sec:level2}\textbf{Stepwise hopping deformation} }
A puzzle remains: as DW velocity is monotonically related to the strain gradient [\textbf{Eq. (12)}], synchronized deformation of the DWs within the assembly would lead to nonlinear oscillations, contradicting the linear sawtooth waveform in \textbf{Fig. 14(a)}. To resolve this, we examine the detailed kinetics during expansion and contraction (\textbf{Fig. 14(b)}). Unexpectedly, the DWs undergo a stepwise hopping deformation, a behavior with no analogue in conventional elasticity. For an initially compressed $\text{DWSS}_{\text{100}}$ with \(W_{\text{ini.DW}} =100\,\mathrm{nm}\), release of the left-end constraint triggers sequential expansion from left to right (Stage I, 0-12 ns): each DW rapidly expands from 100 nm to a pseudo-equilibrium width of 115.8 nm, acquiring a steady velocity of 138.9 m/s. In Stage II (12-26 ns), inertial motion drives the entire assembly leftward, but the rightmost DW is soon halted by the fixed right boundary, undergoing a secondary expansion to 138.5 nm. This process propagates backward from right to left. Upon maximum expansion at \(t=26 \,\mathrm{ns}\), elastic recovery—aided by the free left end—initiates sequential contraction from left to right (Stage III, 26-44 ns), with DWs attaining a rightward speed of 134.3 m/s. By \(t=44 \,\mathrm{ns}\), the system returns to the pseudo-equilibrium width. Finally, in Stage IV (44-55 ns), inertia drives further contraction, restoring the initial compressed state. 

\subsection{\label{sec:level2}\textbf{Natural frequency and its tuning} }
Returning to \textbf{Fig. 14(a)}, oscillations with markedly different amplitudes share nearly the same frequency ($\sim 18 \,\mathrm{MHz}$), indicating a natural frequency. Given the stepwise hopping deformation and the regular triangular waveform, this natural frequency implies a fixed proportionality $\chi$ between DW velocity and the initial DW compression/expansion (or initial spin strain). To tune it, one may either vary the oscillation amplitude for a given initial strain or modify the initial strain while holding the amplitude fixed. \textbf{Fig. 14(c)} demonstrates successful frequency modulation by varying the number of DWs and the nanostrip width (hence the DW width; see \textbf{Fig. 6(a)}). Width tuning has a weaker effect, partly because the normalized DW width $\frac{W_{\text{DW}}}{W_{\text{DW,def}}}$ (with default \(W_{\text{DW,def}}=117.3 \,\mathrm{nm}\)) varies less than the normalized strip width $\frac{W_{\text{strip}}}{W_{\text{strip,def}}}$ (\(W_{\text{strip,def}}=128 \,\mathrm{nm}\)) (refer back to \textbf{Fig. 6(a))}—as illustrated by the gray dashed line, which assumes constant $\chi$ under varying strip width. An additional factor can be attributed to that $\chi$ itself increases (decreases) for topologically enlarged (reduced) DWs as the strip widens (narrows).

\subsection{\label{sec:level2}\textbf{Damping sensitivity} }
Finally, these oscillations are highly sensitive to damping. As shown in \textbf{Fig. 14(d)}, for a conventional damping constant \(\alpha=0.01\), the DWSS is overdamped. The critical damping coefficient is \(\alpha=0.0013\).

\subsection{\label{sec:level2}\textbf{Resonance spectrum} }
The natural frequency endows the DWSS with a resonant response. \textbf{Fig. 15} shows the resonance spectrum of a $\text{DWSS}_{\text{100}}$ driven by a sinusoidal spin-polarized current $j_{\text{sp}}$. A distinct peak appears at $\sim 18 \,\mathrm{MHz}$, coinciding with the natural frequency identified earlier. As expected, the resonance amplitude increases with decreasing damping.

\begin{figure}
    \centering
    \includegraphics[width=0.84\linewidth]{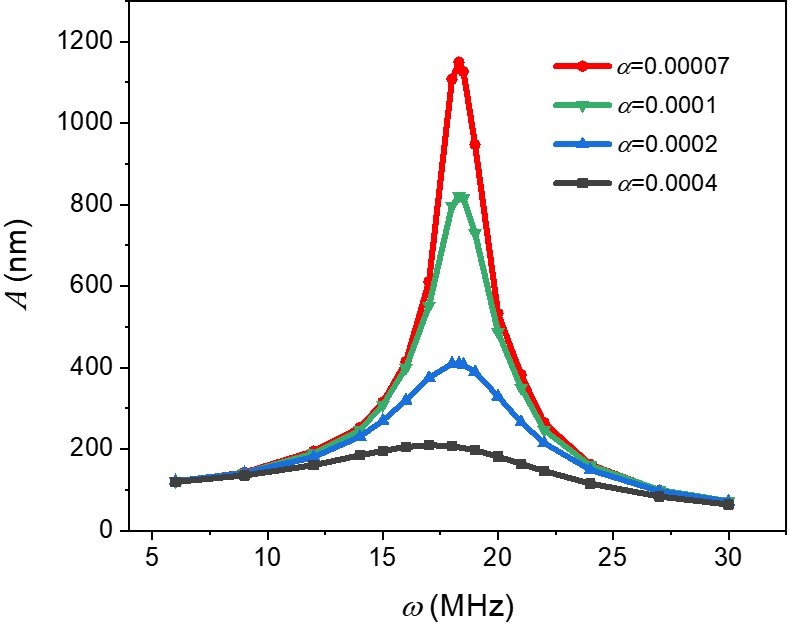}
    \caption{Resonance spectra of a $\text{DWSS}_{\text{100}}$ for different damping constants: \(\alpha=0.00007, 0.0001, 0.0002, 0.0004\). Forced oscillations are driven by a sinusoidal spin polarized current with a peak amplitude \(j_{\text{sp}} =2.0\times 10^{9} \,\mathrm{A/m^2}\).}
    \label{fig:placeholder}
\end{figure}

\section{\label{sec:level14}\textbf{\textbf{SPIN STRESS WAVES} } }

Elasticity and inertia—the two prerequisites for stress waves—are now established in spin elastomers, implying the existence of their dynamic counterpart: \textit{spin stress waves}. To test this prediction, we apply an \textit{x}-directional ac magnetic field to a narrow region (\(\mu_y\in(-1,1)\)) at the center of a DWSS. Spin waves are excited and propagate along the assembly, as evidenced by the $m_y$ profile recorded 1650 nm from the source (\textbf{Fig. 16(a)}). 

Remarkably, these spin waves carry with them a companion: a dynamically oscillating spin stress—a spin stress wave (\textbf{Fig. 16(b)})—at the same location, leading the local spin oscillation by $\pi/4$ in phase. That is, the maximum stressed state coincides with the equilibrium position of the local spin. A snapshot of the DW width distribution during propagation (\textbf{Fig. B.16}) vividly captures the spatial spread of periodic expansion and contraction—the hallmark of a spin stress wave. 

The concomitant transmission of spin elastic energy is shown in \textbf{Fig. 16(c)}. Notably, the energy density oscillates at the same frequency as the spin stress wave—rather than twice that frequency—and is $\pi/2$ out of phase, a departure from conventional expectations. This anomaly arises because, in a spin elastomer hosting oscillatory waves, fluctuations in soliton density are no longer negligible; they become the dominant factor governing energy density. Obviously, the spatial expansion and contraction of DWs are synchronized with spin stress wave, yielding identical frequencies (Energy storage and release within an entire DW, however, are expected to occur at twice the frequency). The observed $\pi/2$ phase shift follows naturally from the fact that peak stress corresponds to maximum DW expansion.

Conceptually, a spin stress wave is a subclass of spin waves—a collective excitation of ordered spins—distinguished by the oscillating stress it inherently carries. Its discovery provides an intuitive picture for characterizing the transmission of spin torque, energy, and deformation within spin elastomers. In particular, it enables the elucidation of nonequilibrium dynamic responses under external loading, thereby completing the theoretical edifice of spin elasticity. From an application perspective, the propagation characteristics of spin stress waves encode intrinsic details of the spin texture, providing a powerful probe for spin-texture imaging. Ultimately, spin stress waves themselves emerge as a compelling medium for information processing—a new degree of freedom in the quest for beyond-CMOS technologies.
 
\begin{figure}
    \centering
    \includegraphics[width=1\linewidth]{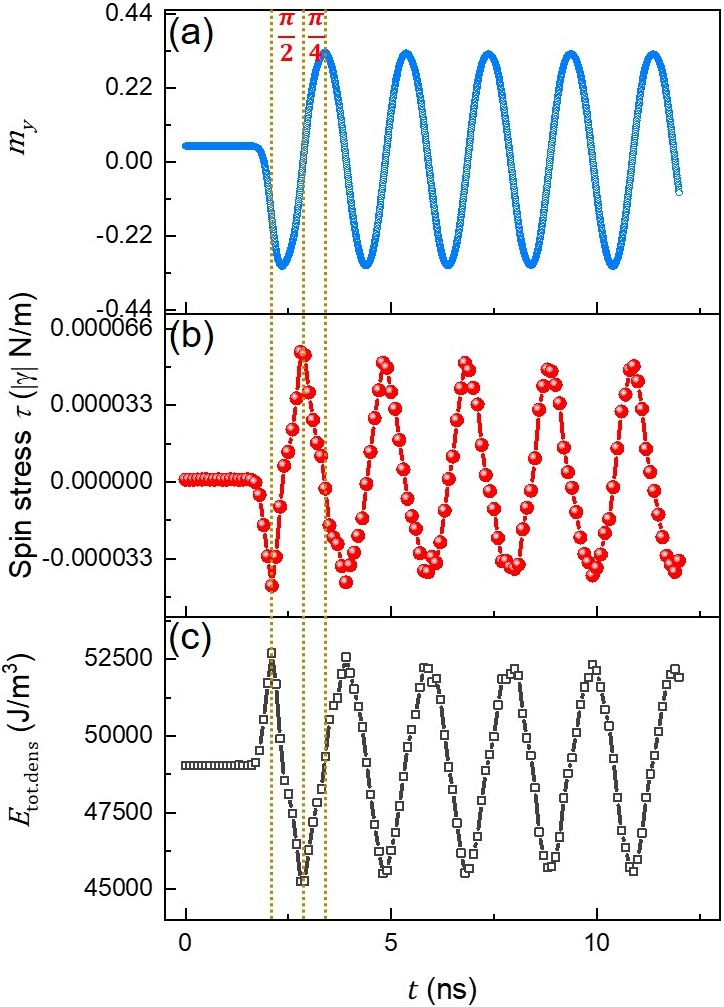}
    \caption{Comparison of (a) spin waves, (b) spin stress waves and (c) energy density waves in a DWSS. Waves are excited by an \textit{x}-directional ac magnetic field \(\textbf{H}_{\text{ac}}=h_0\sin(2\pi ft)\hat{\textbf{x}}\) (\(f=0.5\,\mathrm{GHz}\), \(h_0=50 \,\mathrm{Oe}\)) applied to a narrow region (\(\mu_y\in(-1,1)\)) at the center of a $\text{DWSS}_{\text{136}}$. The spin wave is recorded at a fixed spin site 1650 nm from the source. The spin stress wave is computed from \(\tau_{11}=\bar{E}_{\text{Y},\mu_{x}=1}\cdot D_{11}\), where \(\bar{E}_{\text{Y},\mu_{x}=1}=0.00098\) is the estimated average spin modulus at \(\mu_x=1\) (from \textbf{Fig. 9(e)}). The spin strain $D_{11}$ and the total energy density $ E_{\text{tot.dens}}$ are obtained by tracking the spin state of \(\mu_x=1\) near \(x=1650\,\mathrm{nm}\). The Gilbert damping coefficient is set to \(\alpha=10^{-6}\).}
    \label{fig:placeholder}
\end{figure}

\section{\label{sec:level15}\textbf{\textbf{\textbf{PERSPECTIVES AND PROOF-OF-CONCEPT TRIALS} } } }

The concept of spin elastomers opens a wide spectrum of possibilities across spintronics. Potential applications span spin-texture actuation, detection and metrology, high-frequency communication and microwave devices, energy and data storage, logic elements, electrical manipulations, magnonics, and magnetization engineering. To illustrate the scope and viability of these ideas, \textbf{Fig. 17} presents proof-of-concept designs based on the DWSS platform. These include: 

\begin{figure*}
    \centering
    \includegraphics[width=0.9\linewidth]{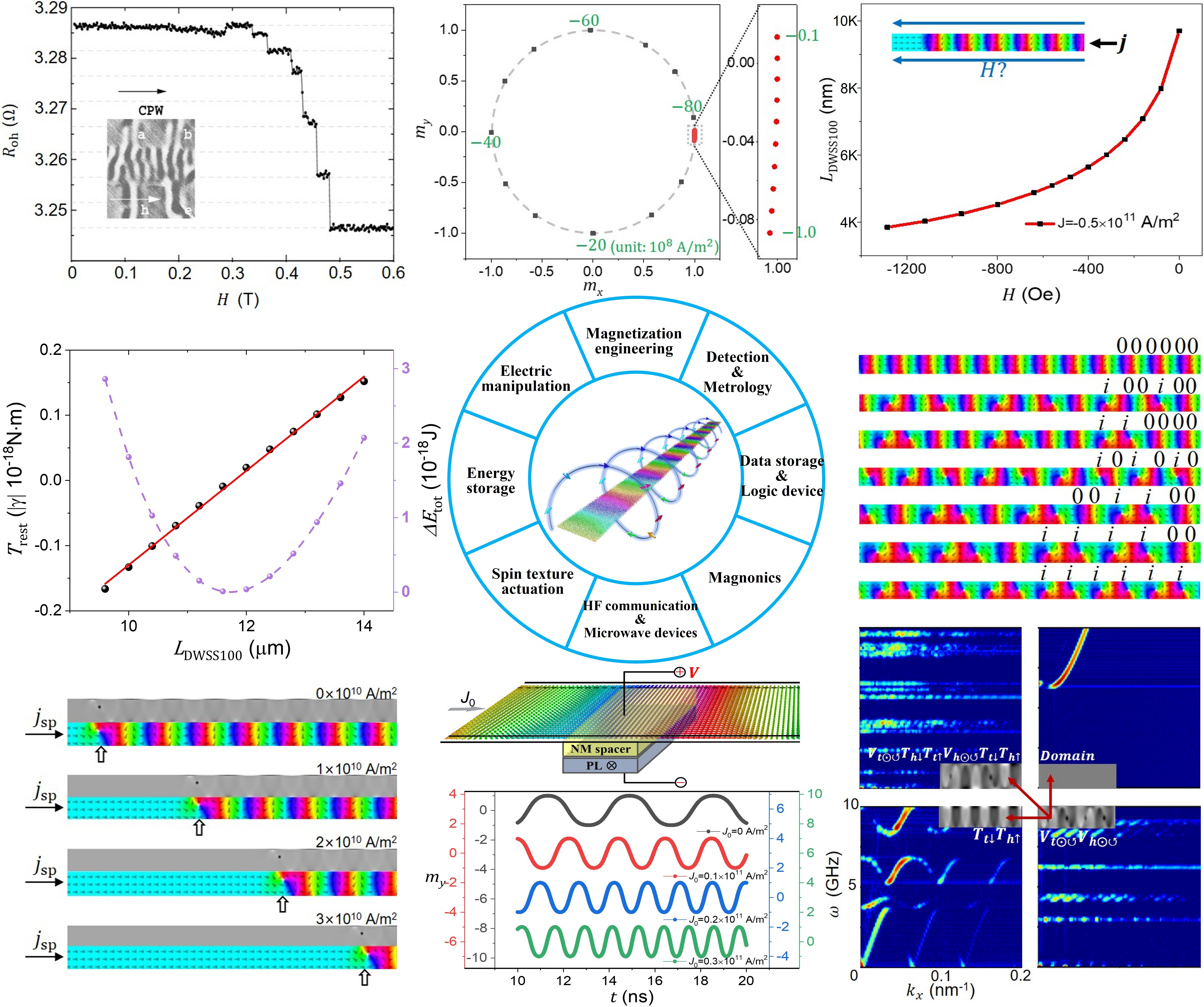}
    \caption{Proof-of-concept applications based on the DWSS platform. Upper-left panel adapted from Ref. [30], with permission from the American Physical Society (APS) (2002).}
    \label{fig:placeholder}
\end{figure*}

$\bullet$ Actuation of vortices and their stray fields

$\bullet$ Detection of magnetic fields, current density, and spin waves

$\bullet$ Tunable-frequency DWSS nano-oscillators

$\bullet$ Spin torque accumulators

$\bullet$ Energy storage units

$\bullet$ High-density racetrack memories

$\bullet$ DWSS-based varistors

$\bullet$ Tunable magnonic crystals

$\bullet$ Sub-nanometer domain control

$\bullet$ High-precision vectorial magnetization manipulation

Detailed modeling and performance are provided in \textbf{Appendix C}. We envision that the principles established here will stimulate the exploration of spin elastomers across a wide range of materials, geometries and configurations, ultimately enabling their integration into functional spintronic devices. Experimental realization of the proposed concepts is the next—and most exciting—step.

\section{\label{sec:level16}\textbf{\textbf{\textbf{\textbf{CONCLUSIONS} } } } }

We have introduced the concept of spin elasticity—a framework that links spin torque to the geometry of spin textures. The key advances reported in this work are summarized as follows:

1. \textbf{Concept and construction. }We proposed the spin elastomer, establishing an assembly pathway rooted in magnetic solitons and magnetization homotopy.

2. \textbf{Intersolitonic interaction.} The first comprehensive interaction diagram for magnetic solitons was mapped, revealing that topological protection transforms magnetostatic attraction into a force landscape reminiscent of atomic potentials.

3. \textbf{Topological Hooke's law.} A large-range linear relation between restoring spin torque and curvilinear deformation of spin elastomer was demonstrated, revealing a distinct topology-aware Hooke's law in the spin degree of freedom.

4. \textbf{Poisson effect. }Lateral contraction/expansion under longitudinal elongation/compression persists in spin elastomers, but with a nonuniform, deformation-dependent Poisson's ratio.

5. \textbf{Spin elastic potential energy. }Spin elastic deformation enables reversible magnetic energy storage, with exchange and dipolar interactions playing distinct roles under compression and tension. 

6. \textbf{Spin elastic theory.} A continuum theory of spin elasticity was developed, introducing spin strain $\textbf{D}$ and spin stress $\bm{\tau}$ as fundamental descriptors that encode the full deformation of spin configurations.

7. \textbf{Spin stress-strain analysis. }Applied to a DWSS, the $\bm{\tau}-\textbf{D}$ framework revealed nonuniform spin strain distribution, non-conservation of spin stress, site-dependent moduli and the spectrum of elastic modes, etc. 

8. \textbf{Electrical control. }Electrical manipulation via spin-polarized currents was demonstrated, uncovering a current-driven spin stress accumulation effect.

9. \textbf{Spin elastodynamics. }Based on high-symmetry points, we derived the equations of spin elastic equilibrium and elastodynamics. 

10. \textbf{Spin elastic oscillations and resonance.} Spin elastomers manifest spontaneous oscillations and resonance, exhibiting a stepwise hopping deformation mode and a tunable natural frequency.

11. \textbf{Spin stress waves.} The existence of spin stress waves was predicted and verified—hallmark of a new class of collective excitations. 

12. \textbf{Applications.} A broad spectrum of potential applications—ranging from actuation and detection to storage and modulation—was proposed and validated through proof-of-concept trials, charting a pathway toward spin elastomer-based technologies. 

Together, these advances establish spin elasticity as a new discipline, unifying the elastic behavior of matter and spin within a single theoretical framework. 

\begin{acknowledgments}
This work was supported by National Key R\&D Program of China (Grants No. 2025YFA1411302 and No. 2022YFA1402802), the National Natural Science Foundation of China (NSFC) (Grants No. 12374103, No. 12434003, No. 12134017, No. 12574131, No. 11974250 and No. U2541261), and Sichuan Science and Technology Program (Grant No. 2025NSFJQ0045). Z. G. conceived the research, developed the theory, performed the numerical calculations, and wrote the paper. T. Z., F. W., J. H., P. Y. and X. H. commented on the manuscript. All authors read and approved the final manuscript. 
\end{acknowledgments}

\appendix

\counterwithin{figure}{section}

\renewcommand{\figurename}{FIG.}

\section{\textbf{MAGNETIC SETUPS AND METHODS} }

Metastable magnetic solitons come in many forms, and spin elastomers built from them are correspondingly diverse. To isolate their signature properties, we consider a soft ferromagnetic material—governed solely by exchange and dipole-dipole interactions—confined to a nanostrip geometry. This confinement restricts deformation to one dimension and supports two types of domain walls (DWs): transverse walls and vortex walls. Depending on the polarity, chirality and surrounding magnetization orientation of the walls, there exists 12 subspecies (illustrated in \textbf{Fig. A.1}): 

$\bullet$ Transverse walls: $T_{h\uparrow}$, $T_{h\downarrow}$, $T_{t\uparrow}$, $T_{t\downarrow}$,

$\bullet$ Vortex walls: $V_{h\odot \circlearrowleft}$, $V_{h\odot \circlearrowright}$, $V_{h\otimes \circlearrowleft}$, $V_{h\otimes \circlearrowright}$, $V_{t\odot \circlearrowleft}$, $V_{t\odot \circlearrowright}$, $V_{t\otimes \circlearrowleft}$, $V_{t\otimes \circlearrowright}$. 

\begin{figure*}
    \centering
    \includegraphics[width=0.9\linewidth]{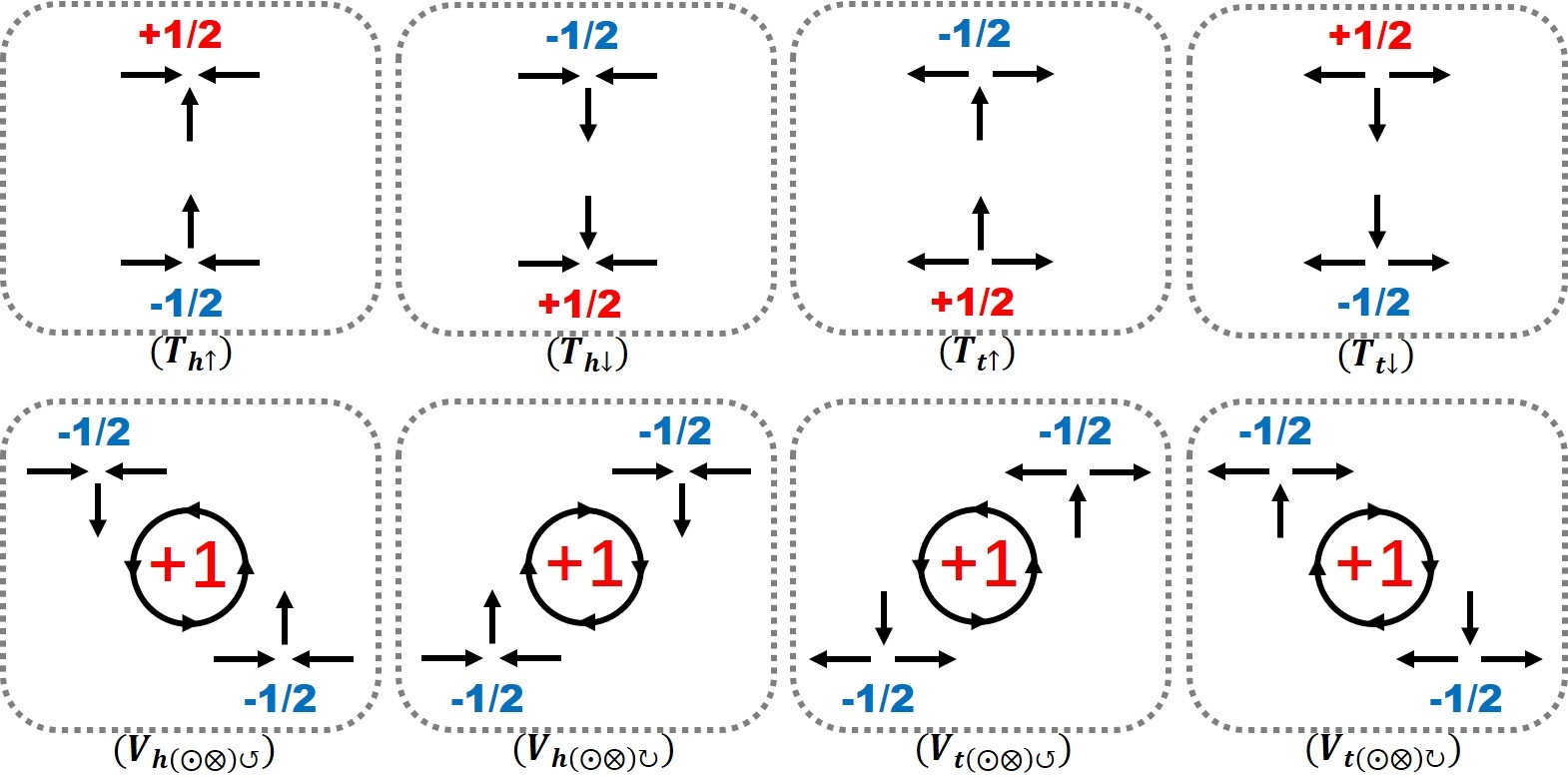}
    \caption{Topological charge distributions (-1/2,+1/2 and +1 [31]) distributions for the 12 domain-wall subspecies in a soft-material nanostrip: $T_{h\uparrow}$, $T_{h\downarrow}$, $T_{t\uparrow}$, $T_{t\downarrow}$, $V_{h\odot \circlearrowleft}$, $V_{h\odot \circlearrowright}$, $V_{h\otimes \circlearrowleft}$, $V_{h\otimes \circlearrowright}$, $V_{t\odot \circlearrowleft}$, $V_{t\odot \circlearrowright}$, $V_{t\otimes \circlearrowleft}$ and $V_{t\otimes \circlearrowright}$ here $h/t$ denotes head-to-head or tail-to-tail type, $\odot/\otimes$ indicates polarity, and $\circlearrowleft/\circlearrowright$ specifies chirality. }
    \label{fig:placeholder}
\end{figure*}

Applying the topological principle introduced in \textbf{Sec. III}, one can determine the complete set of domain-wall pairs that remain stable against annihilation upon direct contact (\textbf{Fig. A.2}). By permuting and combining these viable pairs, a virtually limitless family of domain-wall spin springs (DWSSs) can be constructed. \textbf{Fig. A.3} displays several representative DWSSs, together with their topologically metastable winding trajectories in order-parameter space $\mathbb{S}^2$.
\begin{figure*}
    \centering
    \includegraphics[width=0.5\linewidth]{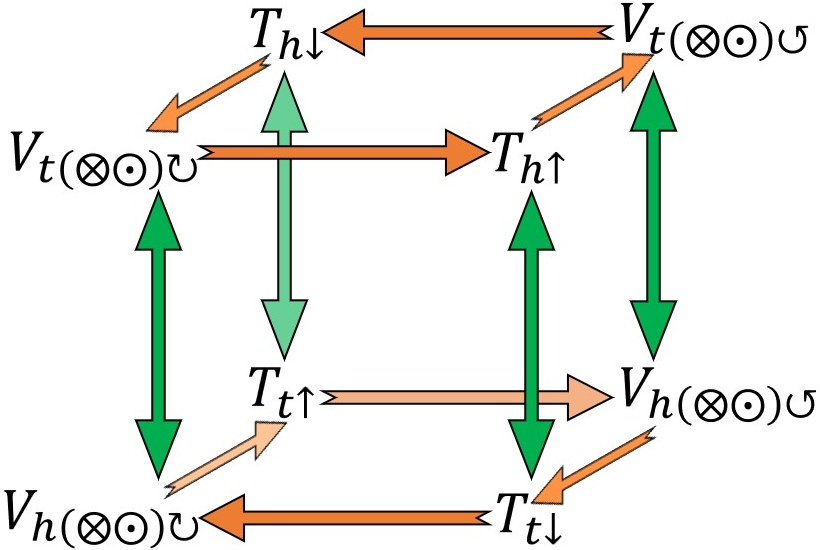}
    \caption{Allowed adjacent pairings between transverse (\textit{T}) and vortex (\textit{V}) domain walls. Arrows indicate permitted left-to-right arrangements without topological annihilation }
    \label{fig:placeholder}
\end{figure*}

\begin{figure*}
    \centering
    \includegraphics[width=0.9\linewidth]{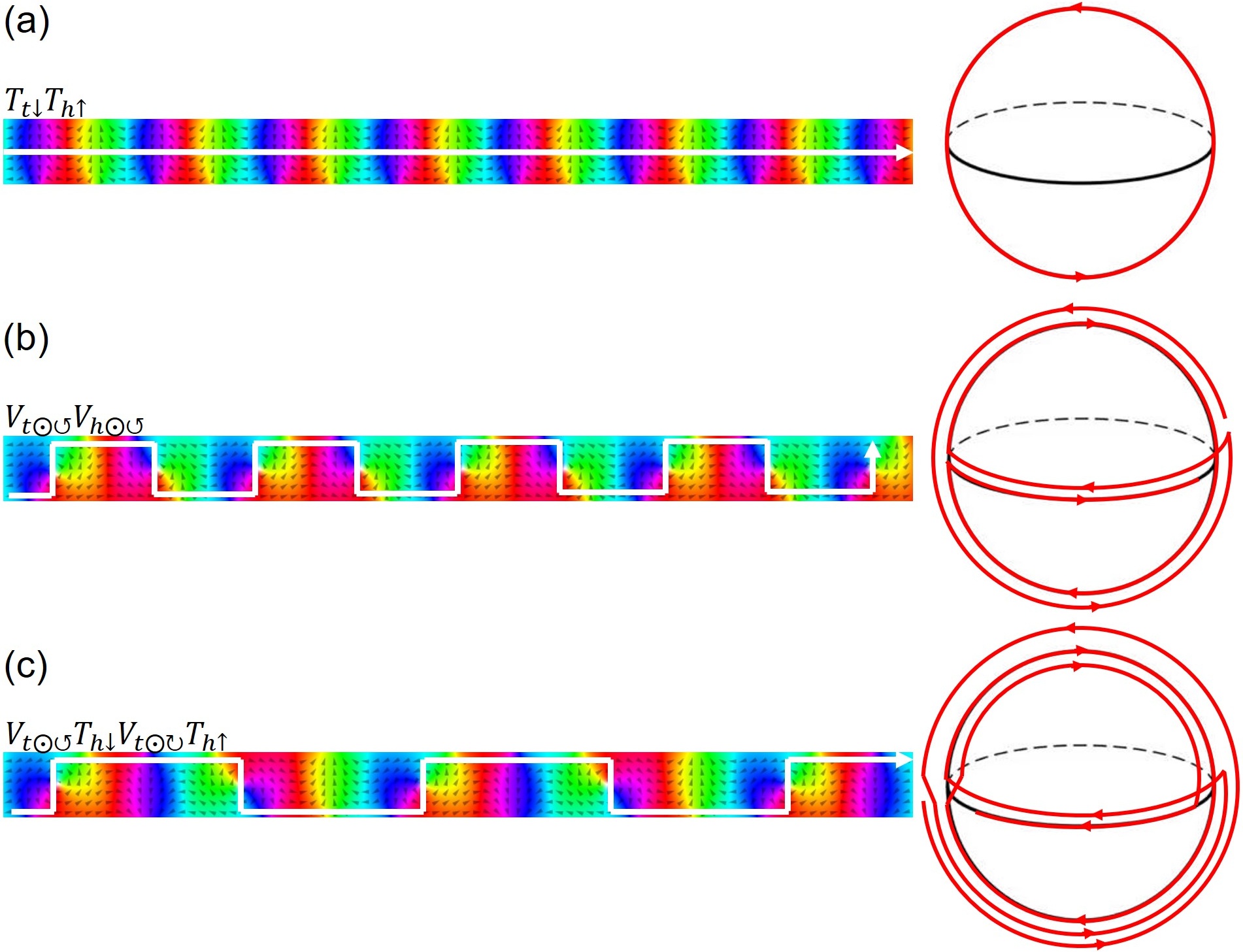}
    \caption{Domain-wall spin springs (DWSSs) assembled from different wall sequences: (a) $T_{t\downarrow}T_{h\uparrow}$, (b) $V_{t\odot \circlearrowleft}V_{h\odot \circlearrowleft}$, and (c) $V_{t\odot \circlearrowleft}T_{h\downarrow}V_{t\odot \circlearrowright}T_{h\uparrow}$. White polylines trace continuous spin rotations along the spring. Their corresponding topologically nontrivial winding trajectories in order-parameter space $\mathbb{S}^2$ are shown to the right.}
    \label{fig:placeholder}
\end{figure*}

\begin{figure*}
    \centering
    \includegraphics[width=0.8\linewidth]{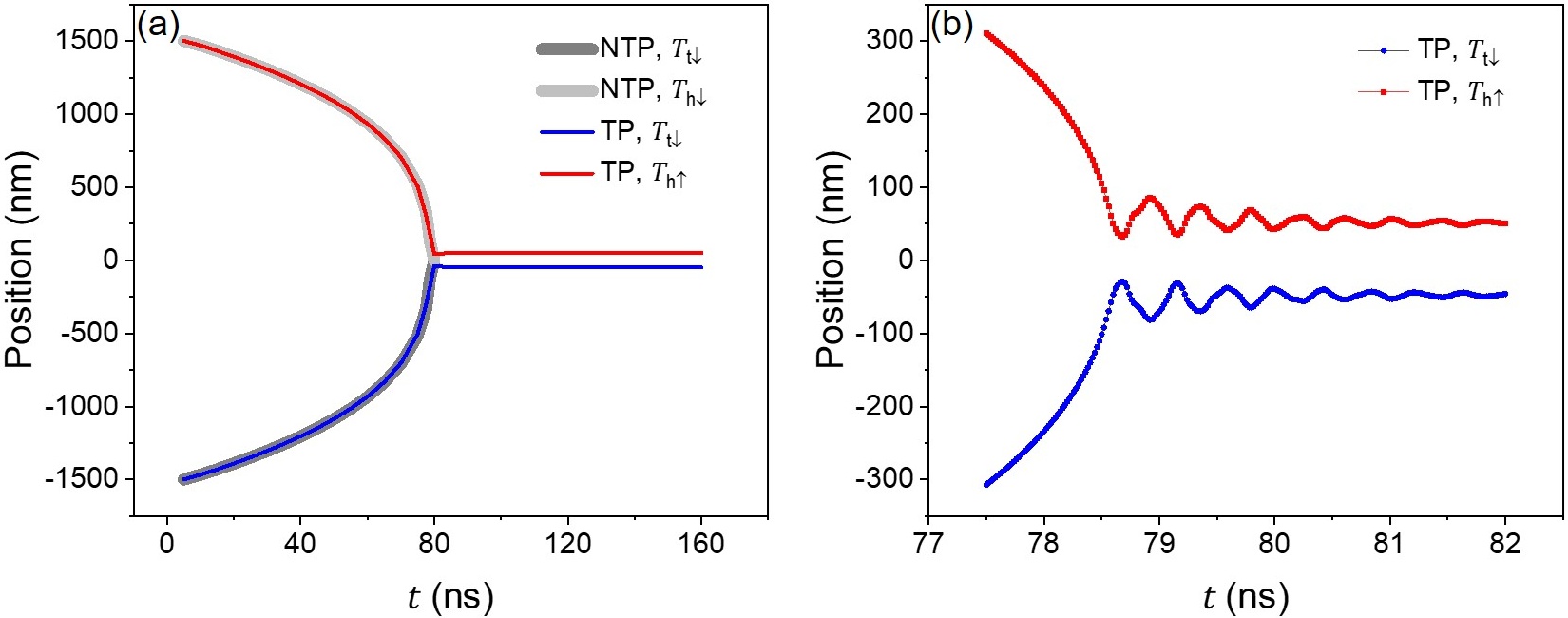}
    \caption{(a) Collisions between two 180° domain walls. TP (NTP) denotes cases with (without) topological protection. (b) Enlarged view of the interaction between two topologically protected 180° walls: $T_{t\downarrow}$ and $T_{h\uparrow}$. Simulations are performed using the magnetic setups described below, with the two walls initially separated by 3000 nm.}

        \label{fig:placeholder}
\end{figure*}

\begin{figure*}
    \centering
    \includegraphics[width=0.8\linewidth]{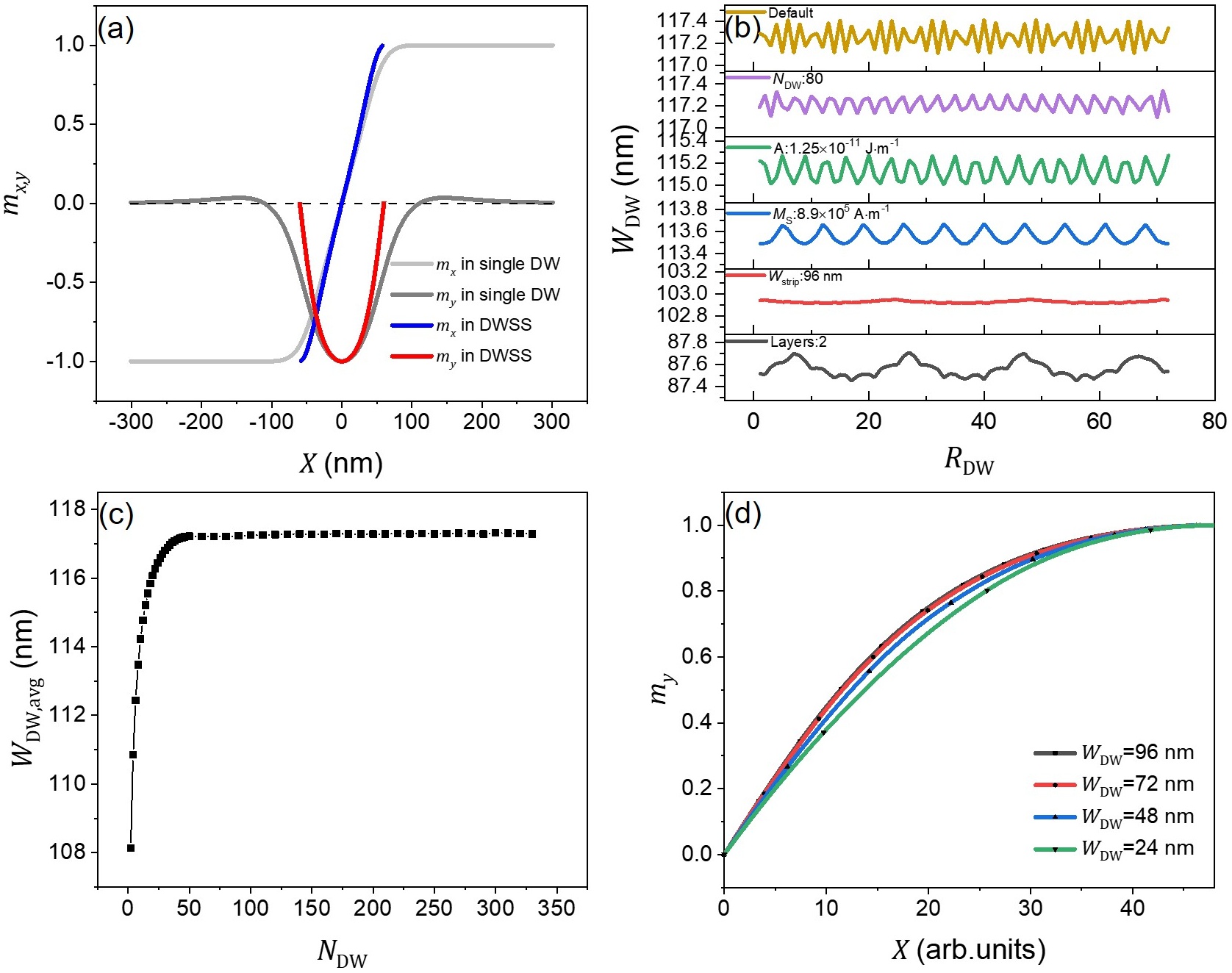}
    \caption{Basic characteristics of the DWSS. (a) Comparison of $m_x$ and $m_y$ profiles in a DWSS (\(N_{\text{DW}}=100\)) with those of an isolated DW. The equilibrium derivative of the in-plane spin angle $\theta$ at the DW boundary is \(\partial_x \theta|_{\text{eq.bound}}\approx0.0377\,\mathrm{rad/nm}\). (b) DW width distribution patterns for different magnetic setups; parameters modified from the default settings are highlighted. (c) Averaged DW width $W_{\text{DW,avg}}$ as a function of the number of DWs $N_{\text{DW}}$. A steady width of $\sim 117.3 \,\mathrm{nm}$ is attained above a critical number \(N_{\text{crit.DW}}=40\). (d) $m_y$ profiles of DWs of different widths: 24, 48, 72 and 96 nm.}
    \label{fig:placeholder}
\end{figure*}

Technologically, DWSSs can be prepared by consecutive injection of domain walls [32,33]; they then self-assemble under the combined action of long-range magnetostatic attraction and short-range topological repulsion. \textbf{Fig. A.4} captures the collisional process of two 180° walls, revealing that the inter-solitonic attraction remains effective even at micrometer separations, uniting the two walls within tens of nanoseconds. This property is particularly valuable, as it exempts the DWSS from tensile limitations during assembly. Once formed, domain walls can be set into motion by a variety of stimuli that carry spin torque—magnetic fields [34,35], spin-polarized currents [36,37], spin waves [38], laser pulses [39], and temperature gradients [40]—all of which are expected to actuate DWSSs with similar effectiveness. Conversely, walls can be pinned by cutting notches [41], modulating local material properties (e.g., anisotropy) [42], or introducing inhomogeneities such as impurities, doping, disorders, or interfacial defects [43]. Such pinning prevents bodily movement of the DWSS and enables controlled deformation. These features make the DWSS an excellent platform for the initial exploration of spin elastomers. In this work, we focus on a simple DWSS built from the pair $T_{t\downarrow}T_{h\uparrow}$ (see \textbf{Fig. A.2}). Notably, spins along the transverse direction of the nanostrip are nearly uniformly oriented, closely approximating a one-dimensional spin chain.

To resolve nanoscale details of the elastic behavior of DWSSs, we perform micromagnetic simulations using the $\text{MuMax}^3$ code [44]. The time evolution of the magnetization is governed by the Landau-Lifshitz-Gilbert (LLG) equation:
\begin{equation}
\frac{\partial \textbf{m}}{\partial t}=-\gamma \textbf{m}\times \textbf{h}_{\text{eff}}+\alpha \textbf{m} \times \frac{\partial \textbf{m}}{\partial t},  
\end{equation}
where $\textbf{m}$ is the normalized magnetization, $\gamma$ the gyromagnetic ratio, $\textbf{h}_{\text{eff}}$ the effective ﬁeld and $\alpha$ the Gilbert damping constant. To isolate systems dominated solely by exchange and dipole-dipole interactions, material parameters are taken from permalloy (Py), which also exhibits excellent DW mobility [35]: saturation magnetization \(M_{\text{S}}=8.6\times 10^5 \,\mathrm{Am^{-1}}\), exchange stiffness \(A=1.3\times 10^{-11} \,\mathrm{Jm^{-1}}\), magnetocrystalline anisotropy \(K=0\), and Gilbert damping \(\alpha=0.01\). For simulations involving spin-polarized currents, the nonadiabatic spin-transfer torque parameter \(\beta=0.1\) and the current polarization \(P=0.56\) are used. The simulation cell size is $4\times4\times5 \,\mathrm{nm^{3}}$. The nanostrip grid is typically: $N\times32\times1$, where $N$ depends on the number and size of the DWs. To mimic an infinitely long strip while providing boundary confinement for the DWSS, spins at the ends are artificially pinned with their surface charge removed [45-48]. These settings constitute the default simulation conditions throughout this work, unless otherwise specified.

\textbf{Fig. A.5} summarizes the basic characteristics of DWSS. Owing to close packing, the asymptotic boundary condition of isolated domain walls is lost, and each DW acquires a well-defined width (\textbf{Fig. A.5(a)}). The width is not uniform along the longitudinal axis but exhibits exquisite periodical patterns depending on sample parameters (\textbf{Fig. A.5(b)}). These patterns reflect a subtle competition between exchange and dipole-dipole interactions; nevertheless, the width variation in each case remains within a narrow range of 0.3 nm. The average DW width as a function of the number of DWs $N_{DW}$ is plotted in \textbf{Fig. A.5(c)}. Unlike a classical spring, the total length $L_{\text{DWSS}}$ does not scale linearly with $N_{DW}$ until a critical number \(N_{\text{crit.DW}}=40\) is reached. For this reason, most simulations in this work use \(N_{\text{DW}}=100\) (hence $\text{DWSS}_{100}$), for which the relaxed average DW width is 117.3 nm. Regarding the detailed DW profile: when only local energy terms (exchange and anisotropy) are present, the profile can be obtained by solving the Euler equations with explicit “functional derivatives” (e.g., the Walker profile [34]). However, for problems concerning wide DWs and their deformation, the global dipole-dipole interaction cannot be neglected; accurate profiles require experimental measurement or numerical calculation. \textbf{Fig. A.5(d)} shows that profiles for different DW widths cannot be described by a single analytical form; as the width narrows, the $m_y$ profile contracts inward. 

\section{\textbf{ADDITIONAL RESULTS AND DISCUSSION}}

This section provides further evidence supporting the main conclusions, detailing the static and dynamic behavior of DWSS through additional simulations and analytical results. 

\begin{figure*}
    \centering
    \includegraphics[width=0.5\linewidth]{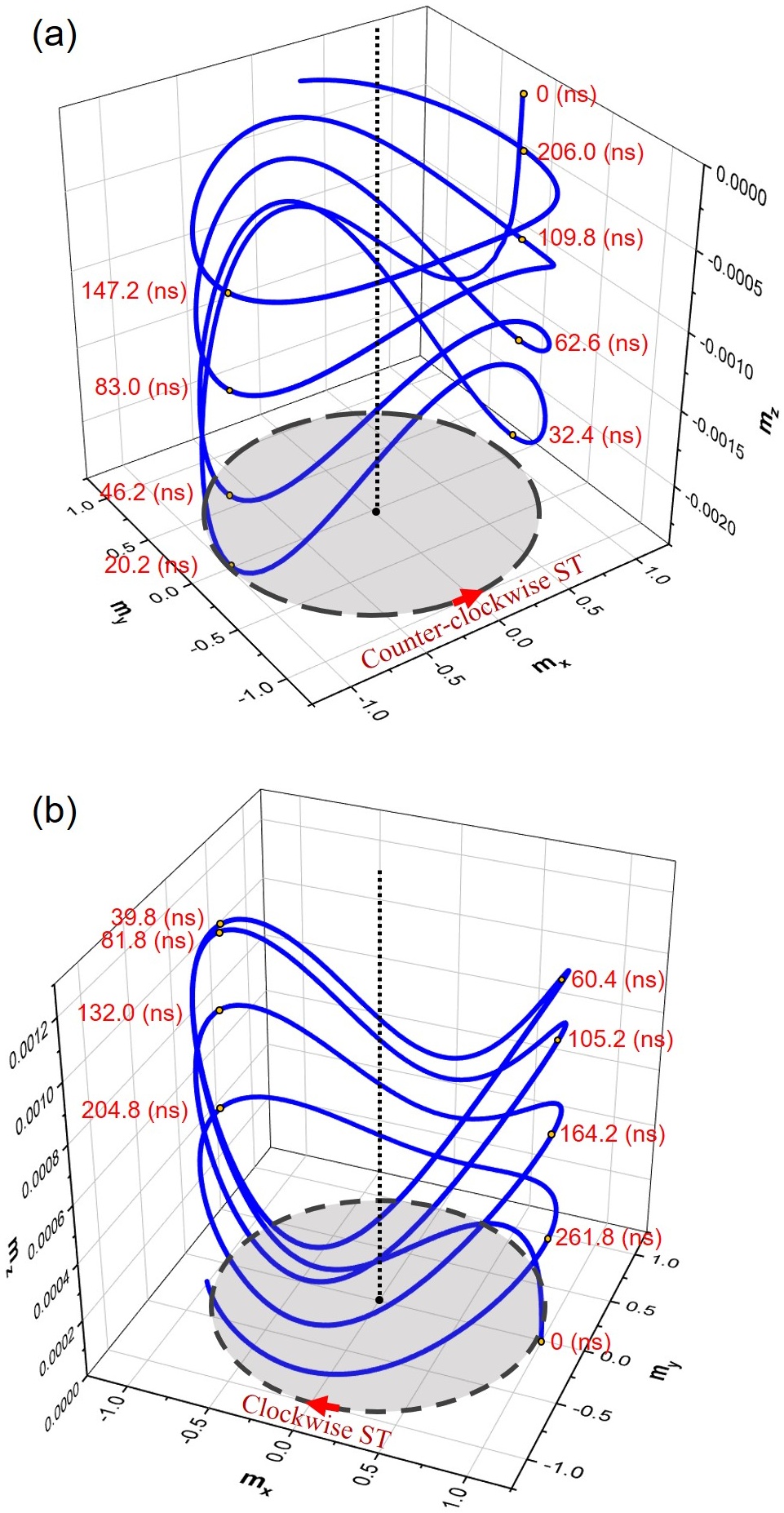}
    \caption{Evolution of magnetization orientation for a local spin initially aligned along \(\textbf{m}=(1,0,0)\) during free relaxation of (a) a compressed $\text{DWSS}_{100}$ (average DW width \(W_{\text{DW}} =96\,\mathrm{nm}\)) and (b) a stretched $\text{DWSS}_{100}$ (average DW width \(W_{\text{DW}} =96\,\mathrm{nm}\)). In the absence of a rigid constraints at the left end (see \textbf{Fig. 4(a)}), the imbalance of spin torque induces a minute but rapid out-of-plane tilt—downward for compression, upward for tension. This tilting and the strong in-plane effective field, through the precessional torque $- \lvert \gamma \rvert \textbf{m} \times \textbf{H}$, drives counter-clockwise (compression) or clockwise (tension) in-plane rotation, leading to expansion or contraction of the DWSS. Notably, a smaller $\lvert m_z \rvert $ results in a smaller in-plane spin torque and thus a longer precession period, as highlighted by the marked time points.}
    \label{fig:placeholder}
\end{figure*}

\begin{figure*}
    \centering
    \includegraphics[width=0.55\linewidth]{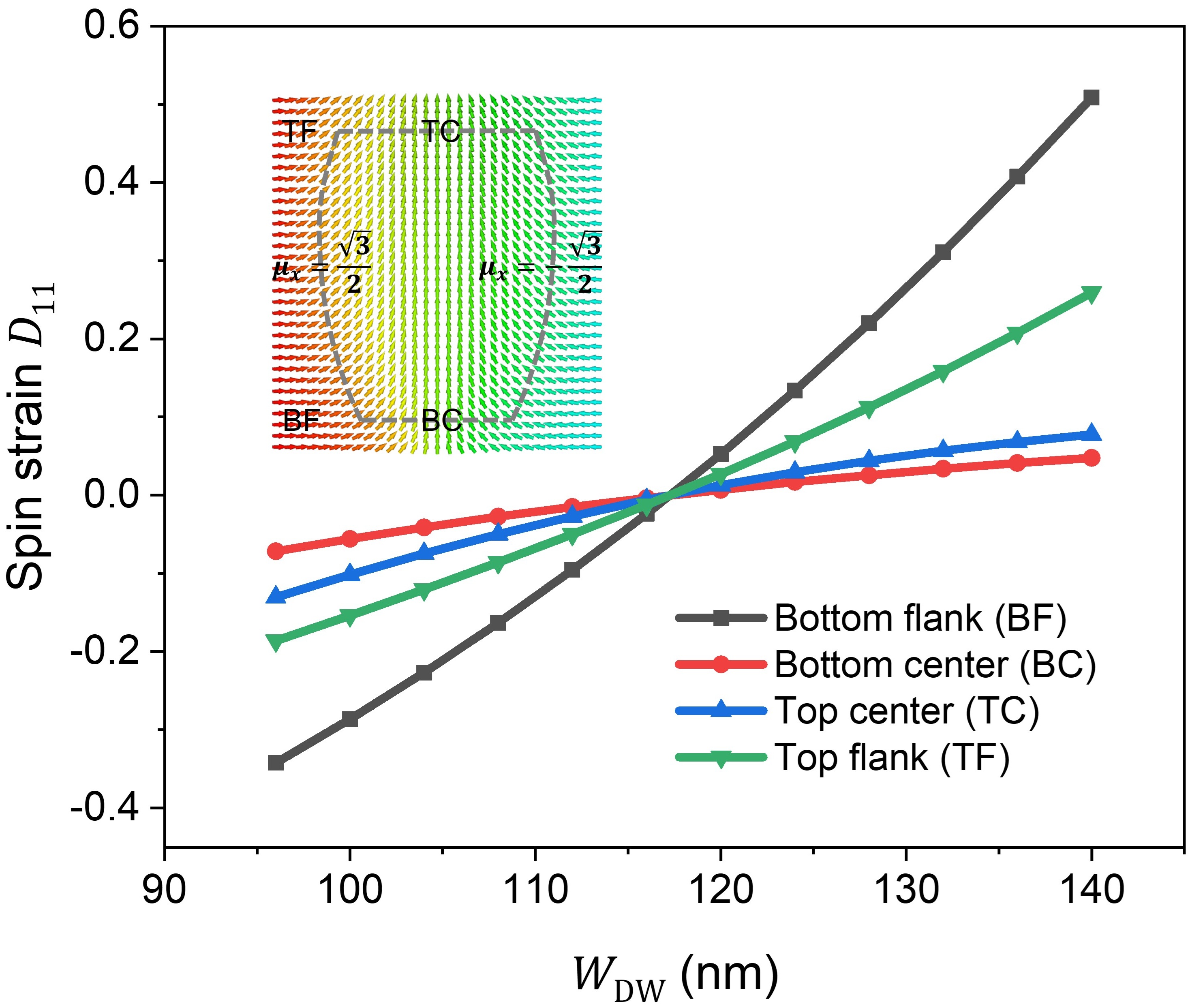}
    \caption{Spin strain $D_{11}$ as a function of DW width $W_{\text{DW}}$ at four representative locations within the DW: bottom flank (\(\mu_x =1\),\(y =18\,\mathrm{nm}\)), bottom center (\(\mu_x =0\),\(y =18\,\mathrm{nm}\)), top center (\(\mu_x =0\),\(y =110\,\mathrm{nm}\)), and top flank (\(\mu_x =1\),\(y =110\,\mathrm{nm}\)).}
    \label{fig:placeholder}
\end{figure*}

\begin{figure*}
    \centering
    \includegraphics[width=0.55\linewidth]{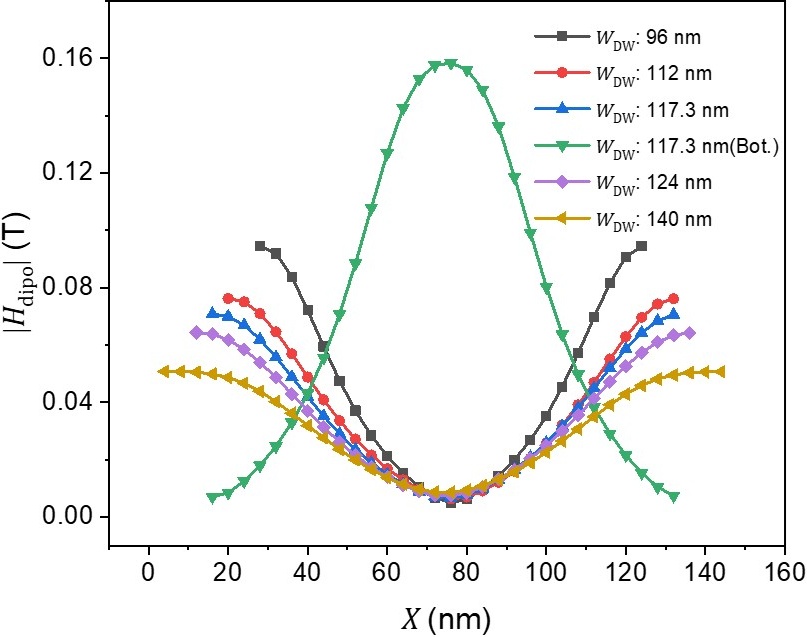}
    \caption{Magnitude of the dipolar field $\lvert \textbf{H}_{\text{dipo}} \rvert$ along the central axis of DWs for different DW widths: \(W_{\text{DW}} =96, 112, 117.3, 124, 140\,\mathrm{nm}\). The green curve corresponds to the profile at the bottom of a DW with width \(W_{\text{DW}} =117.3\,\mathrm{nm}\).}
    \label{fig:placeholder}
\end{figure*}

\begin{figure*}
    \centering
    \includegraphics[width=0.9\linewidth]{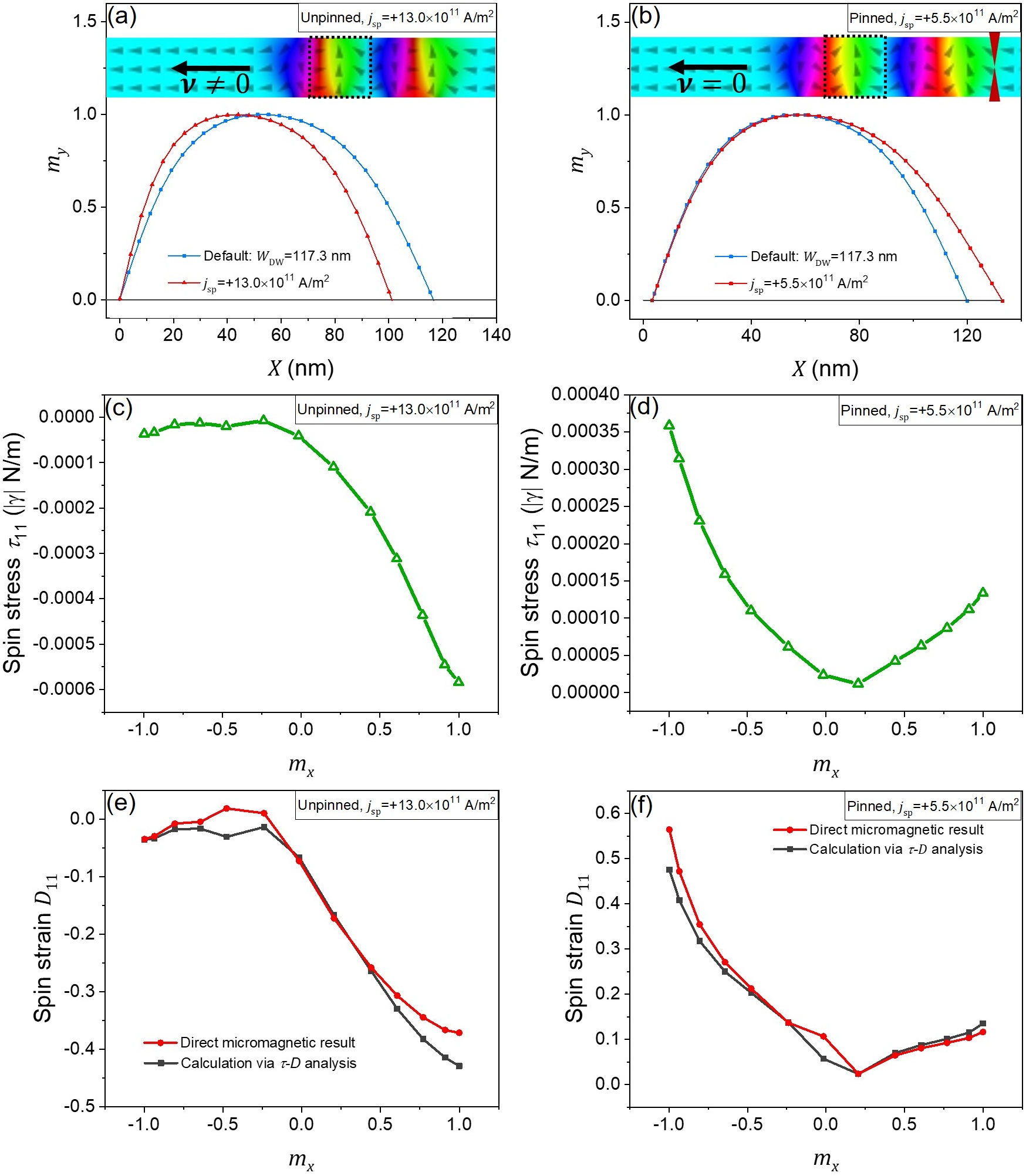}
    \caption{Asymmetric DW deformation and $\tau-D$ validation. (a) Asymmetric deformation of the second DW in an unpinned $\text{DWSS}_4$ under a large spin-polarized current \(j_{\text{sp}} =+13.0\times 10^{11} \,\mathrm{A/m^2}\), compared with the relaxed profile. (b) Corresponding deformation for a pinned DWSS4 under \(j_{\text{sp}} =+5.5\times 10^{11} \,\mathrm{A/m^2}\). (c,d) Spin stress transfer curves for cases (a) and (b). (e,f) Comparison of spin strain $D_{11}$ from $\tau-D$ analysis and direct micromagnetic simulations for (a) and (b). The spin modulus profiles $E_{\text{Y}}$ used in $\tau-D$ analysis accommodate the asymmetric deformation: for (e), \(W_{\text{DW}} =86.6\,\mathrm{nm}\) (left half) and \(W_{\text{DW}} =115.8\,\mathrm{nm}\) (right half); for (f), \(W_{\text{DW}} =117.3\,\mathrm{nm}\) (left half) and \(W_{\text{DW}} =140\,\mathrm{nm}\) (right half). The slight discrepancy in $D_{11}$ may arise from the coarse assignment of $E_{\text{Y}}$.}
    \label{fig:placeholder}
\end{figure*}

\begin{figure*}
    \centering
    \includegraphics[width=0.8\linewidth]{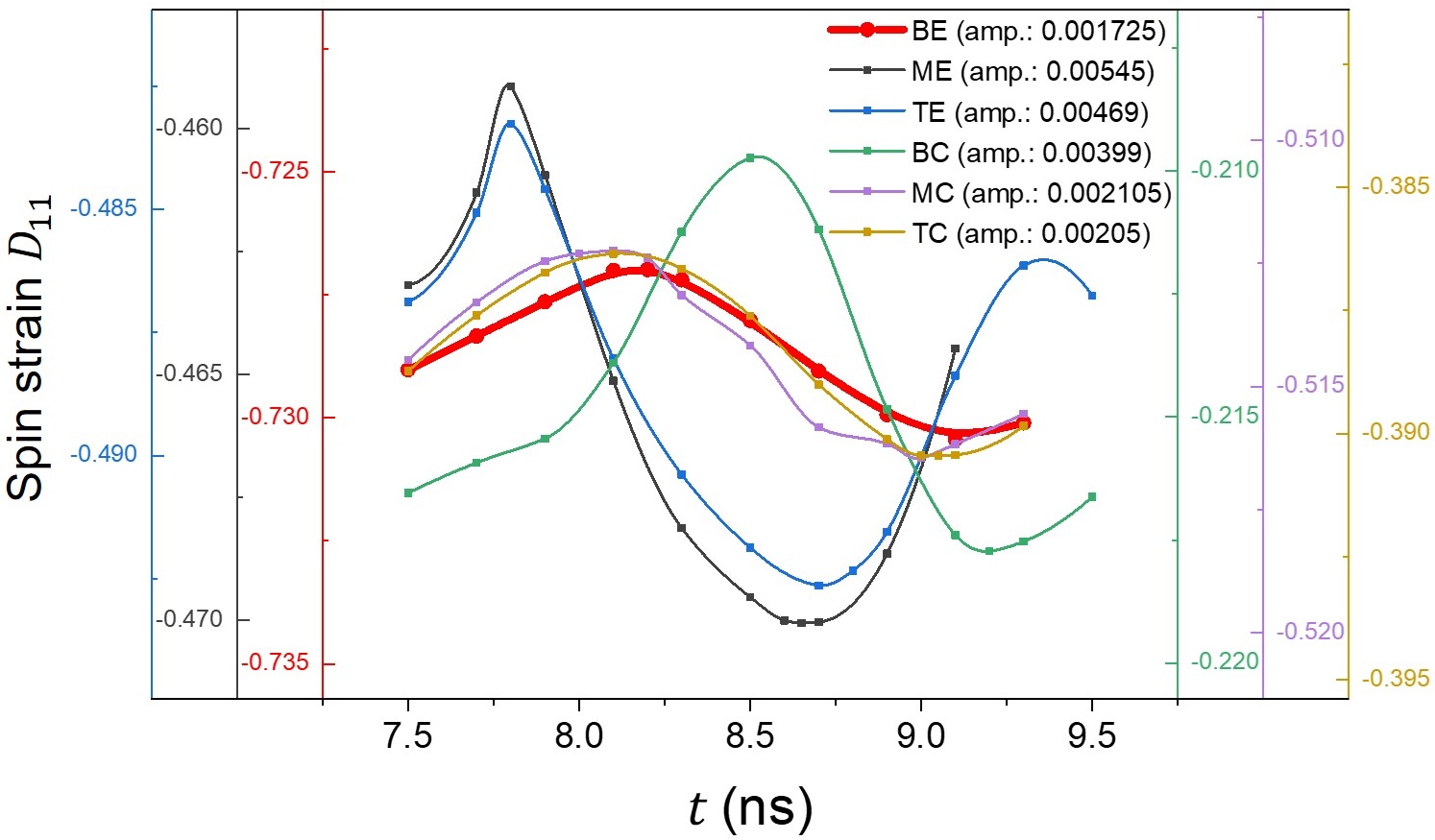}
    \caption{Strain oscillations recorded at six locations within a DW under propagation of a spin stress wave. The $\text{DWSS}_{100}$ is geometrically confined to \(W_{\text{DW}} =60\,\mathrm{nm}\). Measurements are taken near spin state \(\mu_x =1\) at the bottom edge (BE), middle edge (ME), and top edge (TE), and near \(\mu_x =0\) at the bottom center (BC), middle center (MC), and top center (TC). The DW under study is located 1770 nm from the wave source. The spin stress wave is excited using the same magnetic parameters, method, and ac field $\textbf{H}_{\text{ac}}$ as described in \textbf{Sec. XIV}.}
    \label{fig:placeholder}
\end{figure*}

\begin{figure*}
    \centering
    \includegraphics[width=0.93\linewidth]{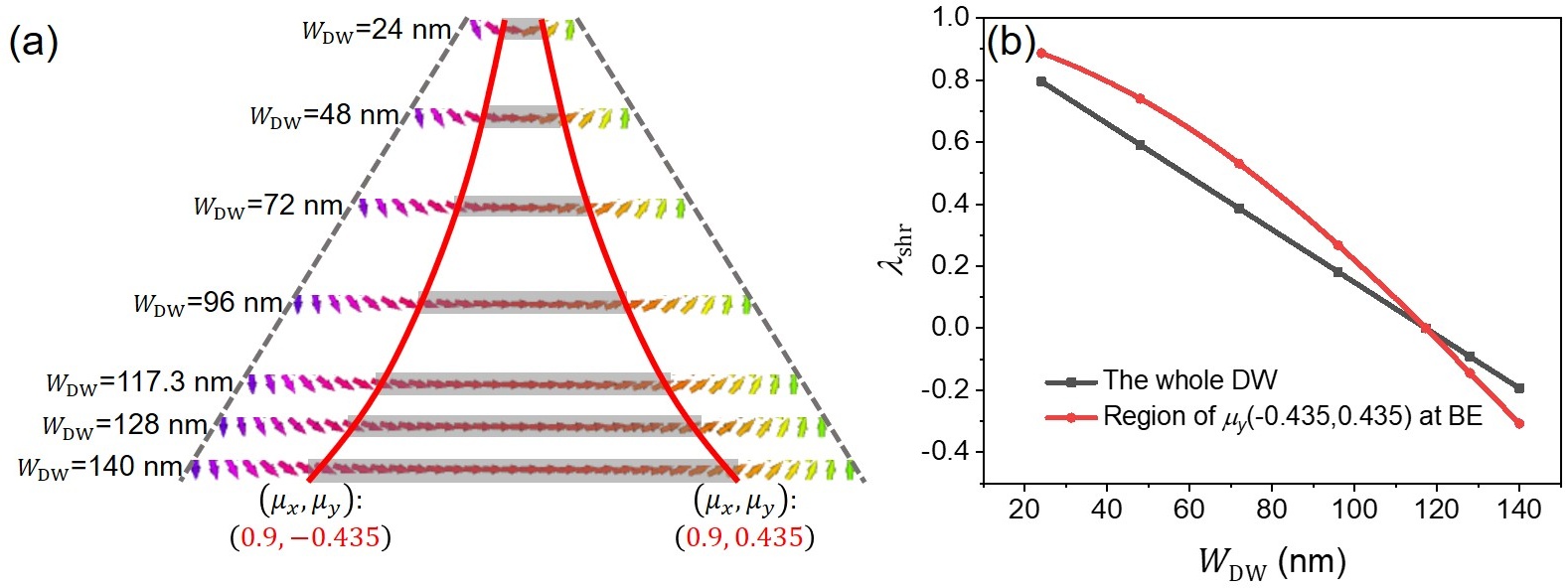}
    \caption{(a) Schematic of the region \(\mu_y\in(-0.435,0.435)\) at the bottom edge (BE) of domain walls with different widths. (b) Shrinkage ratio $\lambda_{\text{shr}}$ of this region compared with that of the entire DW.}
    \label{fig:placeholder}
\end{figure*}

\begin{figure*}
    \centering
    \includegraphics[width=0.8\linewidth]{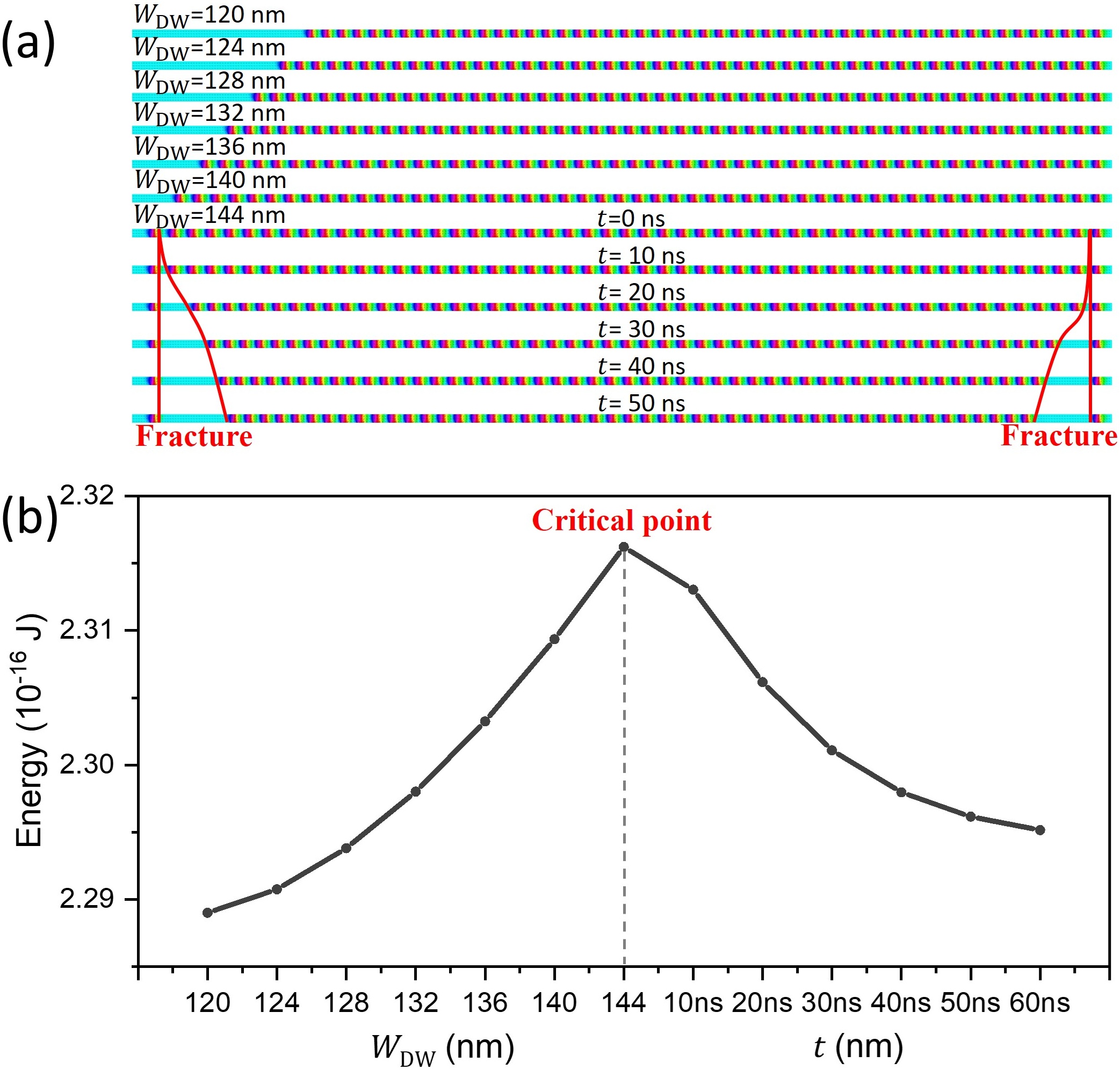}
    \caption{(a) Tensile expansion of a $\text{DWSS}_{100}$ under increasing load at boundary, leading to fracture when the DW width reaches a critical value \(W_{\text{DW}} =144\,\mathrm{nm}\). (b) System energy as a function of $W_{\text{DW}}$ (before fracture) and $t$ (after fracture).}
    \label{fig:placeholder}
\end{figure*}

\begin{figure*}
    \centering
    \includegraphics[width=0.55\linewidth]{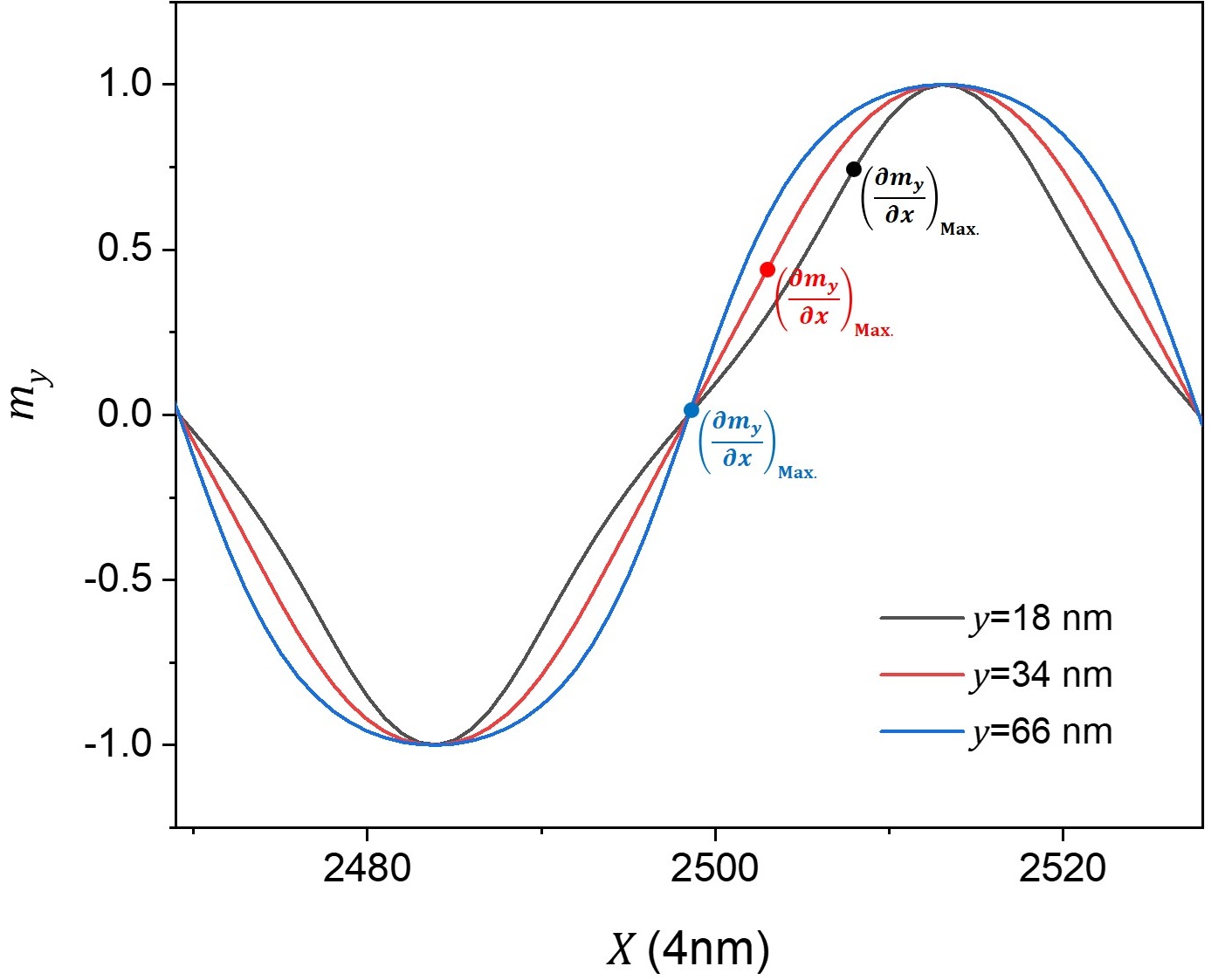}
    \caption{Positions of the maximum gradient $\frac{\partial {m_y}}{\partial x}$ in selected spin rows: \(y=18, 34, 66\,\mathrm{nm}\).}
    \label{fig:placeholder}
\end{figure*}

\begin{figure*}
    \centering
    \includegraphics[width=0.6\linewidth]{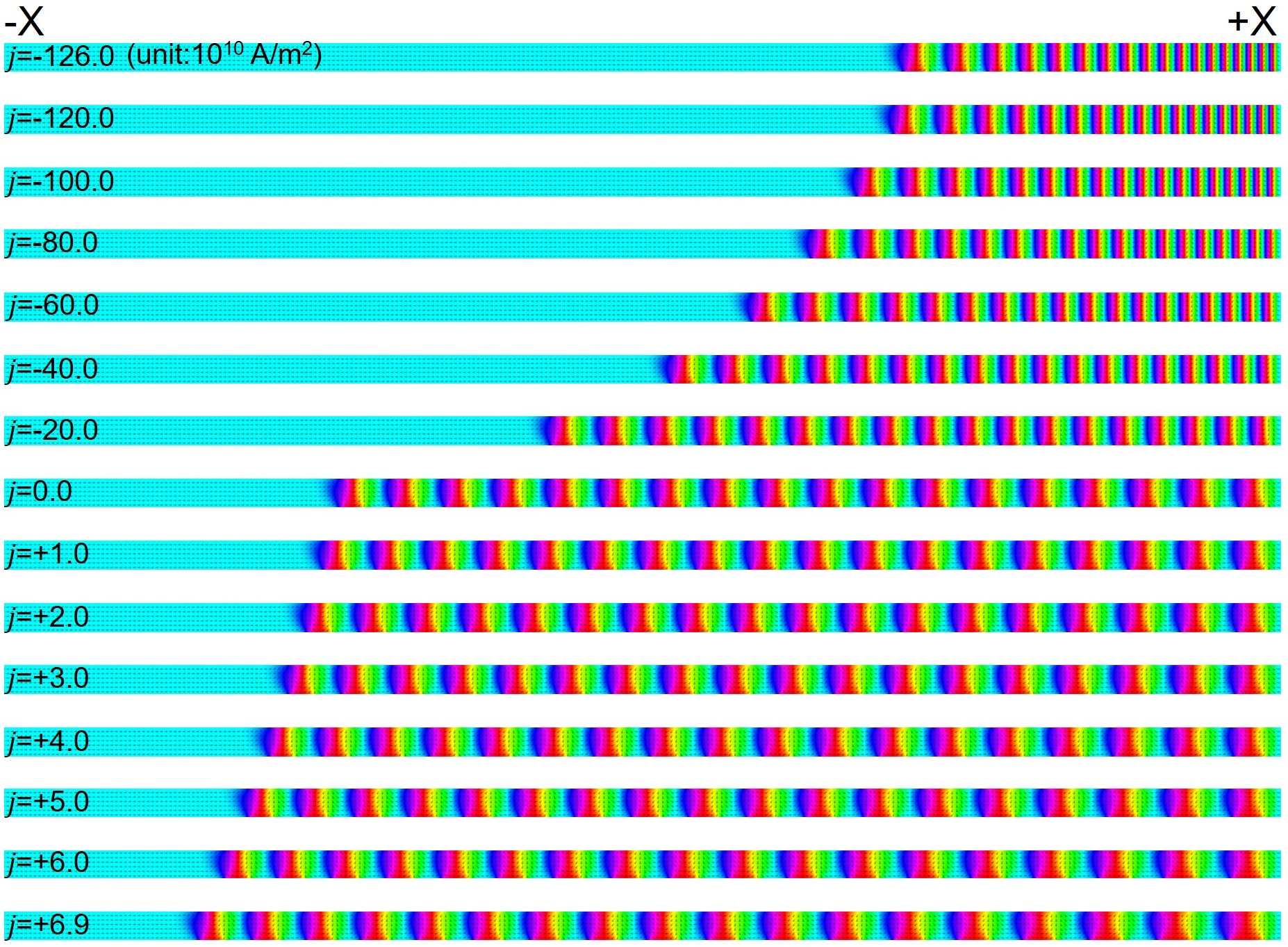}
    \caption{Schematic of $\text{DWSS}_{36}$ deformation under different spin-polarized current $j_{\text{sp}}$ injected along the DWSS axis.}
    \label{fig:placeholder}
\end{figure*}

\begin{figure*}
    \centering
    \includegraphics[width=0.8\linewidth]{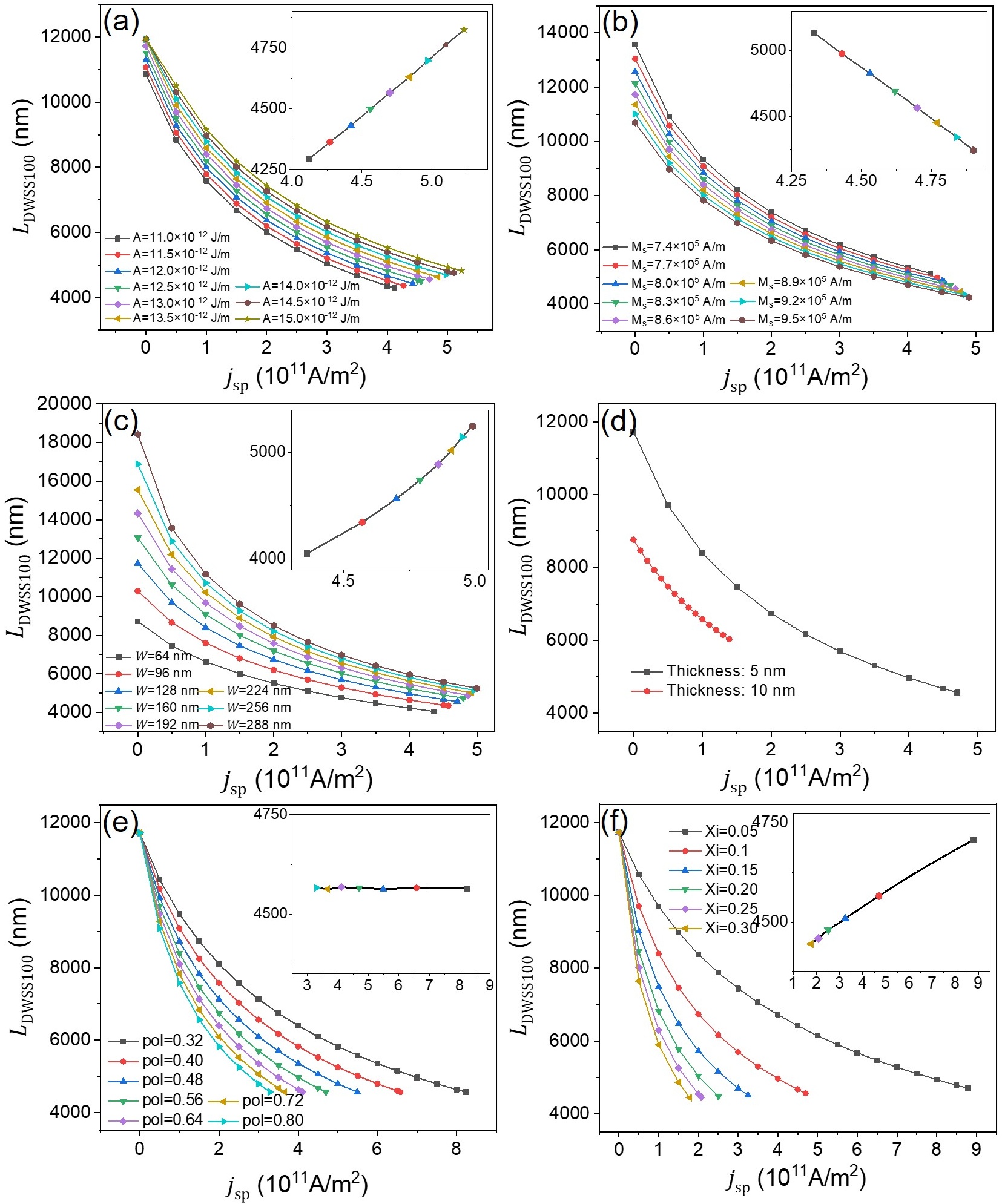}
    \caption{Overall length $L_{\text{DWSS100}}$ as a function of spin-polarized current density $j_{\text{sp}}$ for a $\text{DWSS}_{100}$ with modified material and geometric parameters: (a) exchange stiffness $A$, (b) saturation magnetization $M_{\text{S}}$, (c) nanostrip width $W_{\text{strip}}$, (d) nanostrip thickness, (e) current polarization $P$, and (f) non-adiabaticity STT coefficient $\text{X}_{\text{i}}$ (or $\beta$). $L_{\text{DWSS100}}$ denotes the overall length of $\text{DWSS}_{100}$ and $j_{\text{sp}}$ denotes spin-polarized current density. Insets in each panel highlight the critical length versus the threshold current density for compressive fracture.}
    \label{fig:placeholder}
\end{figure*}

\begin{figure*}
    \centering
    \includegraphics[width=0.55\linewidth]{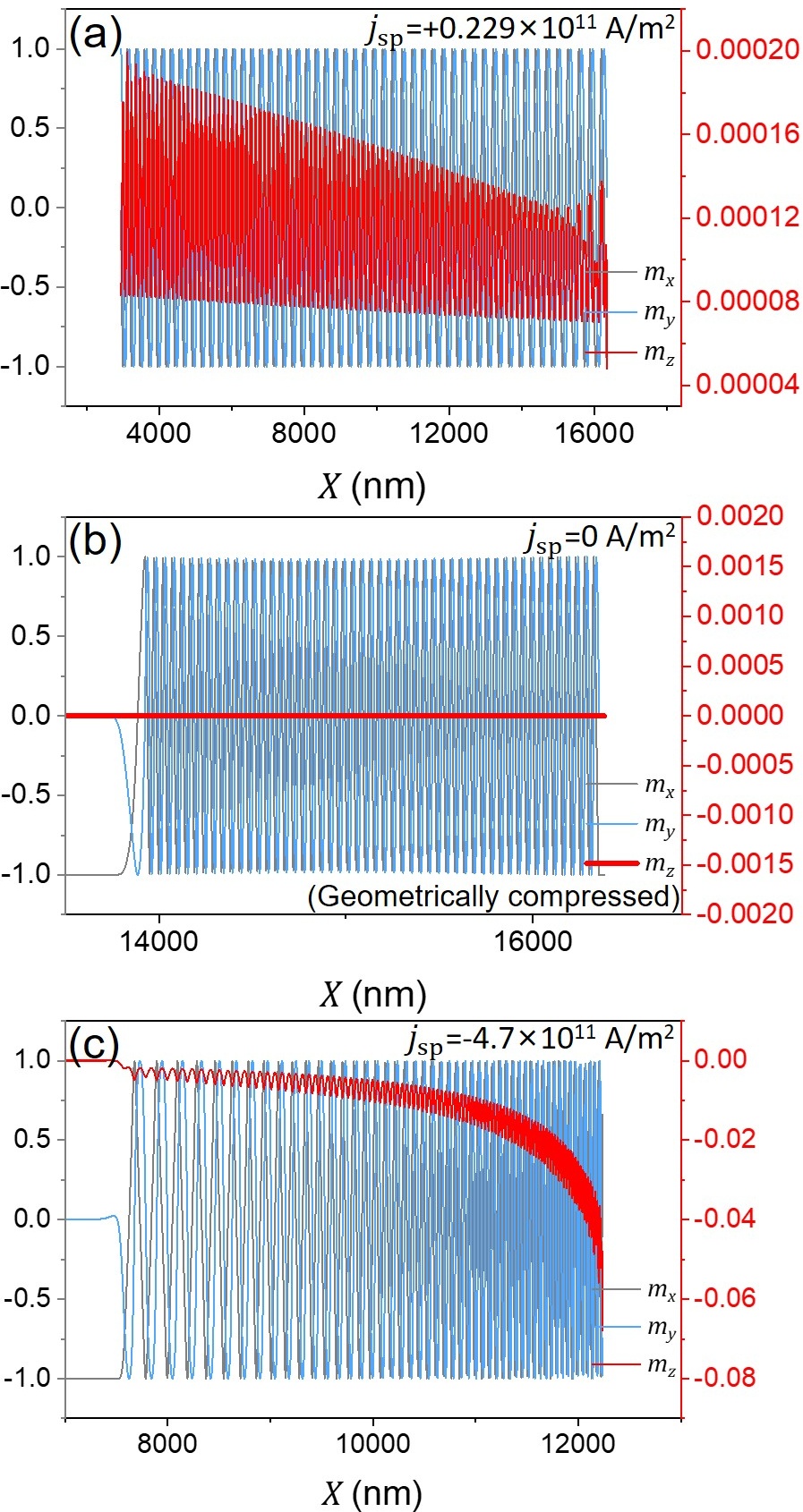}
    \caption{Out-of-the-plane tilting of magnetization in a $\text{DWSS}_{100}$ under different spin-polarized currents: (a) positive tilting ($+m_z$) for \(j_{\text{sp}} =+0.229\times 10^{11} \,\mathrm{A/m^2}\), (b) no tilting (\(m_z=0\)) for \(j_{\text{sp}} =0 \,\mathrm{A/m^2}\), and (c) negative tilting ($-m_z$) for \(j_{\text{sp}} =-4.7\times 10^{11} \,\mathrm{A/m^2}\).}
    \label{fig:placeholder}
\end{figure*}

\begin{figure*}
    \centering
    \includegraphics[width=0.5\linewidth]{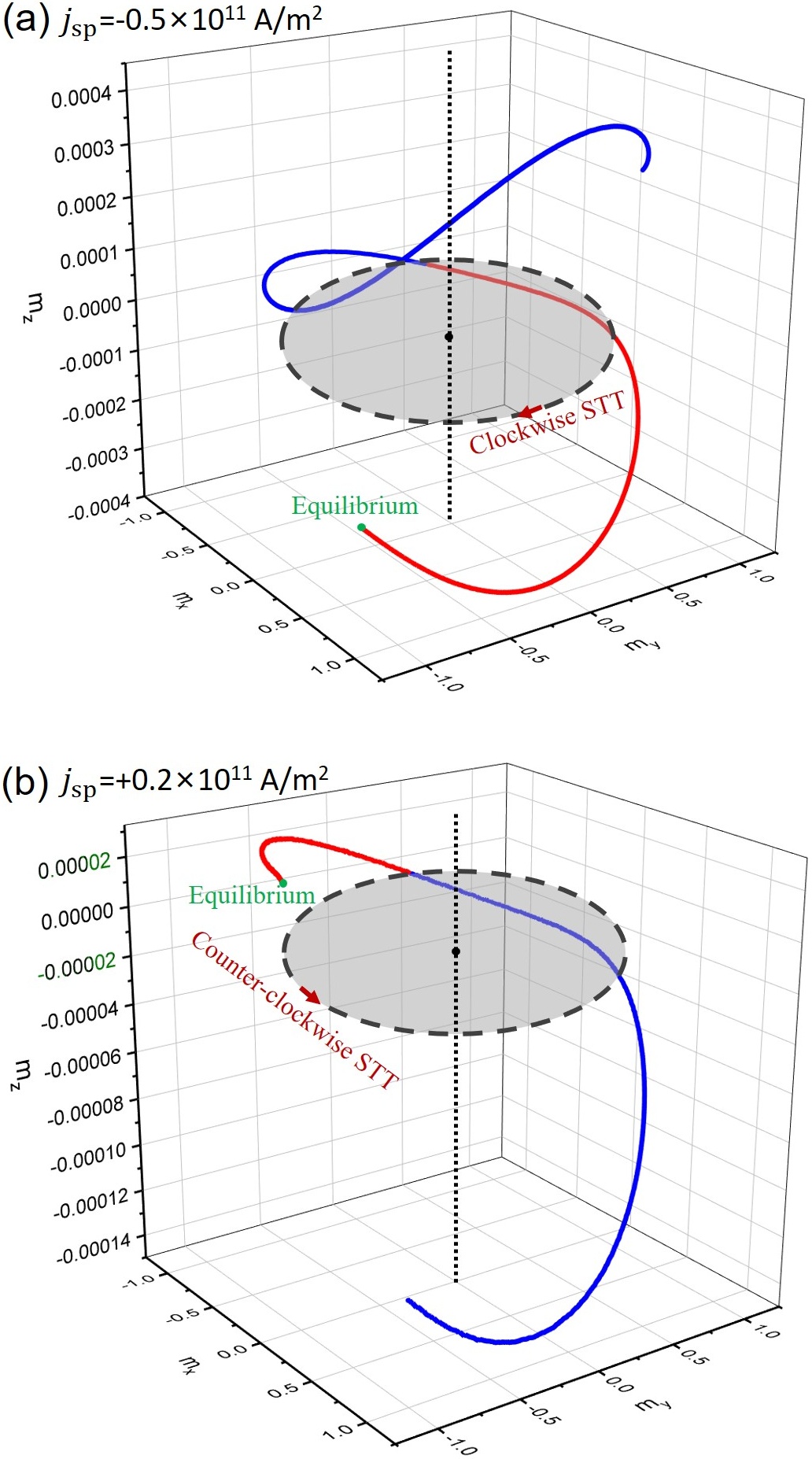}
    \caption{Dynamical evolution of magnetization tilting in a $\text{DWSS}_{100}$ under spin-polarized currents. The spin site is located at the center of the $\text{DWSS}_{100}$ with initial orientation \(\textbf{m}=(1,0,0)\). (a) Negative tilting ($-m_z$) under \(j_{\text{sp}} =-0.5\times 10^{11} \,\mathrm{A/m^2}\), and (b) positive tilting ($+m_z$) under \(j_{\text{sp}} =+0.2\times 10^{11} \,\mathrm{A/m^2}\), both evolving until the field-like torque $-m_z \times \textbf{H}_{\text{ip}}$ counterbalances the STT induced by the current. The STT-driven in-plane rotation dominates throughout the process: upon reaching negative (positive) $m_z$ in the left panel (the right panel), the spin continues rotating clockwise (counterclockwise).}
    \label{fig:placeholder}
\end{figure*}

\begin{figure*}
    \centering
    \includegraphics[width=0.65\linewidth]{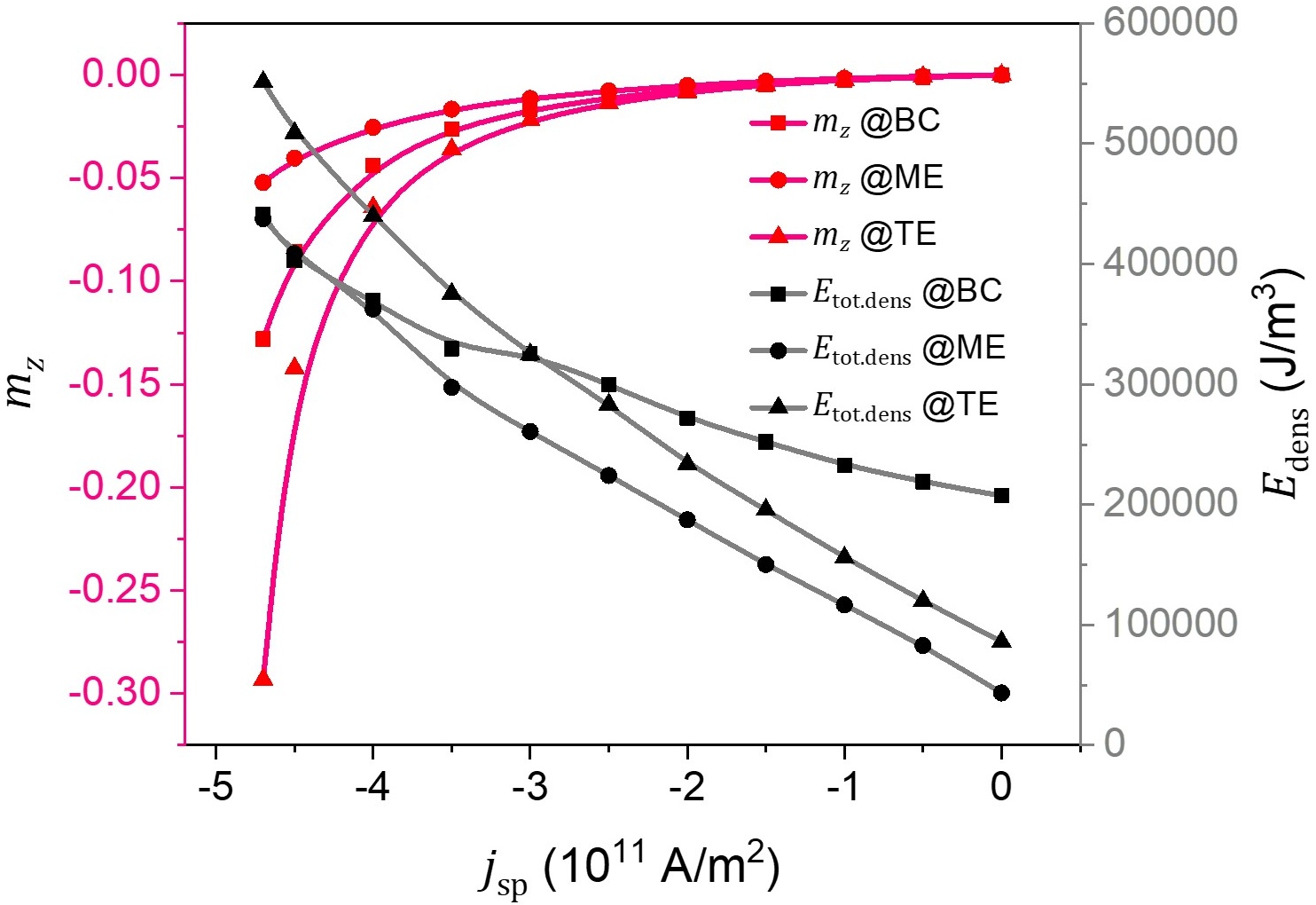}
    \caption{Out-of-plane magnetization component $m_z$ (left axis) and total energy density $E_{\text{tot.dens}}$ (right axis) as functions of spin-polarized current density $j_{\text{sp}}$ at three locations within the $100^{\text{th}}$ DW at the right end of a $\text{DWSS}_{100}$: bottom center (BC), middle edge (ME) and top edge (TE). Under large negative currents, spins at the TE exhibits the most pronounced negative tilting and the highest energy density, triggering vortex nucleation at this site.}
    \label{fig:placeholder}
\end{figure*}

\begin{figure*}
    \centering
    \includegraphics[width=0.9\linewidth]{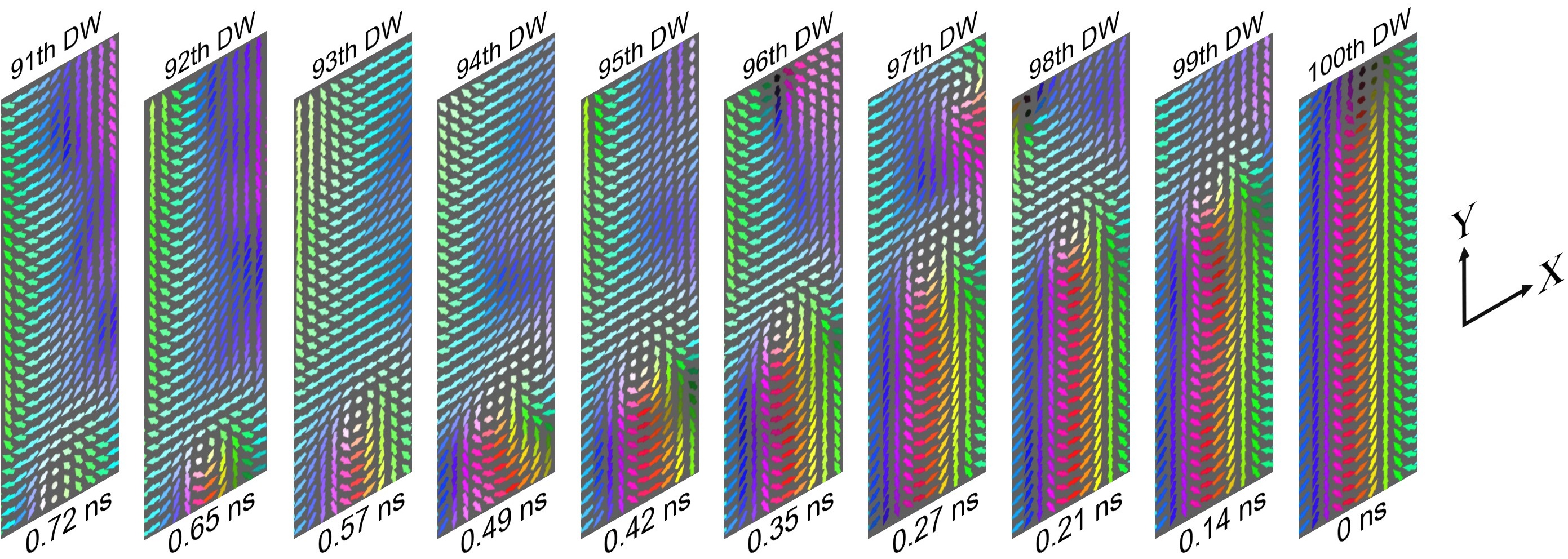}
    \caption{Vortex nucleation at the top edge (TE) of the $100^{\text{th}}$ DW in a $\text{DWSS}_{100}$ under the critical spin-polarized current \(j_{\text{crit}} =-4.7\times 10^{11} \,\mathrm{A/m^2}\), followed by its bottom-leftward migration, which triggers an avalanche collapse of the entire $\text{DWSS}_{100}$.}
    \label{fig:placeholder}
\end{figure*}

\begin{figure*}
    \centering
    \includegraphics[width=0.95\linewidth]{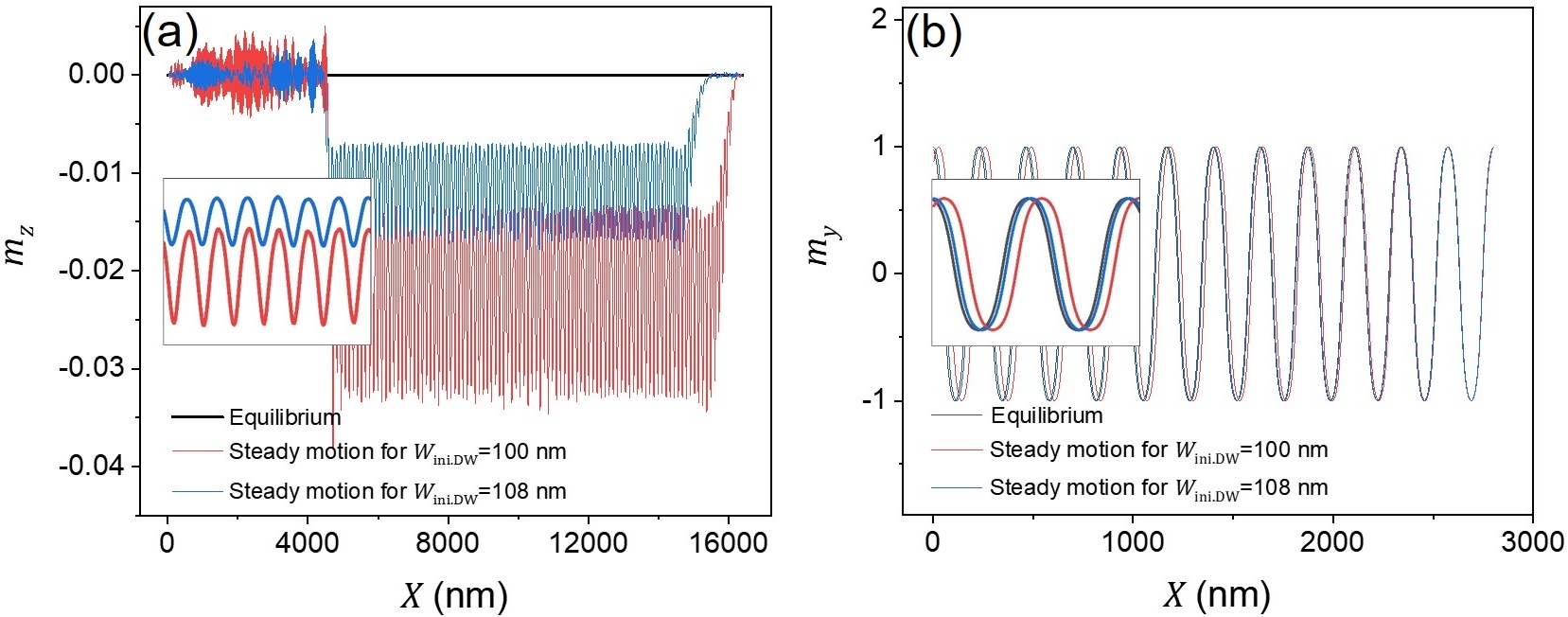}
    \caption{(a) Out-of-plane magnetization profiles $m_z$ of a $\text{DWSS}_{100}$ at equilibrium and under motion at different velocities, for two initial compression states: \(W_{\text{ini.DW}} =100\,\mathrm{nm}\) and \(W_{\text{ini.DW}} =108\,\mathrm{nm}\). The inset highlights the spatially oscillatory distribution of $m_z$. (b) Corresponding in-plane magnetization profiles $m_y$ under the same conditions. The inset shows the width variation of DWs in different dynamical states.}
    \label{fig:placeholder}
\end{figure*}

\begin{figure*}
    \centering
    \includegraphics[width=0.55\linewidth]{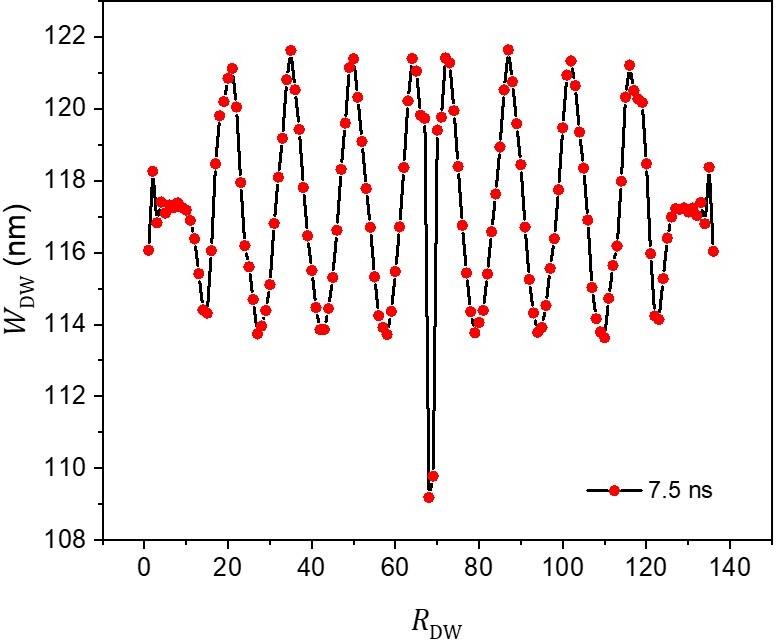}
    \caption{DW width distribution in a $\text{DWSS}_{136}$ at \(t=7.5\,\mathrm{ns}\) under an \textit{x}-directional ac magnetic field  \(\textbf{H}_{\text{ac}}=h_0\sin(2\pi ft)\hat{\textbf{x}}\) (\(f=0.5\,\mathrm{GHz}\), \(h_0=50 \,\mathrm{Oe}\)) applied to a narrow central region ($\mu_y \in(-1,1)$).}
    \label{fig:placeholder}
\end{figure*}

\clearpage
\vspace{0pt}

\section{\textbf{DWSS-PLATFORM APPLICATION PROPOSALS}}

This section demonstrates the versatility of the DWSS platform by introducing a series of proof-of-concept device designs. Their operating principles and simulated performance are examined, validating the feasibility of spin elastomer-based applications and paving the way for future experimental exploration.

\begin{figure}[htbp]
    \centering
    \includegraphics[width=0.9\linewidth]{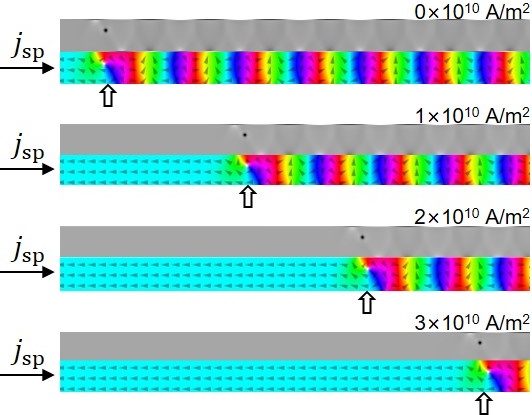}
    \caption{Vortex actuation and stray-field control in a modified $\text{DWSS}_{100}$. The leftmost transverse wall is replaced by a vortex wall following the admissible pairings in \textbf{Fig. A.2}. Grayscale images show magnetic force microscopy (MFM) maps acquired 5 nm above the nanostrip. The black dot marks the strong out-of-plane polarization of the vortex core.}
    \label{fig:placeholder}
\end{figure}

\begin{figure}
    \centering
    \includegraphics[width=0.9\linewidth]{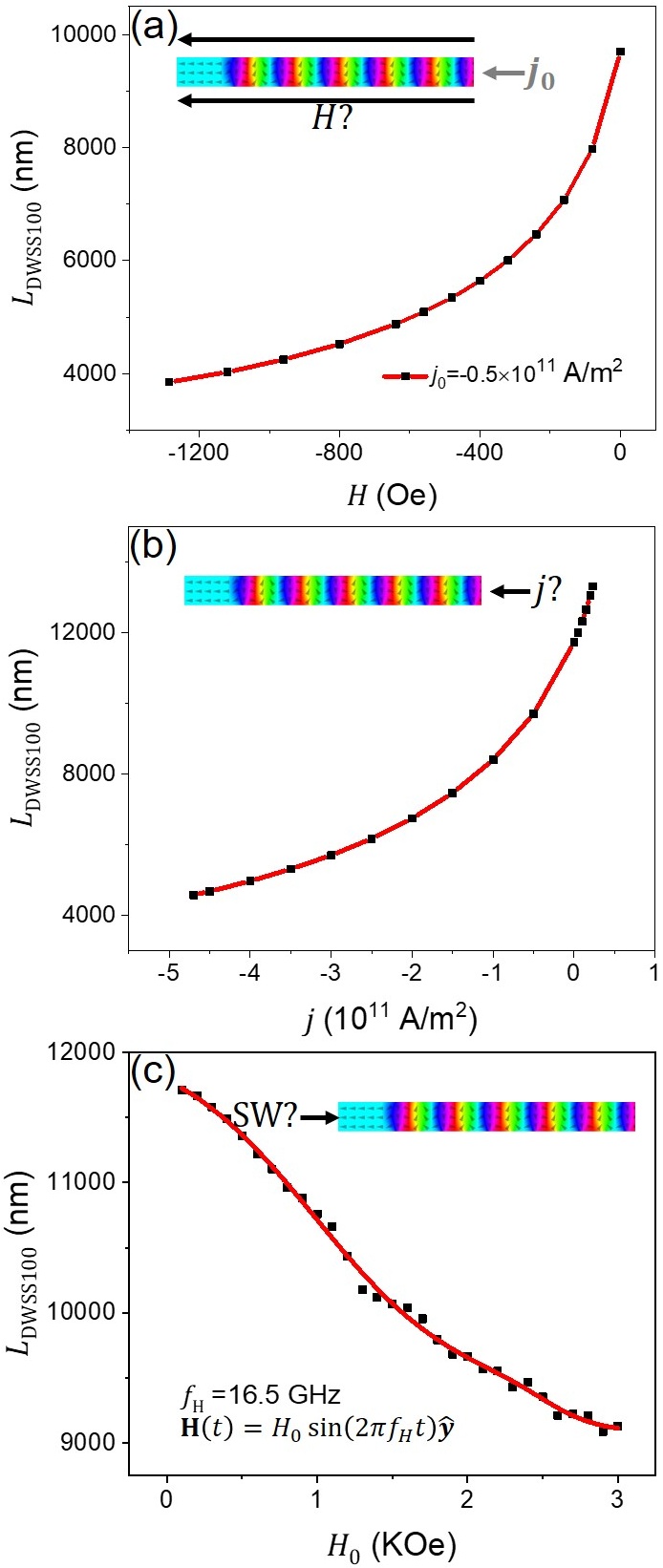}
    \caption{Length modulation of a $\text{DWSS}_{100}$ by different external stimuli. (a) Overall length $L_{\text{DWSS100}}$ as a function of an externally applied magnetic field $\textbf{H}$ aligned parallel to the DWSS axis. A spin-polarized current \(j_{\text{sp}} =-0.5\times 10^{11} \,\mathrm{A/m^2}\) is injected along the strip to maintain structural integrity. (b) $L_{\text{DWSS100}}$ as a function of spin-polarized current $j_{\text{sp}}$. (c) $L_{\text{DWSS100}}$ as a function of the amplitude $H_{0}$ of a spin-wave-exciting field \(\textbf{H}(t)=H_0\sin(2\pi f_{\text{H}}t)\hat{\textbf{y}}\) with \(f_{\text{H}}=16.5 \,\mathrm{GHz}\), applied locally 500 nm from the left end of the DWSS.}
    \label{fig:placeholder}
\end{figure}

\begin{figure*}
    \centering
    \includegraphics[width=0.7\linewidth]{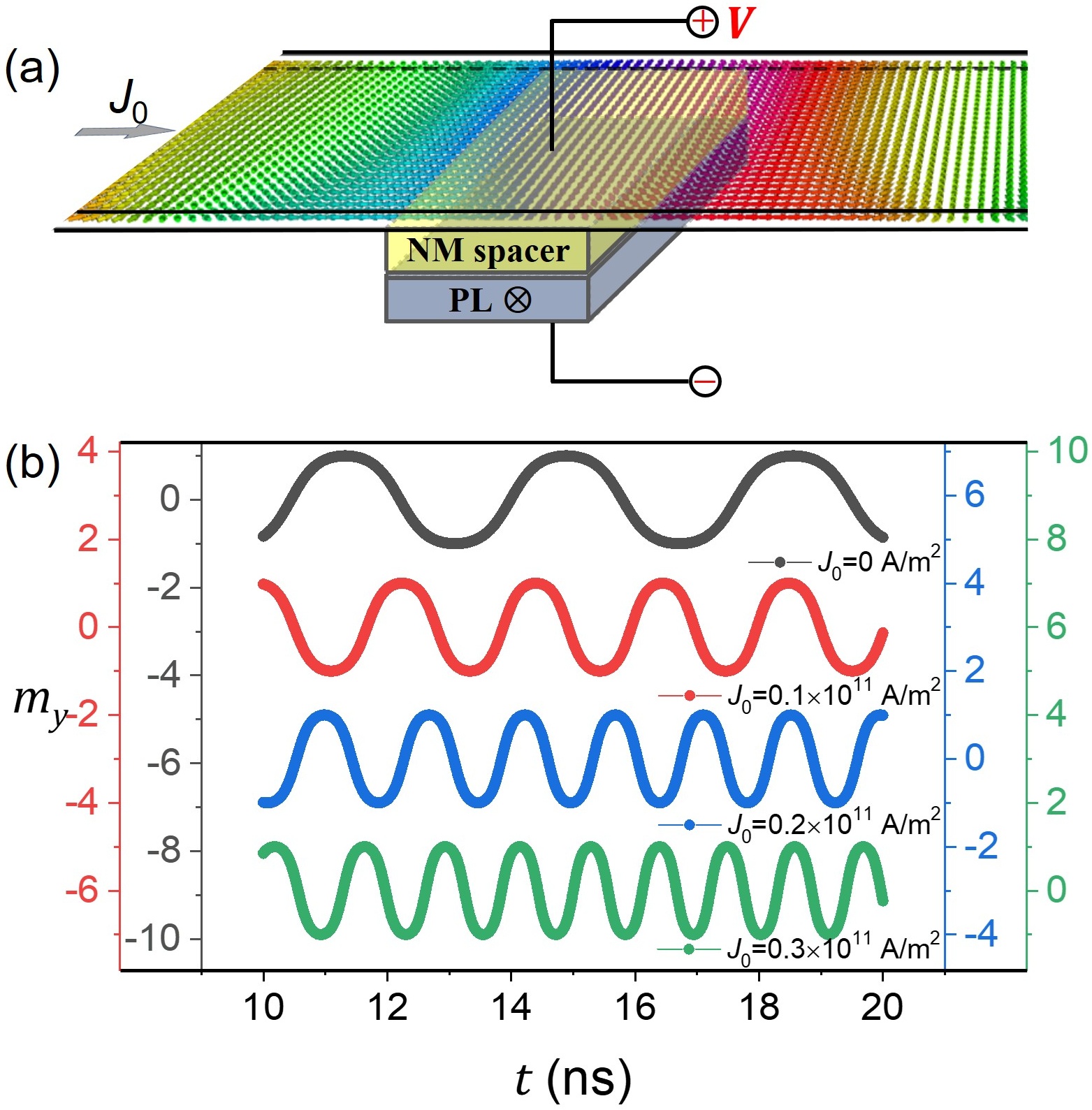}
    \caption{(a) Schematic of a DWSS-based nano-oscillator. The device consists of a fixed polarizer (PL) at the bottom, a non-magnetic (NM) spacer in the middle, and a free layer containing a DWSS on top. Injection of a spin-polarized current $J_0$ into the top layer induces oscillations of the magnetization (and hence the voltage) above the NM layer. The oscillation frequency can be dynamically tuned in real time by adjusting $J_0$. (b) Temporal oscillations of the $m_y$ component at a spin site above the NM spacer for different values of $J_0$.}
    \label{fig:placeholder}
\end{figure*}

\begin{figure*}
    \centering
    \includegraphics[width=0.5\linewidth]{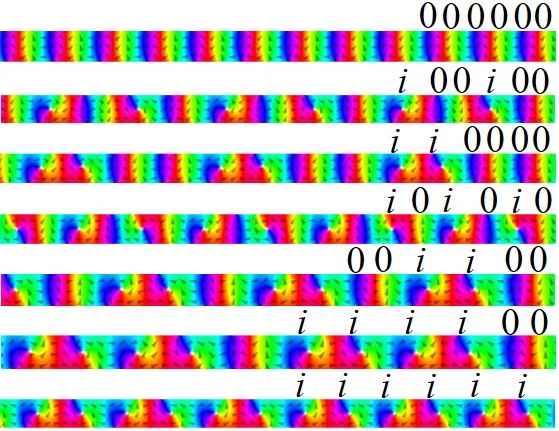}
    \caption{Schematic of a high-density DWSS racetrack memory. Data bits are encoded by domain-wall type (e.g., \(V\equiv i\), \(T\equiv 0\)). By exploiting vortex wall polarization, state $i$ can be further bifurcated into $+1$ and $-1$, enabling ternary encoding. Data rewriting is achieved through reversible vortex-transverse wall interconversion [49,50].}
    \label{fig:placeholder}
\end{figure*}

\begin{figure*}
    \centering
    \includegraphics[width=0.6\linewidth]{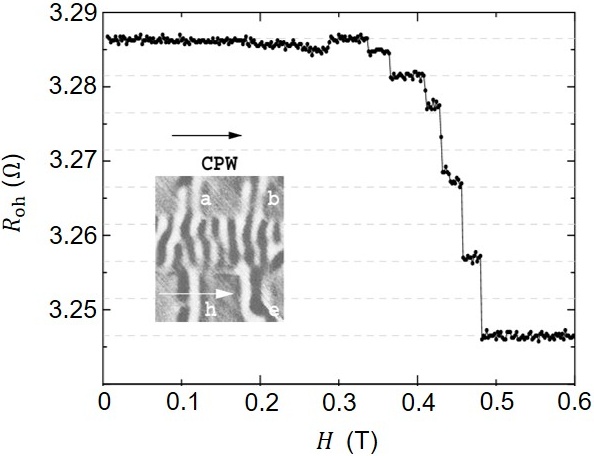}
    \caption{Excess resistance above the saturation value due to the presence of domain walls. Steps in the resistance curve correspond to the disappearance of individual DWs under increasing magnetic field. Adapted with permission from Ref. [30], Copyright (2002) American Physical Society (APS). By tuning the number and width of DWs, the resistance of a DWSS can be effectively modulated [30,51].}
    \label{fig:placeholder}
\end{figure*}

\begin{figure*}
    \centering
    \includegraphics[width=0.7\linewidth]{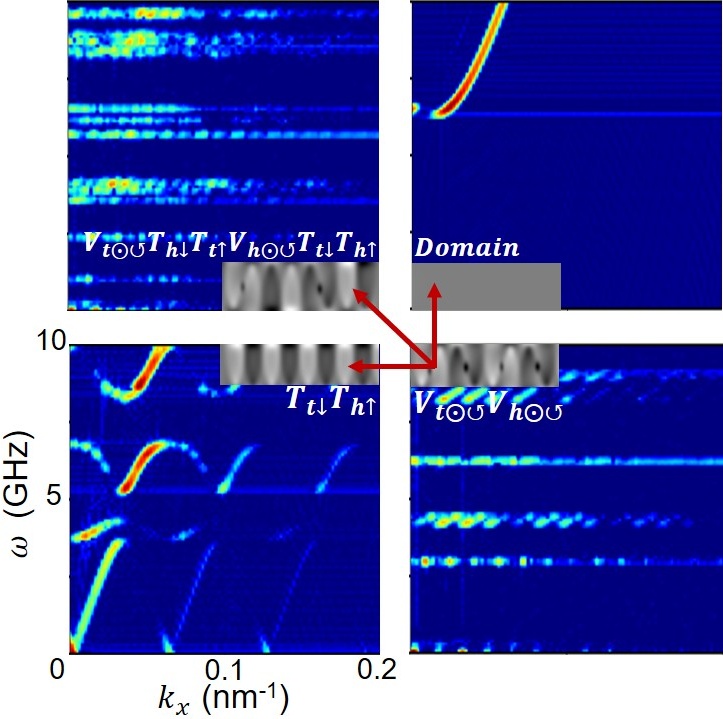}
    \caption{Magnonic dispersion engineering via phase transitions of a DWSS, enabled by topological soliton transformation and annihilation. Shown are the phase transitions among the following unit-cell configurations:  $V_{t\odot \circlearrowleft}V_{h\odot \circlearrowleft}$, $T_{t\downarrow}T_{h\uparrow}$, $V_{t\odot \circlearrowleft}T_{h\downarrow}T_{t\uparrow}V_{h\odot \circlearrowleft}T_{t\downarrow}T_{h\uparrow}$ and a single domain.   }
    \label{fig:placeholder}
\end{figure*}

\begin{figure*}
    \centering
    \includegraphics[width=0.9\linewidth]{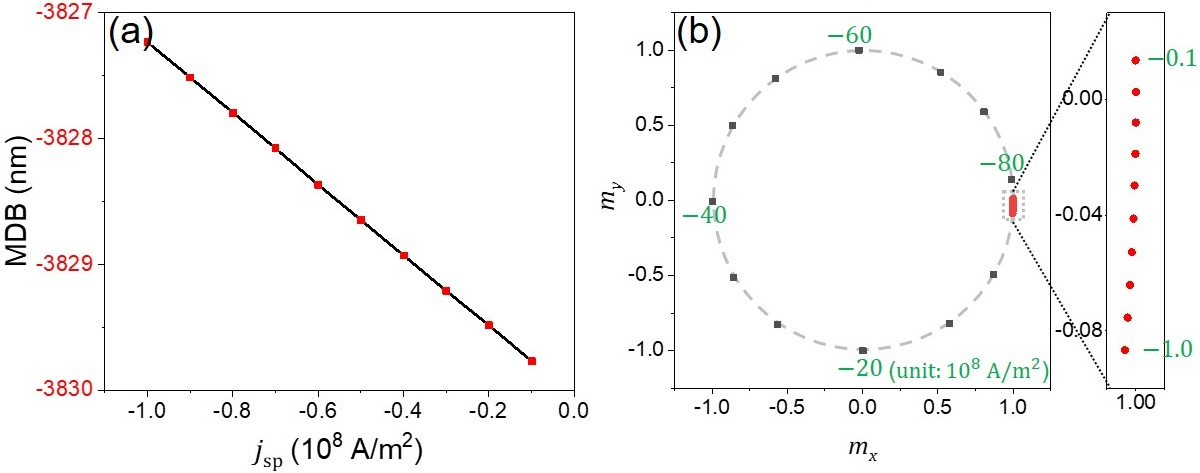}
    \caption{(a) High-precision positioning ($<1\,\mathrm{nm}$) of a magnetic domain boundary (MDB)—equivalently the left boundary of a $\text{DWSS}_{100}$. (b) High-precision vectorial magnetization control (\(\Delta m_y <0.01\)) at specific spin sites using a spin-polarized current of density \(j_{\text{sp}} =1\times 10^{8} \,\mathrm{A/m^2}\). In the absence of thermal and quantum fluctuations, no fundamental limit to the achievable modulation precision is expected.}
    \label{fig:placeholder}
\end{figure*}

\clearpage
\vspace{0pt}

\nocite{*}

\bibliography{apssamp}

\end{document}